\documentclass[twocolumn, pra, superscriptaddress]{revtex4-1}
\usepackage{upgreek}
\usepackage{tikz}
\usepackage{dcolumn}
\usepackage{color,tcolorbox}
\usepackage{amsmath,mathrsfs,amssymb,esint}
\usepackage{dcolumn}
\usepackage{bm}
\usepackage{blindtext}
\usepackage{natbib}
\usepackage{graphicx}
\usepackage{rotating}
\usepackage{titlesec}
\usepackage{multirow}
\usepackage{mathtools}
\usepackage{soul}
\usepackage{amsmath}
\usepackage[pdfborder={0 0 0},colorlinks=true,linkcolor=blue,urlcolor=blue,citecolor=blue]{hyperref}
\colorlet{orange1}{green!10!orange!90!}
\usepackage{enumitem}

\definecolor{ForestGreen}{rgb}{0,0.5,0}

\def\be{\begin{equation}}
\def\ee{\end{equation}}
\def\bea{\begin{eqnarray}}
\def\eea{\end{eqnarray}}
\def\bi{\begin{itemize}}
\def\ei{\end{itemize}}

\usepackage[T1]{fontenc}

\definecolor{ao(english)}{rgb}{0.0, 0.5, 0.0}
\begin{document}
\title{Kibble-Zurek Dynamics in a Trapped Ultracold Bose Gas}
\author{I-Kang Liu}
\affiliation{Joint Quantum Centre (JQC) Durham--Newcastle, School of Mathematics, Statistics and Physics, Newcastle University,Newcastle upon Tyne, NE1 7RU, United Kingdom}
\affiliation{Department of Physics and Graduate Institute of Photonics, National Changhua University of Education, Changhua 50058, Taiwan}
\author{Jacek Dziarmaga} 
\affiliation{Institute of Theoretical Physics, Jagiellonian University, 
ul. \L{}ojasiewicza 11, 30-348 Krak\'ow, Poland}
\author{Shih-Chuan Gou}
\affiliation{Department of Physics and Graduate Institute of Photonics, National Changhua University of Education, Changhua 50058, Taiwan}
\author{Franco Dalfovo}
\affiliation{INO-CNR BEC Center and Dipartimento di Fisica, Universit\`a di Trento, 38123 Trento, Italy}
\author{Nick P. Proukakis}
\affiliation{Joint Quantum Centre (JQC) Durham--Newcastle, School of Mathematics, Statistics and Physics, Newcastle University,Newcastle upon Tyne, NE1 7RU, United Kingdom}

\begin{abstract}
The dynamical evolution of an inhomogeneous ultracold atomic gas quenched at different controllable rates through the Bose-Einstein condensation phase transition is studied numerically in the premise of a recent experiment in an anisotropic harmonic trap. Our findings based on the stochastic (projected) Gross-Pitaevskii equation are shown to be consistent at early times with the predictions of the homogeneous Kibble-Zurek mechanism. This is demonstrated by collapsing the early dynamical evolution of densities, spectral functions and correlation lengths for different quench rates, based on an appropriate characterization of the distance to criticality felt by the quenched system. The subsequent long-time evolution, beyond the identified dynamical critical region, is also investigated by looking at the behaviour of the density wavefront evolution and the corresponding phase ordering dynamics.
\end{abstract}

\maketitle

\section{Introduction}

The Kibble-Zurek (KZ) mechanism originated from the scenario for defect creation in cosmological symmetry-breaking phase transitions \cite{K-a, *K-b, *K-c}. As the Universe cools, causally disconnected regions choose symmetry-breaking vacuum independently. The randomly oriented domains result in topologically nontrivial configurations that survive as topological defects. This general scenario was substantiated with a dynamical theory \cite{Z-a, *Z-b, *Z-c, Z-d} that predicts the size of the domains, and therefore also the initial density of defects, employing critical exponents of the transition and the quench time $\tau_Q$. The KZ mechanism has been numerically studied across diverse condensed matter systems, including superconducting junction arrays and holographic superconductors, superfluid $^{3}$He and $^{4}$He, and driven-dissipative exciton-polaritons \cite{KZnum-a,KZnum-b,KZnum-c,KZnum-d,KZnum-e,KZnum-f,KZnum-g,inhomo_classical-c,QKZteor-i,KZnum-h,KZnum-i,KZnum-j,KZnum-k,KZnum-l,KZnum-m,KZexp-w,Su13,McDonald2015,Liu_2016,bland_marolleau_20,zamora_20}. There have been numerous supportive laboratory experiments \cite{KZexp-a,KZexp-b,KZexp-c,KZexp-d,KZexp-e,KZexp-f,KZexp-g,KZexp-i,KZexp-j,KZexp-k,KZexp-l,KZexp-m,KZexp-n,KZexp-o,KZexp-p,KZexp-q,KZexp-s,KZexp-t,KZexp-x}, including recent ones in the context of ultracold atomic gases across different geometries and dimensionalities~\cite{KZexp-gg,KZexp-h,lamporesi2013spontaneous,KZexp-r,dalibard-ring-KZ-2014,KZexp-v,
KZexp-t,
KZexp-u,donadello2016creation,shin-KZ-Fermi-2019}. In recent years, the KZ mechanism has been generalized to quantum phase transitions \cite{Polkovnikov2005,QKZ1,QKZ2,d2005,d2010-a, d2010-b}. Theoretical developments~\cite{QKZteor-a,QKZteor-b,QKZteor-c,QKZteor-d,QKZteor-e,QKZteor-f,QKZteor-g,QKZteor-h,QKZteor-i,QKZteor-j,QKZteor-k,QKZteor-l,QKZteor-m,QKZteor-n,QKZteor-o,QKZteor-p,QKZteor-q,QKZteor-r,QKZteor-s,QKZteor-t,Keita2018a,Keita2018b,Keita2018c,Lamacraft2007,Lee2009,Sabbatini2011,Swistock2013,Saito2013,Witkowska2013} and experimental tests \cite{KZexp-gg,
QKZexp-b, QKZexp-c, QKZexp-d, QKZexp-e, QKZexp-f, QKZexp-g,deMarco2,Lukin18,Qiu2020} of the quantum KZ mechanism followed, with a recent experiment \cite{Lukin18} emulating a quantum Ising chain in the transverse field using Rydberg atoms being fully consistent with the predicted scaling \cite{QKZ2,d2005}. 


\begin{figure}[b!]
\centering
\includegraphics[width=1\linewidth]{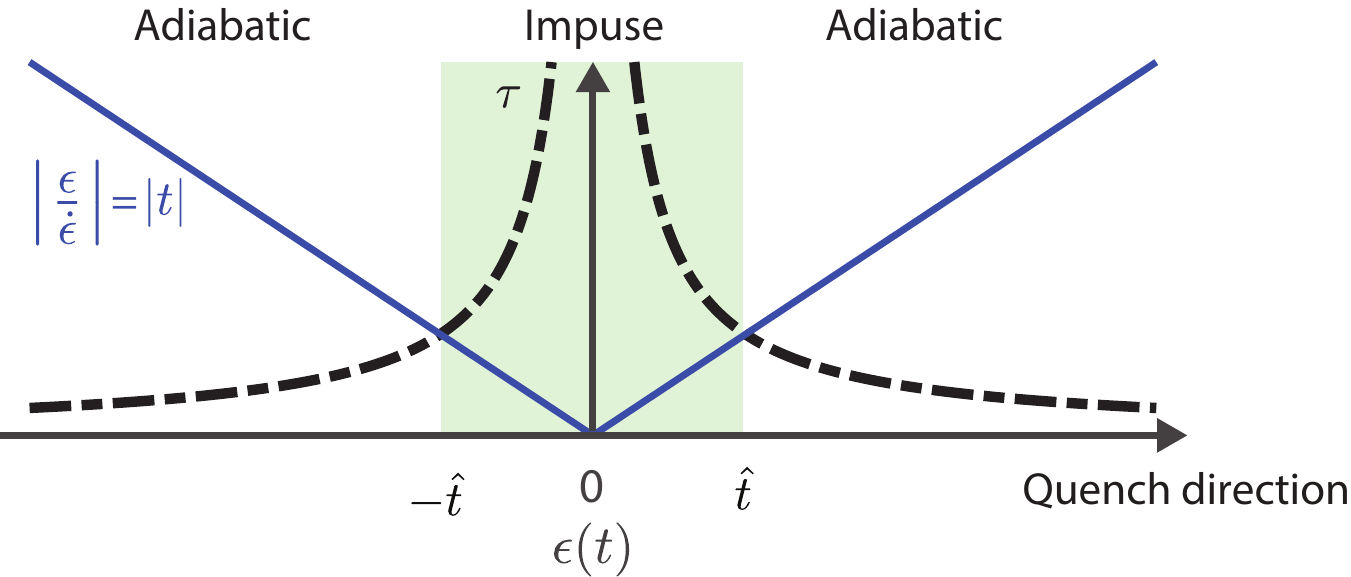}
\caption{
Schematic of the homogeneous KZ mechanism, marking the interplay between the diverging system relaxation time $\tau$ (black dashed line) and the time $|t|=|\epsilon/\dot{\epsilon}|$ to the transition (solid blue lines). The intersection points of these two curves mark the crossover times $-\hat t$ and $+ \hat t$.
}
\label{fig_1}
\end{figure}

In this paper we focus on the evolution of a trapped ultracold atomic gas across the transition to a Bose-Einstein condensate. We perform a detailed numerical analysis of externally-driven spontaneous symmetry breaking and dynamical growth of an elongated, harmonically confined, three-dimensional (3D) condensate by solving the stochastic (projected) Gross-Pitaevskii equation (SPGPE) in realistic experimental parameter regimes, previously identified in our quantitative analysis of the late-time relaxation dynamics probed experimentally \cite{lamporesi2013spontaneous,donadello2016creation,KZexp-r,KZexp-w}. 

Our numerical results are interpreted in terms of the homogeneous KZ mechanism by comparing the solutions of the full 3D stochastic nonlinear equation against analytical predictions of the linearized limit of the same equation. At short times from the transition, where the system is close to criticality, we find excellent agreement with the KZ scaling laws predicted by the linearized theory, with our numerical curves for different quench timescales appropriately collapsing onto a unified curve. In particular, the growth of the condensate is delayed with respect to the critical point by a delay time proportional to the KZ timescale. Remarkably, we also find that the KZ delay persists at later times as long as the system is ramped linearly in time. Specifically, density growth is found to occur along elliptically-expanding regions of phase-space which mimic the underlying trap geometry, with the rescaled expanding wavefronts collapsing to a single (non-universal) curve for different quench rates. 

Despite the inhomogeneous nature of the harmonically trapped gas, our present work seems to indicate that the temperature quenches probed in the experiments \cite{lamporesi2013spontaneous,donadello2016creation,KZexp-r,KZexp-w} were such that transition effectively occurs within the remit of the `homogeneous' KZ mechanism; the predicted modifications due to the interplay of causality and geometry \cite{inhomo_classical-a,inhomo_classical-c,inhomo_classical-f} seem not to be needed in this case.

\section{Quenched Protocol and Modeling}

\subsection{Temperature quench and KZ mechanism}

The gas is initially prepared in a thermal state above the critical temperature $T_c$ and then ramped across the phase transition, where a symmetry breaking occurs and an order parameter appears. During such evolution, the effective distance from the critical point can be measured by a dimensionless parameter $\epsilon=1-T/T_c$. Close to the critical point this parameter can be assumed to be linear in time, as
\begin{equation}
\epsilon(t)=\frac{t}{\tau_Q}\ , 
\label{epsilont}
\end{equation}
where $\tau_Q$ is referred to as the quench time. While the system approaches $T_c$ from above, but still far from it, the evolution is adiabatic, i.e., the gas follows its adiabatic thermal equilibrium state. Such adiabaticity fails at a characteristic time, $-\hat t$, when the relaxation time becomes longer than the instantaneous timescale $|\epsilon/\dot \epsilon| = |t|$ at which $\epsilon$ is ramped. The relaxation time diverges as $|\epsilon|^{-z\nu}$, where $\nu$ and $z$ are the equilibrium (correlation length) and dynamical critical exponents, respectively. From the equation $|t|\simeq |t/\tau_Q|^{-z\nu}$ one obtains 
\be 
\hat t\propto \tau_Q^{z\nu/(1+z\nu)} \ ,
\label{hattgeneral}
\ee 
which corresponds to a deviation from criticality
\begin{equation}
\hat\epsilon=\frac{\hat t}{\tau_Q}\propto\tau_Q^{-1/(1+z\nu)} \ .
\end{equation}

In the `cartoon' version of the homogeneous KZ mechanism (see Fig.~\ref{fig_1}), during the system evolution, started at large negative initial value of $\epsilon$, the state of the gas freezes-out at $-\hat t$ and subsequently remains unchanged until a time $+\hat t$, when the adiabatic evolution starts again. During that period, the correlation length $\xi$ is frozen at the value $\hat{\xi}$ of the equilibrium correlation length at $-\hat\epsilon$, given by
\begin{equation}
\hat\xi \propto \hat\epsilon^{-\nu} \propto \tau_Q^{\nu/(1+z\nu)} \ . 
\label{hatxigeneral}
\end{equation}
The above scenario (adiabatic-impulse-adiabatic approximation) is of course a simplification of the actual dynamics, as physical quantities still evolve during the time the system spends in the critical (impulse) region, as qualitatively demonstrated e.g.~in Ref.~\cite{KZexp-w}, and characterized in detail in Ref.~\cite{sonic}. However, the notable importance of the simplistic KZ mechanism, which also explains its broad applicability to a range of different physical systems, is that it correctly predicts the scaling of the characteristic lengthscale $\hat\xi$ and the timescale $\hat t$ with the quench time $\tau_Q$. It is noteworthy that the two scales are related by
\be 
\hat t\simeq \hat\xi^z.
\label{hatthatxi}
\ee 
They both diverge in the adiabatic limit, $\tau_Q\to\infty$, where they become the unique relevant scales in the KZ scaling ansatz~\cite{KZscaling1,KZscaling2,Francuzetal}. For instance, a two-point correlation function $C_R(t)$, between two points separated by a distance $R$, should satisfy
\be 
\hat\xi^{d-2+\eta}\, C_R(t) = G\left(t/\hat\xi^z,R/\hat\xi\right).
\label{CRscaling}
\ee
Here $d$ is the number of dimensions, $\eta$ is a universal critical exponent, and $G$ is a non-universal scaling function. Equation~\eqref{CRscaling} is expected to be accurate in the long-wavelength and low-frequency limit. The adiabatic-impulse-adiabatic approximation is consistent with the scaling hypothesis , Eq.~\eqref{CRscaling}, but it implies a particular form of the scaling function $G$ that does not depend on $t/\hat\xi^z$ during the freeze-out between $-\hat t$ and $+\hat t$.

\subsection{Stochastic Projected Gross-Pitaevskii Equation}

The dynamical quench of a weakly interacting ultracold Bose gases across the phase transition can be modelled by classical-field simulations \cite{Bradley2008,blakie2008dynamics,proukakis2008finite,Rooney2013,Proukakis13,berloff_brachet_14,Brewczyk2007,KZnum-j}, and, in particular, by the stochastic (projected) Gross-Pitaevskii equation~\cite{
Stoof01,Su13,KZexp-h,KZexp-w,KZnum-k,Proukakis03,Proukakis_Schmiedmayer_2006,Proukakis09,inhomo_classical-c,QKZteor-i,KZnum-k,De_2014,Liu_2016,gallucci2016engineering,Kobayashi_16a,Kobayashi_16b,Eckel_2018,comaron2019quench,bland_marolleau_20},
with many such works demonstrating notable quantitative success when directly compared against experimental observations of quenched phase transitions~\cite{Stoof01,KZexp-h,
KZexp-w,bland_marolleau_20}, and related dynamical settings \cite{blakie2008dynamics,Rooney2013}, making this approach ideal in the present context.

The stochastic projected Gross--Pitaevskii equation (SPGPE) models the dynamics of the low-lying highly-populated `classical field' $\psi$, through
\cite{blakie2008dynamics}
\begin{equation}
d\psi=\mathcal{P}\left[
-\frac{i}{\hbar}\mathcal{L}+\frac{\gamma}{\hbar}\left(\mu-\mathcal{L}\right)
\right]\psi dt+dW \ ,
\label{eq:SPGPE}
\end{equation}
where
\begin{equation}
\mathcal{L}=-\frac{\hbar^2\nabla^2}{2M}+V_\textrm{trap}(\mathbf{r})+g|\psi|^2 \ ,
\label{calL}
\end{equation}
is the Gross-Pitaevskii term including single-particle evolution and mean field potential (nonlinearity), and $dW$ denotes complex Gaussian white noise with a correlator
\begin{equation}
\langle dW^\ast(\mathbf{r},t)dW(\mathbf{r}^\prime,t^\prime)\rangle=\frac{2\gamma k_BT}{\hbar}\delta(\mathbf{r}-\mathbf{r}^\prime)\delta(t-t^\prime)dt \ .
\end{equation}
The projection operator $\mathcal{P}$ in Eq.~(\ref{eq:SPGPE}) restricts the dynamics below the energy cutoff $E_{\rm cut}$, which is fixed here as $2.5\mu_f$, where $\mu_f$ is the chemical potential of the gas at the end of the quench. The interaction strength $g=4\pi\hbar^2a_s/M$ is set by the $s$-wave scattering length $a_s$. The dimensionless parameter $\gamma$ controls the rate of relaxation of the classical field modes to the equilibrium state set by the chemical potential $\mu$ and temperature $T$ of the reservoir of atoms located above the cutoff, which are treated as a heat bath.

The detailed numerical simulations performed in this work are based on earlier stochastic dynamics simulated by some of us in different geometries, dimensionalities, platforms and systems~\cite{KZexp-w,bland_marolleau_20,Liu_2016,Ota_2018,comaron_18,comaron2019quench,zamora_20} (and related work~\cite{Proukakis03,Proukakis_Schmiedmayer_2006,Proukakis09,Cockburn_2012,gallucci2016engineering}).

\subsection{Parameter choice and quench protocol}
\label{qp}

Our study is performed for the parameters corresponding to a recent experiment \cite{KZexp-w,donadello2016creation}, performed with few$\times 10^{7}$ $^{23}$Na atoms in the $|F, \,m_F \rangle = |1,\,-1 \rangle$ state (with $a_s=2.91$~nm), trapped in an anisotropic harmonic potential, $V_\textrm{trap}(\mathbf{r})=(1/2)M[\omega_x^2x^2+\omega_\perp^2(y^2+z^2)]$, with longitudinal and transversal trap frequencies $\omega_x = 2 \pi \times 13$~Hz, $\omega_\perp = 2 \pi \times 131.4$~Hz, yielding a highly elongated 3D system.
In the experiment, after creating a thermal cloud above the critical temperature, evaporative cooling is used to ramp the temperature down in an approximately linear manner to much below $T_c$, where the system exhibits significant condensation. The experiment performed a detailed study of the late-time evolution of vortex defects originally generated during the symmetry-breaking phase transition~\cite{donadello2016creation}, finding a power-law decay within the range expected by the KZ mechanism. 
Our previous SPGPE simulations (conducted by means of the quench protocol discussed in Eq.~(\ref{protocol}) below) \cite{KZexp-w} were found to be in good agreement with observations in the late-time regime where experimental data were available; however, experimental limitations could not facilitate such quantitative analysis of the system dynamics at earlier times. Here we focus on the early-time regime of the condensate formation. Our starting point is a better numerical estimate of the distance to criticality $\epsilon$ which enables us to cast the dynamics in the standard language of the KZ mechanism and characterize it in terms of $\hat t$.

In our SPGPE simulations, after initially equilibrating the system via Eq.~(\ref{eq:SPGPE}) to the desired initial thermal state defined by its chemical potential $\mu$ and temperature $T$, we linearly vary $T$ and $\mu$ over a timescale 2$\tau_Q$, with the ramp initiated at $t=-\tau_Q$ and finished at $t=\tau_Q$, based on the imposed quench protocol
(for $|t|\leq\tau_Q$)
\begin{equation} 
T(t)=T_0 - \Delta T \, \frac{t}{\tau_Q} \ \ {\rm and }\ \ 
\mu(t)=\mu_f \, \frac{t}{\tau_Q} \ .
\label{protocol}
\end{equation}
This `hybrid' quench protocol was already discussed and implemented in
Ref.~\cite{KZexp-w}, where it was found to lead to good agreement with available experimental observations~\cite{donadello2016creation}. The rationale for such a protocol is to account for both the experimentally observed decreasing temperature and decreasing atom number numerically within the context of the SPGPE:
following Ref.~\cite{KZexp-w}, initial and final values for both these parameters were chosen to match typical experimental numbers, giving $T_0=500$~nK, $\Delta T=290$~nK, and $\mu_f=22\hbar\omega_\perp$. These correspond to initial and final temperature and atom number combinations ($T_i=790$nK, $N_i=22 \times 10^6$) and ($T_f=210$nK, $N_f=6.6 \times 10^6$), noting that the above atom numbers refer to total atoms, i.e.~also explicitly including above cut-off
atoms (under the usual assumption that they are static -- see below).
Note that, after the end of this linear ramp at $t=\tau_Q$, the `input' parameters $T$ and $\mu$ remain fixed at their final values. 

Equation~(\ref{eq:SPGPE}) is solved dynamically in a $\approx314.1\times34.9^2$~$\mu$m$^{3}$ cuboid box with $1170\times130^2$ grid points (with a grid size $\approx0.27\;\mu$m in all directions) using a plane-wave basis, with nearly 1.3 million modes 
below the cutoff, with occupation at the cutoff $n_{\rm cut} \gtrsim 1$.

\begin{figure*}[t!]
\centering
\includegraphics[width=1\linewidth]{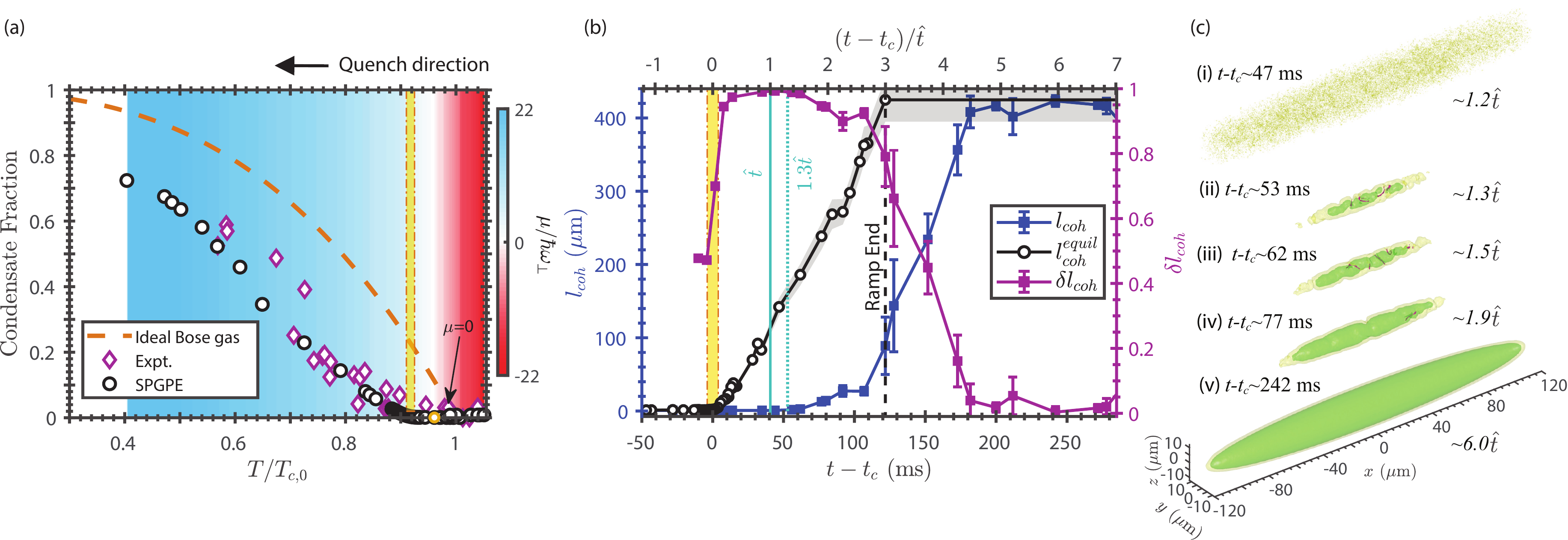}
\caption{
(a) SPGPE results (black circles) and experimental data~\cite{Ferrari_private} (purple diamonds) for the equilibrium condensate fraction $N_0/N$ vs. $T_{c,0}$. The dashed orange curve shows the ideal Bose gas prediction as a reference. The vertical yellow band marks the location of the numerically-identified critical region in SPGPE simulations, $T_c\sim445.5\pm7.3$~nK; in this range, the total particle number is $(5.6 \pm 0.1)\times10^6$, which corresponds to the ideal gas critical temperature $T_{c,0}=(488 \pm 5)$~nK. 
Background colour indicates the value of the chemical potential at each temperature during a quench, with $\mu(t)$ and $T(t)$ proceeding from the rightmost to the leftmost point.
(b) Filled blue squares: longitudinal correlation length $l_{coh}$, extracted as in Eq.~(\ref{lcoh}), during a SPGPE simulation of a quench with $\tau_Q=150$ms. Time is measured from the equilibrium critical time $t_c$, i.e., the centre of the yellow vertical band (the same as in panel (a)). Open black circles: same quantity calculated in SPGPE simulation for equilibrium states with input values $\mu(t-t_c)$ and $T(t-t_c)$. 
During the quench the growth of $l_{coh}$ is delayed with respect to the instantaneous equilibrium: we find such delay to correspond to $(t-t_c) \sim 1.3 \hat{t}$ (dotted vertical cyan line), where $\hat{t}$ (solid vertical cyan line) is the timescale predicted by the KZ mechanism. The scaled deviation, $\delta l_{coh}$, between dynamical and equilibrium correlation lengths, defined by Eq.~(\ref{eq:deltacoh}), is shown by the purple squares and exhibits a very rapid increase in the critical region, followed by a slower decay during the re-equilibration process, which reflects the phase ordering process. The end of the ramp is denoted by the vertical black dashed line. (c) Corresponding characteristic single-trajectory evolution of the Penrose-Onsager condensate density profiles ($\tau_Q=150$ms): plotted yellow and green regions respectively map out the density isosurfaces for $0.1\%$ and $3\%$ of the peak value of the final post-quench equilibrium condensate density; purple filaments denote region of high velocity field, corresponding to the location of spontaneously-generated vortices.
}
\label{fig_2}
\end{figure*}

\subsection{Equilibrium phase diagram}
\label{equilibrium}

We first characterize the precise location of the equilibrium critical point ($\epsilon =0$). For this purpose, we numerically calculate the equilibrium configuration of the gas for a given set of $T$ and $\mu$, in order to construct the corresponding equilibrium phase diagram of condensate fraction vs. temperature.

Equilibrium calculations are performed within the self-consistent Hartree-Fock approximation~\cite{blakie2008dynamics,Rooney2010,ncut}. 
The total atom number $N_\textrm{tot}=N_c+N_{\rm I}$ includes atoms in the c-field, $N_c=\int \, {\mathrm d} \mathbf{r}\, \langle |\psi(\mathbf{r})|^2 \rangle$, and in the thermal bath, $N_{\rm I}$. 
The atoms $N_{\rm I}$ located above $E_{\rm cut}$ are assumed to be in a thermal reservoir at the given $T$ and $\mu$, with $N_{\rm I}=(2\pi)^{-3}\iint_{\epsilon(\mathbf{r},\mathbf{k})>E_{\rm cut}}^\infty d\mathbf{k}d\mathbf{r} \, F_{\rm I}(\mathbf{r},\mathbf{k})$, where
\begin{equation}
F_{\rm I}(\mathbf{r},\mathbf{k})=\frac{1}{\exp[\{\epsilon(\mathbf{r},\mathbf{k})-\mu\}/k_BT]-1}
\end{equation}
and $\epsilon(\mathbf{r},\mathbf{k})=\frac{\hbar^2k^2}{2M}+V_\textrm{trap}(\mathbf{r})$. For a given $N_\textrm{tot}$, one can estimate the transition temperature for the corresponding ideal Bose gas, given by $T_{c,0}\approx0.94\hbar\tilde{\omega}N_\textrm{tot}^{1/3}/k_B$ 
where $\tilde{\omega}\equiv(\omega_x\omega_\perp^2)^{1/3}$. This can only serve as a useful reference value for an actual interacting system, due to the competition between thermal fluctuations and inter-particle interactions~\cite{Andersen2004}, and the relevance of finite size effects~\cite{Giorgini1996,Bezett2009}. 

Following a standard procedure \cite{blakie2008dynamics}, we can calculate the condensate atom number, $N_0$, from the classical-field wavefunction $\psi$ by means of the Penrose-Onsager criterion, through identification of the largest eigenvalue, and the corresponding eigenfunction $\psi_0(\mathbf{r})$, of the single-particle density matrix 
\begin{equation}
\rho(\mathbf{r},\mathbf{r}^\prime)=\langle\psi^\ast(\mathbf{r})\psi(\mathbf{r}^\prime)\rangle \ ,
\label{eq:density_matrix}
\end{equation}
where $\langle\cdots\rangle$ denotes a short-time average over $100$ samples~\cite{average_time}. The resulting condensate fraction $N_0/N_\textrm{tot}$ is plotted in Fig.~\ref{fig_2}(a) together with experimental data of the Trento group~\cite{Ferrari_private}. Our equilibrium simulations, in agreement with the experiment, clearly reveal that the condensate fraction arises at $T/T_{c,0}\sim0.9$ rather than $1$, with the corresponding critical chemical potential $\mu$ shifting to a positive value due to the finite-size and interaction effects~\cite{Damle1996,Giorgini1996,Dalfovo1999}.

In order to better identify the critical region, in addition to the condensate fraction we calculate three further quantities exhibiting critical behaviour in a narrow temperature region~\cite{Campostrini2001,Davis2003,Davis2006,Bezett2009,wright_proukakis_11,Kobayashi_16a,Kobayashi_16b,comaron2019quench}, namely the correlation length, $l_{coh}$, the Binder cumulant, $C_b$, and the order parameter, $m$.
The longitudinal correlation length can be extracted by an appropriate fit to the first-order correlation function, via~\cite{KZexp-w}
\begin{eqnarray}
G^{(1)}(d_x)&=&\iint \! {\rm d}y\,{\rm d}z\int_{-L_x/2}^{L_x/2}\! {\rm d}x\, w_{\rm PO}(\mathbf{r})\langle\psi^\ast(\mathbf{r}+d_x\hat{\mathbf{x}})\psi(\mathbf{r})\rangle \nonumber \\
&\approx& \left(1-\frac{d_x}{L_x}\right)e^{-d_x/l_{\rm coh}}\;.
\label{lcoh}
\end{eqnarray}
Here $L_x \approx 54.4\;\mu$m denotes a central portion of the axial extent of the inhomogeneous system over which the correlation function is evaluated (for comparison, the final equilibrium condensate spans the range $\approx[-114,\,114]\mu$m). The weighting function $w_{\rm PO}$ is introduced here to reduce the contribution of low-density regions in the transverse direction. Details of our procedure to calculate $l_{coh}$, $C_b$, and $m$ are given in Appendix~\ref{Appendix-A}, with different extraction protocols showing excellent agreement between them. As a result, we identify the equilibrium transition temperature in the range $T_c\sim445.5\pm7.3$~nK, roughly corresponding to $T/T_{c,0} \sim [0.91,\,0.93]$. The corresponding critical chemical potential is $\mu_c=(4.13\pm0.55)\hbar\omega_\perp >0$. 
Based on such values we can extract, for each quench $\tau_Q$, the dynamical
critical time, $t_c$, during the quench when $\mu(t)$ and $T(t)$ reach their corresponding critical values $\mu_c$ and $T_c$. Specifically, we find 
\be 
t_c= \frac{\mu_c}{\mu_f} \, \tau_Q \sim(0.188\pm0.026)\, \tau_Q\ . 
\ee
We can identify the time $t_c$ as the reference time from where to measure the distance $\epsilon$, which enables us to cast all dynamical behaviour and relation to the KZ mechanism in terms of the shifted time $(t-t_c)$ from the equilibrium phase transition.
\subsection{Quenched dynamics}

The quenched dynamical growth of the system can be visualized by means of a particular simulation example. Building on our earlier work which focused on the (late-time) re-equilibration dynamics of a quenched Bose gas~\cite{KZexp-w}, Fig.~\ref{fig_2}(b)-(c) shows the evolution of the correlation length and density profiles for the particular case of $\tau_Q = 150$~ms.

Examining the evolution of the correlation length during a quench as a function of ($t-t_c$), shown in Fig.~\ref{fig_2}(b), we notice that the growth of the dynamical correlation length $l_{coh}(t)$ starts, as expected, at a later time to that of the corresponding equilibrium correlation length $l_{coh}^{equil}$, evaluated at equilibrium with the same $\mu(t)$ and $T(t)$. In accordance with KZ mechanism, our simulations indicate a delay proportional to $\hat t$.

To complement our findings, Fig.~\ref{fig_2}(c) shows corresponding 3D single-trajectory density profile snapshots during the quenched evolution. Shortly after $\hat{t}$, the system remains dominated by fluctuations as shown in Fig.~\ref{fig_2}(c)(i). The first evidence of condensation onset appears around $1.3\hat{t}$, in the form of a localized elongated higher density condensate region containing multiple spontaneously-generated defects (purple filaments) as in Fig.~\ref{fig_2}(c)(ii). Subsequent dynamics [Fig.~\ref{fig_2}(c)(iii)-(v)] are dominated by the interplay between condensate growth (driven by the increasing $\mu(t)>\mu_c$ and decreasing $T(t)<T_c$) and phase-ordering through defect relaxation, which was previously shown to lead to a decoupling of number and coherence growth~\cite{KZexp-w}. Fig.~\ref{fig_2}(c)(v) shows a typical long-term profile after both density and coherence have saturated to their equilibrium values, which for the particular example displayed here occurs after the end of the external driving.

The evolution of coherence during the quench can also be visualized through the `auxiliary' variable~\cite{KZexp-w}
\be
\delta l_{coh}(t)=\frac{l_{coh}^{equil}(t)-l_{coh}(t)}{l_{coh}^{equil}(t)}
\label{eq:deltacoh}
\ee
where $l_{coh}^{equil}(t)$ is the equilibrium correlation length at time $t$. Early on in the quench, during the adiabatic regime, the dynamical correlation length closely follows the corresponding equilibrium one, until the system enters the critical region and $\delta l_{coh}(t)$ quickly increases from $0$ to $1$. The value of $\delta l_{coh}(t)$ remains $\approx 1$ until $(t-t_c) \sim 1.3 \hat{t}$, after which time it clearly starts decreasing, but at a much slower rate that its initial increase: the latter decay, previously characterized in Ref.~\cite{KZexp-w} is evidence of defect relaxation and phase ordering, until reaching values $\delta l_{coh}(t) \sim 0$, at which late time the dynamical system has grown sufficiently to become practically indistinguishable from the corresponding equilibrium one.

\section{Linearized SPGPE}
\label{linearizedSPGPE}

In the symmetric phase before the phase transition, when $\mu(t)<\mu_c$, there are small thermal fluctuations around the symmetric vacuum $\psi=0$. In the non-interacting limit, $\mu_c=0$. We expect that during the non-equilibrium linear quench these fluctuations remain small until some time after the critical point.
Therefore, the out of equilibrium evolution near the critical point can be reasonably described by a linearized version of Eq.~(\ref{eq:SPGPE}) where the interaction term $g|\psi|^2$ in Eq.~(\ref{calL}) is neglected. Furthermore, as the initial growth occurs around the centre of the trap, we assume here for simplicity that $V_\textrm{trap}({\bf r})\approx 0$.

With these two approximations (and omitting the projector in our analytical considerations) 
\begin{equation}
d\psi = 
\left[
(\gamma+i)
\frac{\hbar\nabla^2}{2M}
+\frac{\gamma}{\hbar} \mu(t) \right]\psi dt + dW \ . 
\label{linearized}
\end{equation}
In this framework the small fluctuations become dynamically unstable towards exponential growth of a condensate when $\mu(t)$ crosses $0$ towards positive values. This can be understood by noticing that the dissipative terms on the right hand side of Eq.~(\ref{eq:SPGPE}), that are proportional to $\gamma$, include a minus gradient of a Mexican-hat-like potential, $-\partial V/\partial\psi^*$, where
\be 
V\left(|\psi|\right) = \frac12 g|\psi|^4 - \mu(t) |\psi|^2 \ .
\label{eq:Mhat}
\ee
When $\mu(t)>0$ the symmetry is broken and the potential has instantaneous minima at a ring
\begin{equation}
|\psi|^2_{\rm eq}(t)=\frac{\mu_f}{g} \left(\frac{t}{\tau_Q}\right) \ ,
\label{eq:psi_eq}
\end{equation}
with the dissipation driving $\psi$ towards this instantaneous vacuum manifold. 

Before proceeding with such analytical treatment below, which will guide our subsequent numerical analysis of the full nonlinear dynamics, we make two important comments. Firstly, the presented analytical discussion implies that the critical point arises exactly at $t=0$. However, the experimentally relevant equilibrium phase diagram of Fig.~\ref{fig_2}(a) has already revealed a shift in time, which we will subsequently account for by replacing $t$ by $(t-t_c)$. Secondly, the linearized discussion neglects the role of the nonlinearity $g|\psi|^2\psi$ up to $\hat t$, at which point it will be argued to slow down the exponential blow-up of $|\psi|$. However, its effect is not completely negligible even before $\hat t$ in the simulation. As shown in Sec. \ref{equilibrium}, in equilibrium there is a range of values $0< \mu < \mu_c$ where the symmetry breaking Mexican hat potential is too shallow to prevent restoring the symmetry by thermal fluctuations. In this way, the actual symmetry breaking transition is shifted from the simplistic approximation of $\mu=0$ to the more appropriate $\mu=\mu_c >0$. 

With those `caveats' in mind, we proceed next with our analytical predictions, initially conducted for a homogeneous system, and subsequently generalized to modes beyond $\mathbf{k}=\mathbf{0}$.

\subsection{Uniform field}

Let us first consider a uniform field $\psi(t)$ when Eq.~(\ref{linearized}) becomes 
\begin{equation}
d\psi = \frac{\gamma}{\hbar} \mu(t) \psi dt + dW \ . 
\label{linearuniform}
\end{equation}
Here $dW$ is also assumed uniform.
When $\mu<0$ then $\psi=0$ is stable and its instantaneous relaxation time~\cite{McDonald2015} is 
\begin{equation}
\tau=\left(\frac{\gamma}{\hbar} |\mu(t)|\right)^{-1} \ .
\label{tau}
\end{equation}
Below the critical temperature, when $\mu>0$ and the symmetric state becomes dynamically unstable, Eq.~(\ref{tau}) is the characteristic time scale on which small perturbations grow exponentially.
In general $\tau$ is a time-scale on which the system can adjust to the time-dependent $\mu(t)$. The time-scale diverges at the critical point $\mu=0$. Near the critical point the system is too slow to adjust, no matter how long $\tau_Q$ is, and its state is effectively frozen between the two crossover times, $\mp \hat t$, when the reaction time of the system equals the time to the transition (see Fig.~\ref{fig_1}):
\begin{equation}
\tau=|t|_{t=\hat{t}}\ , 
\label{KZeq}
\end{equation}
Solution of this equation with respect to $\hat t$ yields the crossover time:
\begin{equation}
\hat t = \tau_Q^{1/2} \sqrt{\frac{\hbar}{\gamma\mu_f}}\ .
\label{hatt}
\end{equation}
This is the KZ time-scale. Near $-\hat t$ the uniform $\psi$ goes out of equilibrium with the instantaneous $\mu(t)$, hence its fluctuations do not diverge at $\mu=0$ -- as might be suggested by Eq.~(\ref{linearuniform}) -- but remain small in consistency with the linearized approximation. The linearization remains self-consistent until near $+\hat t$ when $\psi$ begins to catch up with the varying $\mu(t)$ again and the dynamical instability begins to blow up exponentially.

\subsection{Reciprocal space}
\label{sec:reciprocal_space}

In order to go beyond the uniform case, ${\bf k}=0$, we consider a (modified) Fourier transform,
\be 
\tilde\psi(t,{\bf k})=
e^{\frac{i\hbar k^2}{2M}t}
\int\frac{d^3k}{(2\pi)^{3/2}}
e^{-i{\bf k \cdot r}}
\psi(t,{\bf r}) \ ,
\ee
with an extra dynamical phase pre-factor included. In the reciprocal space the linearized SPGPE, Eq.~(\ref{linearized}), becomes a Wiener-like stochastic equation
\begin{equation}
\dot{\tilde\psi} = 
\frac{\gamma}{\hbar}\left[ \mu(t) - \frac{\hbar^2k^2}{2M} \right] \tilde\psi + 
\tilde\zeta,
\label{linearSPGPE}
\end{equation}
where $\tilde\zeta(t,{\bf k})$ is a Gaussian white noise with a correlator
\be 
\langle \tilde\zeta^*(t,{\bf k})\tilde\zeta(t',{\bf k'}) \rangle =
\frac{2\gamma k_B T}{\hbar}
\delta(t-t')
\delta({\bf k-k'})\ .
\label{zetazeta}
\ee
For $\mathbf{k=0}$, Eq.~(\ref{linearSPGPE}) becomes the uniform Eq.~(\ref{linearuniform}).

When $\mu>0$ then all modes $\tilde\psi(t,{\bf k})$ with $\frac{\hbar^2k^2}{2M}<\mu$ are dynamically unstable. At $+\hat t$, when the dynamical instability begins to blow up, all modes with $k$ up to
\begin{equation}
\hat k = \tau_Q^{-1/4}\left(\frac{4M^2\mu_f}{\gamma\hbar^3}\right)^{1/4}
\label{hatk}
\end{equation}
are already unstable. This borderline $\hat k$ is a solution of $\frac{\hbar^2\hat k^2}{2M}=\mu(\hat t)$.
They are amplified by the dynamical instability and dominate the power spectrum near and after $+\hat t$. An inverse of $\hat k$, 
\begin{equation}
\hat \xi = \tau_Q^{1/4}\left(\frac{4M^2\mu_f}{\gamma\hbar^3}\right)^{-1/4}\ ,
\label{hatxi}
\end{equation}
is the KZ correlation length. The power spectrum is dominated by modes with wave lengths longer than $\hat\xi$.

The power laws in Eq.~(\ref{hatt}), (\ref{hatk}) and (\ref{hatxi}) are consistent with the general KZ predictions in Eq.~(\ref{hattgeneral})-(\ref{hatxigeneral}) involving the critical exponents $z$ and $\nu$. Indeed, Eq.~(\ref{tau}) implies that $\tau$ is proportional to an inverse of the distance from the critical point, here measured by $|\mu|$, hence $z\nu=1$. At the critical point, $\mu=0$, the right hand side of 
Eq.~(\ref{linearSPGPE}) implies relaxation with a rate $\propto k^2$, hence $z=2$. Therefore, the general KZ formulas (\ref{hattgeneral})-(\ref{hatxigeneral}) predict
\be
\hat t\propto\tau_Q^{1/2} \hspace{1.0cm} \mathrm{and} \hspace{1.0cm} 
\hat\xi\propto\tau_Q^{1/4} \ ,
\ee
in agreement with Eq.~(\ref{hatt}) and (\ref{hatxi}), respectively.

\subsection{KZ scaling hypothesis}
\label{KZscalinghyp}

For large $\tau_Q$, the length-scale $\hat\xi$ and the time-scale $\hat t$ become longer than any other scales and, therefore, they become the only relevant scales in the low frequency and long wave length regime. Therefore, according to the KZ scaling hypothesis, in this regime physical observables depend on time $t$, distance ${\bf r}$, and wave vector ${\bf k}$ through scaled variables $t/\hat t$ and ${\bf r}/\hat\xi$, and $\hat\xi{\bf k}$, respectively. Here we verify the hypothesis for the linearized Eq.~(\ref{linearSPGPE}).

A formal solution of stochastic Eq.~(\ref{linearSPGPE}) is
\be 
\tilde\psi(t,{\bf k}) 
=
\int_{-\infty}^t dt'
\tilde\zeta(t',{\bf k})
e^{ 
-\frac{\gamma\hbar k^2}{2M} \left(t-t'\right)
+\frac{\gamma\mu_0}{2\hbar\tau_Q} \left(t^2-t'^2\right)
}\ .
\label{fs}
\ee
An equal-time correlator of these Gaussian fluctuations follows from 
Eq.~(\ref{zetazeta}) as
\be 
\langle \tilde\psi^*(t,{\bf k}) \tilde\psi(t,{\bf k}') \rangle \equiv
\delta({\bf k-k'})
f(t,{\bf k})\ ,
\ee
where the spectral function is
\bea
f(t,{\bf k})
&=&
\frac{2\gamma k_{B}T}{\hbar}
e^{\frac{\gamma\mu_f}{\hbar\tau_Q}(t-t_k)^2}
\int_{-\infty}^t 
dt'~
e^{-\frac{\gamma\mu_f}{\hbar\tau_Q}(t'-t_k)^2}
\nonumber\\
&=&
\frac{2\gamma k_{B}T}{\hbar}~
\hat t~
e^{u^2}
\int_{-\infty}^{u}
du'~
e^{-u'^2}\ .
\label{s}
\eea
Here 
$ 
t_k=\tau_Q ( \hbar^2k^2/2M) / \mu_f 
$
is the time when the Fourier mode $\tilde\psi(t,{\bf k})$ becomes dynamically unstable. 

The spectral function depends on $t$ and ${\bf k}$ through a single variable
\be 
u=t/\hat t-\hat\xi^2 k^2\ .
\label{u}
\ee 
This demonstrates not only the anticipated KZ scaling in the form
\begin{equation}
f(t,\mathbf{k})=\hat t ~F\left(t/\hat t,\hat{\xi}\mathbf{k}\right)
\label{eq:scalable}
\end{equation}
but an even stronger relation
\begin{equation}
f(t,\mathbf{k})=\hat t ~{\cal F}\left(t/\hat t-\hat{\xi}^2\mathbf{k}^2\right)\ .
\label{eq:strongscalable}
\end{equation}
Here $F$ and ${\cal F}$ are non-universal scaling functions.

\subsection{Near $+\hat t$}

The spectral function of Eq.~(\ref{s}) is monotonically increasing with $u$.
Consequently, for any time it peaks at ${\bf k}=0$ and
for any ${\bf k}$ it is increasing with time.
The peak value begins to blow up like $e^{(t/\hat t)^2}$ near $t/\hat t\approx 1$:
\bea
f(t,{\bf 0})&=&
\frac{2\gamma k_{B}T}{\hbar}~
\hat t~
e^{(t/\hat t)^2}
\int_{-\infty}^{t/\hat t}
du'~
e^{-u'^2}
\nonumber\\
&\approx &
\frac{2\sqrt{\pi}\gamma k_{B}T}{\hbar}~
\hat t~
e^{(t/\hat t)^2}\ .
\label{s0}
\eea
The blow-up enhances the peak of the spectral function in a neighborhood of ${\bf k}=0$ where $u$ becomes large enough for the integral in Eq.~(\ref{s}) to be approximated by $\sqrt{\pi}$:
\bea
f(t,{\bf k})\approx 
\frac{2\sqrt{\pi}\gamma k_{B}T}{\hbar}~
\hat t~
e^{(t/\hat t-\hat\xi^2k^2)^2}\ .
\eea
In its regime of validity $\hat\xi^2k^2\ll t/\hat t$, hence it can be further simplified to a Gaussian:
\bea
f(t,{\bf k})\approx 
\frac{2\sqrt{\pi}\gamma k_{B}T}{\hbar}~
\hat t~
e^{(t/\hat t)^2}
e^{-2(t/\hat t)\hat\xi^2k^2}\ .
\eea
This Gaussian neglects fluctuations with wave lengths shorter than 
$ 
\xi=2\hat\xi(t/\hat t)^{1/2}
$
that have not been enhanced by the blow-up yet. The Gaussian spectral function translates to a coarse-grained equal-time correlation function
\bea 
\hspace{-0.6cm} \langle
\psi^*(t,{\bf r}) \psi(t,{\bf r'})
\rangle 
&=&
\int\frac{d^3k}{(2\pi)^3}
e^{-i{\bf k}({\bf r-r'})}
f(t,{\bf k})
\nonumber \\
&=&
\frac{\gamma k_B T}{2\pi^{3/2}\hbar}
\left( \frac{\hat t}{\xi^3} \right)
e^{(t/\hat t)^2}
e^{-({\bf r-r'})^2/2\xi^2}\ .
\label{g1}
\eea
As anticipated, near $\hat t$ its range $\xi$ becomes the KZ correlation length $\hat\xi$. 

It is noteworthy that for any $t/\hat t$ this correlation function is proportional to 
$\hat t/\hat\xi^3\propto \hat\xi^{-1}$ which is consistent with the general scaling hypothesis, Eq.~(\ref{CRscaling}), given that $d=3$ and, in our linearized Gaussian theory, $\eta=0$.

Setting ${\bf r'}={\bf r}$ we obtain average strength of the coarse-grained fluctuations:
\bea
\langle 
|\psi(t,{\bf r})|^2
\rangle =
\frac{\gamma k_B T}{16\pi^{3/2}\hbar}
\left( \frac{\hat t}{\hat\xi^3} \right)
\left(\frac{\hat t}{t}\right)^{3/2}
e^{(t/\hat t)^2}.
\label{exp}
\eea
accurate near $\hat t$ or later. These are also times when Eq.~(\ref{exp}) blows up and the linearized SPGPE begins to break down. This suggests a scaling behaviour that 
\bea
\tau_Q^{1/4}
\langle 
|\psi(t,{\bf r})|^2
\rangle \propto \left(\frac{\hat t}{t}\right)^{3/2}
e^{(t/\hat t)^2}.
\label{scalingdensity}
\eea
Further growth is halted by the interaction term in the Mexican hat potential, Eq.~(\ref{eq:Mhat}), that was neglected in the linearized equation. The nonlinear interaction begins to be felt already at the inflection point of the potential:
\be 
|\psi|^2=\frac{\mu_f}{3g}\left( \frac{t}{\tau_Q} \right).
\label{eq:psi_infl}
\ee
Therefore, equating
$
\langle 
|\psi(t,{\bf r})|^2
\rangle
$
to Eq.~(\ref{eq:psi_infl}) is a good indicator when the linearized approximation breaks down. Thanks to the exponential nature of the blow-up Eq.~(\ref{exp}) the breakdown time is close to $\hat t$ up to logarithmic corrections.

It is noteworthy that, at $t\approx\hat t$, the KZ correlation length equals the healing length in the instantaneous Mexican hat potential. The healing length is a width of a vortex core, hence it is not possible to stabilize a tangle of vortex lines whose separations are less than the healing length. This justifies {\it a posteriori} our coarse-graining over wave lengths shorter than the KZ coherence length. The shorter fluctuations are not relevant for formation of stable vortex lines.

\subsection{Beyond $+\hat t$}

According to the linearized theory, near $\hat t$ the magnitude $\langle|\psi|^2\rangle$ should come close to the inflection point of the Mexican hat potential. Near the inflection the potential is approximately a linear function, hence its gradient is a constant and the magnitude $|\psi|$ should grow linearly in time. This is a significant slow-down after the initial exponential blow-up. Nevertheless, eventually $|\psi|$ grows enough to get close to the instantaneous equilibrium magnitude, Eq.~\ref{eq:psi_eq}). at the bottom of the potential (true vacuum). This equilibrium depends on time through $t/\tau_Q$, rather than $t/\hat t$ characteristic for the early times before and around $\hat t$, because it follows the linear ramp that depends on $t/\tau_Q$. 

However, as the equilibrium magnitude depends on time, the equilibration cannot be perfect and $\langle|\psi|^2\rangle$ must be delayed with respect to the instantaneous equilibrium, Eq.~(\ref{eq:psi_eq}). The delay time should be proportional to a relaxation time towards the bottom of the Mexican hat potential at the moment when the magnitude's growth slows down near its inflection point. This relaxation time is proportional to the universal KZ timescale, $\hat t$. Therefore, we expect that the instantaneous equilibrium Eq.~(\ref{eq:psi_eq}) should be replaced by a crude formula:
\be 
\langle 
|\psi(t,{\bf r})|^2
\rangle \approx 
|\psi|^2_{eq}\left(t-\alpha\hat t\right)=
\frac{\mu_f}{g}
\left( \frac{t-\alpha\hat t}{\tau_Q} \right)\ ,
\label{eqcrude}
\ee 
where $\alpha$ is a non-universal constant, excepted to be $\sim O(1)$. This is approximately valid long after $\hat t$, when the KZ scaling hypothesis no longer applies, but there is still a delay proportional to the KZ delay time $\hat t$.

It is worth emphasizing that even after the near-equilibration of the magnitude, the phase of $\psi$ should remain random with a characteristic KZ coherence length $\hat\xi$. The phase is the Goldstone mode for this symmetry breaking, hence it is not subject to the aforementioned relaxation. It is only in the subsequent evolution that the phase undergoes slow phase ordering kinetics~\cite{POK} that proceeds by gradual annihilation of the randomly-generated vortex networks. In this sense the KZ coherence length $\hat\xi$ is a more robust imprint of the KZ physics that survives to very late times.

\subsection{Shift of the critical point }

In the proceeding discussion the nonlinearity $g|\psi|^2\psi$ was neglected up to $\hat t$ where it was argued to slow down the exponential blow-up of $|\psi|$. However, its effect is not completely negligible even before $\hat t$ in the simulation. As shown in Sec.~\ref{equilibrium}, in equilibrium there is a range of $\mu >0$ up to $\mu_c \approx 4.13 \hbar \omega_\perp$ where the symmetry breaking Mexican hat potential is too shallow to prevent restoring the symmetry by thermal fluctuations. In this way the actual symmetry breaking transition is shifted to $\mu=\mu_c$. In addition to the shift, the equilibrium universality class is also altered with the mean-field correlation length exponent $\nu=1/2$ replaced by the exact $\nu=0.67$. Correspondingly, given the dynamical exponent $z=2$, the predicted $\hat t\propto\tau_Q^{z\nu/(1+z\nu)}$ should be altered from $\hat t \propto \tau_Q^{0.50}$ to $\hat t \propto \tau_Q^{0.57}$.

In the following we assume validity of the physical picture developed within the Gaussian theory but incorporate the criticality shift from $t=0$ to $t_c$,
into predictions of the linearized S(P)GPE by making a replacement 
$t \to t-t_c$. Regarding the scaling of $\hat t$ with $\tau_Q$, we note that due to the 
statistical uncertainties, it is not possible to discriminate between the similar power laws: $\hat t\propto\tau_Q^{0.50}$ (our Gaussian approximation) and the improved scaling $\hat t\propto\tau_Q^{0.57}$, a point further discussed in Appendix~\ref{Appendix-B}.

\subsection{Homogeneous assumption}
\label{homogeneous}

For the harmonically trapped system considered in this work, the instability addressed above is considered to be occurring in the volume where $\mu(t)-V_\textrm{trap}(\mathbf{r})>0$. Due to the anisotropy, this volume is enclosed in an ellipsoid $x^2/a_x(t)^2+(y^2+z^2)/a_\perp(t)^2=1$ where 
\be
\begin{array}{rl}
a_x(t)=&\displaystyle\sqrt{\frac{2\mu_f}{M\omega_x^2}\left( \frac{t}{\tau_Q} \right) }
\\
a_\perp(t)=&\displaystyle \sqrt{\frac{2\mu_f}{M\omega_x^2\lambda_\perp^2} \left( \frac{t}{\tau_Q} \right) }=
\frac{a_x(t)}{\lambda_\perp} \; \ .
\end{array}
\ee
with $\lambda_\perp=\omega_\perp/\omega_x$. Such an ellipsoid defines a critical volume of the system, $V_c\equiv4\pi a_x(t)a_\perp^2(t)/3$, and expands along its principle semi-axes with velocities
\be
\displaystyle v_{x}=\frac{da_{x}}{dt}=\sqrt{\frac{\mu_f}{2M\omega_x^2\tau_Qt}} \hspace{0.4cm} \textrm{ and } \hspace{0.4cm} v_\perp=\frac{v_x}{\lambda_\perp} \;\ .
\ee
These velocities diverge in the centre of the trap where the instability appears first at $t=0$. 

\begin{figure}[t!]
\centering
\includegraphics[width=1\linewidth]{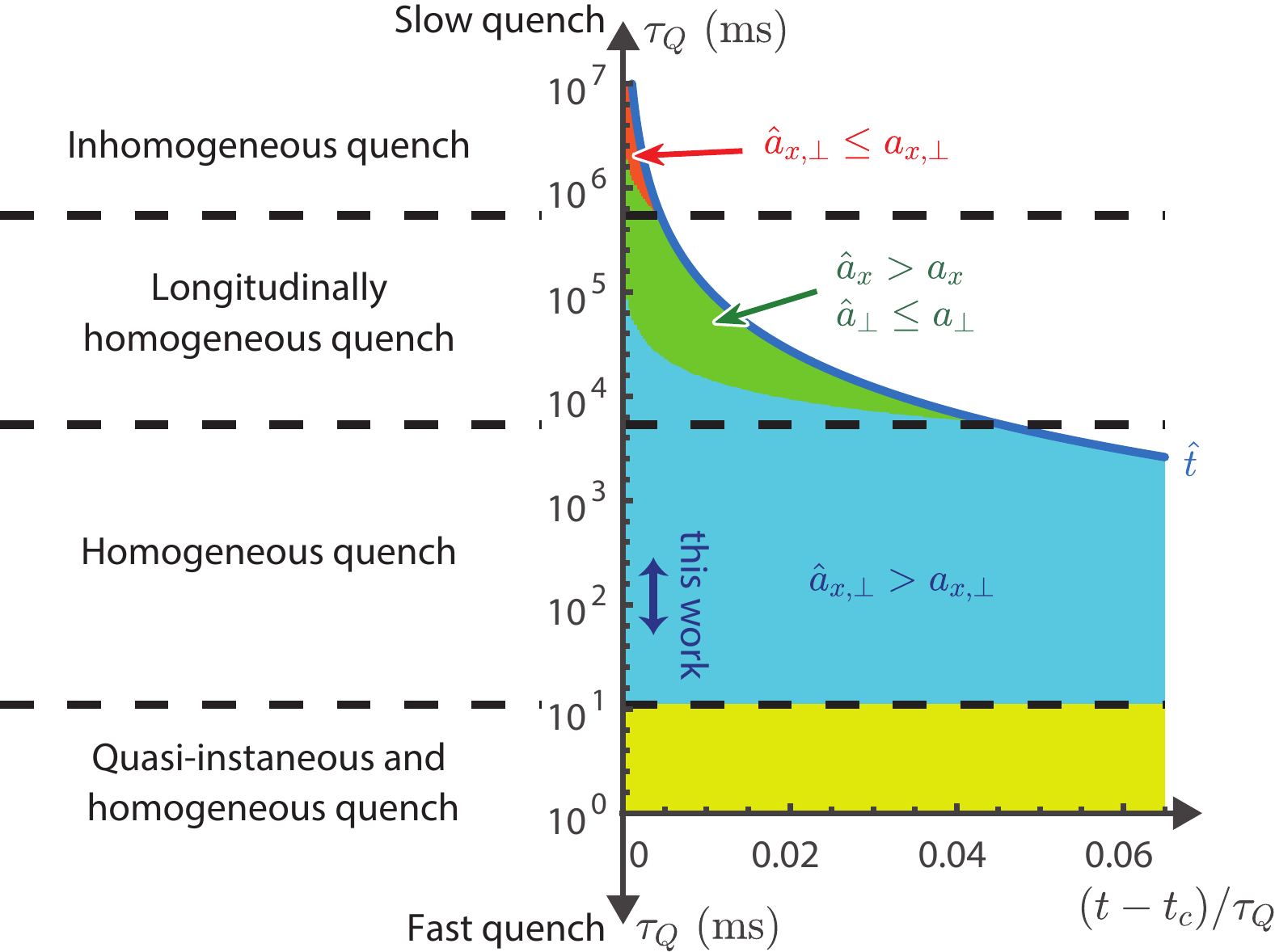}
\caption{
Types of quench classified by comparing the critical wavefront $a_{x,\perp}$ and $\hat{a}_{x,\perp}$ for times up to $\hat{t}$. When the quench duration is comparable with $\hat{t}$, it becomes quasi-instantaneous, with
momenta beyond $1/\hat{\xi}$ excited. 
Increasing $\tau_Q$ leads to a broad homogeneous quench regime (blue), which fully encompasses all experimentally-relevant quenches probed in this work, whose range is marked by the vertical arrow. 
Slower quenches can lead to regimes where the quench is inhomogeneous only in the transverse direction (green), or in both directions (red), but both of these would require ramp durations exceeding 10s for the current trap.
Note that, since the critical wavefront starts from the trap centre at $t_c$, the quench is always homogeneous at $t-t_c=0$.
The black dashed lines are boundaries estimated by Eq.~(\ref{eq:homo_quench1})--(\ref{eq:homo_quench4}).
}
\label{fig_3}
\end{figure}

An investigation of possible corrections due to the system inhomogeneity requires us to compare these velocities of the critical front ($v_x$, $v_\perp$) with the perturbation velocity $\hat v$ within the critical regime given by
\be
\hat{v}\approx \frac{\hat{\xi}}{{\hat{t}}}=
\left(\frac{\gamma^3\hbar\mu_f}{4M^2}\right)^{1/4}\tau_Q^{-1/4}\ .
\ee
The quench is effectively homogeneous when the critical front velocities ($v_x$, $v_\perp$) are larger than $\hat v$. As the critical front velocities diverge in the center of the trap, the quench is effectively homogeneous there. In the longitudinal (transverse) direction the quench remains homogeneous until the moment when $v_x=\hat v$ ($v_\perp=\hat v$). The latter equations can be solved with respect to $a_x$ ($a_\perp$) to respectively define
\be 
\hat a_x \simeq
\left(
\frac{4\mu_f^3}{M^2\omega_x^8\gamma^3\hbar \tau_Q^3}
\right)^{1/4}, \hspace{0.8cm}
\hat a_\perp = \frac{ \hat a_x}{\lambda_\perp^2} \ .
\ee
Inside the ellipsoid with semi axis $\hat{a}_x$ and $\hat{a}_\perp$, 
where both $v_x$ and $v_\perp$ are faster than $\hat v$,
the system is effectively homogeneous, and we find that these two quantities are respectively larger than $a_x$ and $a_\perp$.

The ratio of $\hat{a}_{x(\perp)}/a_{x(\perp)}(t=\hat{t})$ 
for a given $\tau_Q$ provides a guidance of the homogeneity of a quench.
When $\hat{a}_{x(\perp)}/a_{x(\perp)}(t=\hat{t})>1$, the quench is homogeneous in the $x$ (transverse) direction.
We can compare the instability front $a_{x(\perp)}(t)$ and $\hat{a}_{x(\perp)}$ up to $t=\hat{t}$.
The conditions for a quench to be longitudinally/transversally homogeneous thus are
\begin{equation}
\begin{array}{cr}
\displaystyle\tau_Q\lessapprox\hat{t} &\quad\begin{tabular}{c}
for quasi-instaneous quench\\
(homogeneous in both directions)
\end{tabular},
\end{array}\label{eq:homo_quench1}\end{equation}
\begin{equation}
\begin{array}{cr}
\displaystyle\hat{t}<\tau_Q<\frac{1}{\lambda_\perp^2}\frac{\mu_f}{\gamma\hbar\omega_x^2}
& \quad\begin{tabular}{c}for homogeneous quench\\ in both directions\end{tabular},
\end{array}\label{eq:homo_quench2}\end{equation}
\begin{equation}\begin{array}{cr}
\displaystyle\frac{1}{\lambda_\perp^2}\frac{\mu_f}{\gamma\hbar\omega_x^2}<\tau_Q<\frac{\mu_f}{\gamma\hbar\omega_x^2} & \begin{tabular}{c} for longitudinally homogeneous\\ but transversally inhomogeneous\\ quench\end{tabular},
\end{array}\label{eq:homo_quench3}\end{equation}
and
\begin{equation}\begin{array}{cr}
\displaystyle\tau_Q>\frac{\mu_f}{\gamma\hbar\omega_x^2}& \quad\begin{tabular}{c}for inhomogeneous quench\\ in both directions.\end{tabular}
\end{array}\label{eq:homo_quench4}\end{equation}

Identification of different criteria for a homogeneous quench across the longitudinal and transverse directions gives rise to a rich diagram of possible behaviour, based on our quenched input parameter $\mu(t)$.
The types of quenches possible for the considered trapping potential, characterized in terms of their (in)homogeneity up to $\hat t$ are summarized in Fig.~\ref{fig_3}.
%
In the blue region, $\hat{a}_{x(\perp)}>a_{x(\perp)}$ and the quenches are effectively homogeneous.
As $\tau_Q$ is increased to cover the green regime, the quenches are effectively longitudinally homogeneous when $\hat{a}_{x}>a_{x}$ but transversally inhomogeneous as $\hat{a}_\perp<a_{\perp}$.
In the red region, the quenches are effectively inhomogeneous, since $\hat{a}_{x,\perp}<a_{x,\perp}$.
We also note here that, shortly after the transition, the quenches are all effectively homogeneous, as the critical wavefront starts growing outwards from the trap centre at $t=t_c$. 

The experimentally relevant quench parameters investigated in our present study lie well within the homogeneous quench regime, and hence the above linearized SPGPE analysis is expected to be applicable.
When the quench duration becomes comparable to $\hat{t}$, the quenches can be regarded as quasi-instaneous quench (yellow region): in such cases, the presence of the nonlinearity allows for momenta beyond $1/\hat{\xi}$ to be excited after the termination of the fast ramp.
To see the inhomogeneous effects in a quench, one could consider much slower quenches or increase the trapping frequencies and aspect ratio, tuning the boundaries according to Eq.~(\ref{eq:homo_quench1}) to (\ref{eq:homo_quench4}).

\section{Early-time KZ scaling and SPGPE}

Having demonstrated the relevance of the homogeneous KZ mechanism for the parameter regime considered in this work, we now examine the extent to which the linearized SPGPE -- supplemented with the time shift $(t-t_c)$ -- can accurately explain the results of the full nonlinear SPGPE numerical simulations.

\begin{figure}[b!]
\centering
\includegraphics[width=1\linewidth]{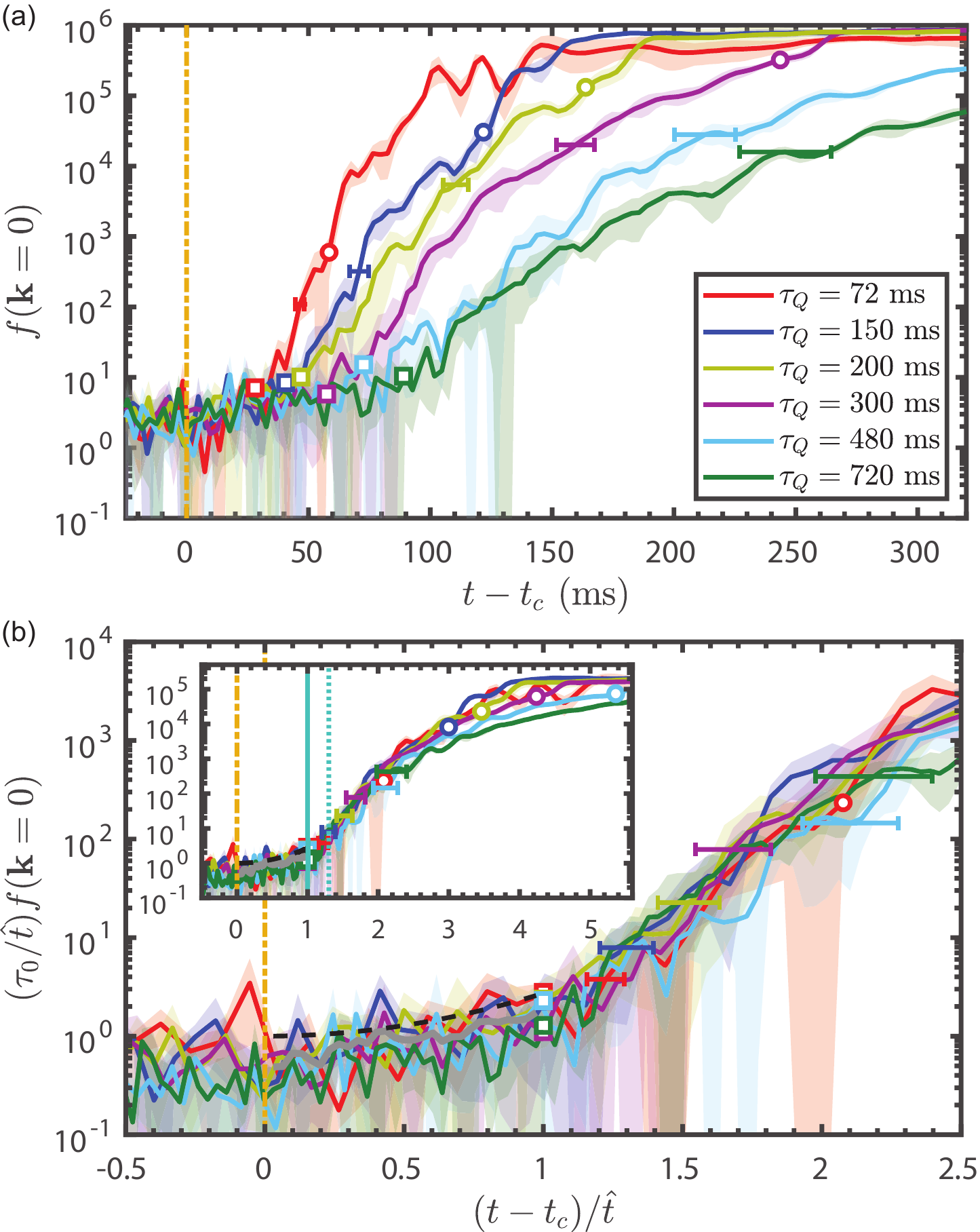}
\caption{
(a) Evolution of the peak spectral function $f(t,\mathbf{k}=0)$ in the SPGPE simulations as a function of $(t-t_c)$; (b) same curves but rescaled according to the KZ scaling predicted by the linearized theory. The scaled curves for different $\tau_Q$ collapse to a common scaling function $F[(t-t_c)/\hat t,{\bf 0}]$ in the approximate regime $(t-t_c)/\hat{t} \in (0,2)$. Furthermore, as predicted by Eq.~(\ref{s0}), in the same regime there is a blow-up that begins near/shortly after $(t-t_c)/\hat t \sim 1$, with 
the light blue vertical solid and dotted lines in the inset respectively marking the positions of $\hat{t}$ and $1.3\hat{t}$.
The hollow squares mark $\hat t$ as defined in Eq.~(\ref{hatt}), while the hollow circles mark the end of the linear ramp. The black dashed line plots the Gaussian divergent trend $\propto$exp$\{[(t-t_c)/\hat{t}]^2\}$ in Eq.~(\ref{s0}).
}
\label{fig_4}
\end{figure}

\begin{figure*}[t!]
\centering
\includegraphics[width=0.8\linewidth]{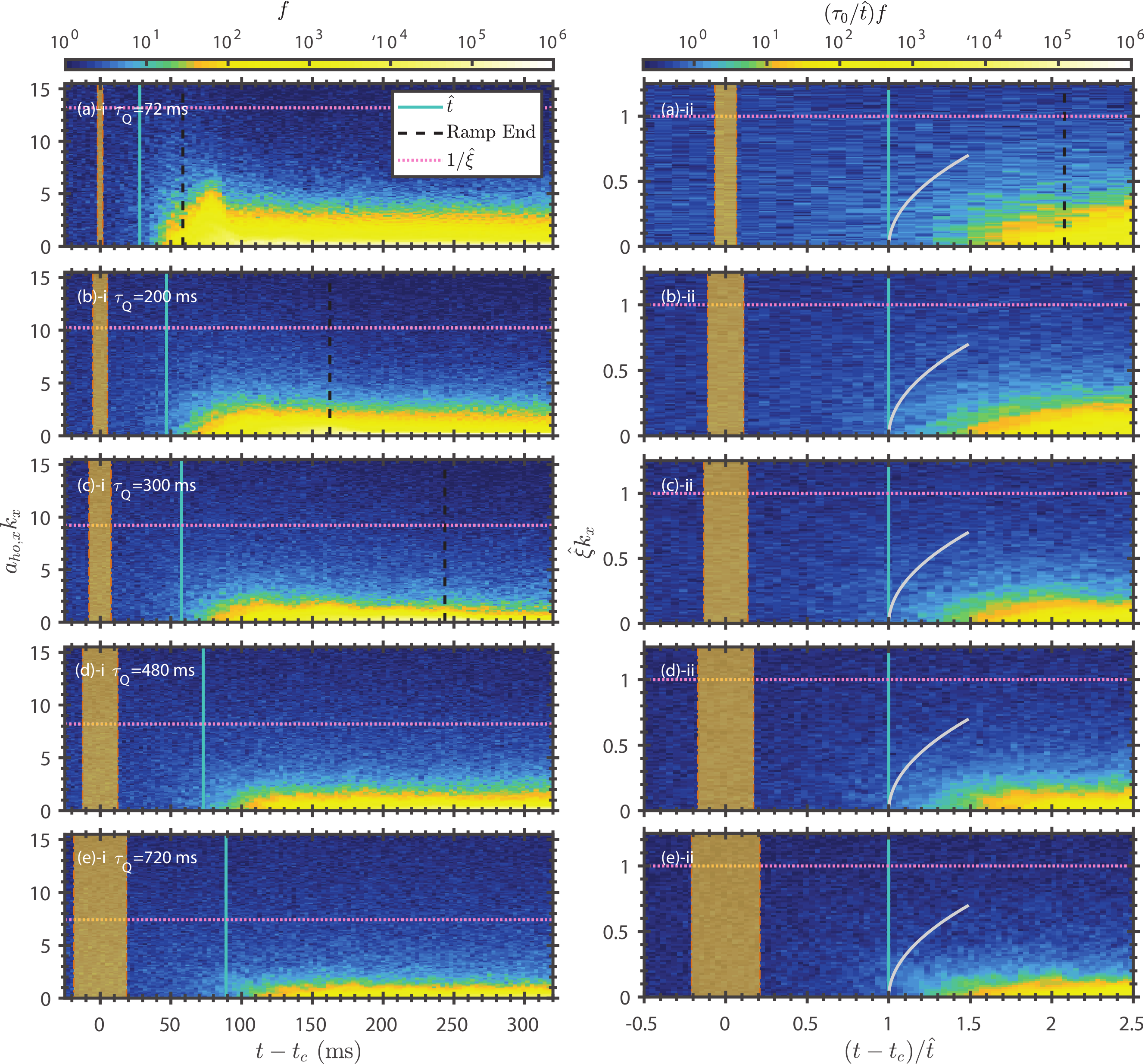}
\caption{
Evolution of the spectral function, Eq.~(\ref{s}), in SPGPE simulations for different quench times. Left columns: raw data. Right column: same results but rescaled according to Eq.~(\ref{eq:scalable}).
Vertical bands mark the uncertainty of $t_c$ obtained from equilibrium analysis. The scaled curves for different $\tau_Q$ (at least for the three slower quenches) collapse to a common scaling function $F[(t-t_c)/\hat t,{\bf 0}]$ in the approximate regime $(t-t_c) \lesssim 2 \hat t$. Grey lines in right column represent the parabola $u=(t-t_c)/\hat t-\hat\xi^2k^2=1$, along which the spectral function is predicted to be constant according to the stronger relation of Eq.~(\ref{eq:strongscalable}), with $t \to t-t_c$, within their regime of validity. The horizontal pink dashed line marks the value of $1/\hat{\xi}$, which corresponds to the largest wave number being excited up to $t-t_c=\hat{t}$ as discussed in Sec.~\ref{sec:reciprocal_space}.
}
\label{fig_5}
\end{figure*}

Firstly, we consider the spectral function, defined in Section~\ref{KZscalinghyp}. The time evolution of the spectral function can be extracted from the full SPGPE simulations. Figure~\ref{fig_4} shows the evolution of the peak value $f(t,{\bf 0})$ as a function of $t-t_c$. The same curves are plotted in panel (b), but rescaled according to the analytic scaling law predicted by the linearized theory, Eq.~(\ref{eq:scalable}). We see that the curves corresponding to different quench rates collapse onto each other in the approximate range $(t-t_c) \lesssim 2 \hat t$, thus demonstrating the validity of the KZ scaling hypothesis. Furthermore, the collapsed curves blow-up near the scaled time $(t-t_c)/\hat t=1$ as predicted by Eq.~(\ref{s0}). At later times, as the field fluctuations approach the inflection point of the Mexican hat potential, the slope of the curves decreases as an effect of the nonlinear interaction term in the SPGPE not included in the linearized theory.

The evolution of the whole spectral function $f(t,{\bf k})$ is investigated in Fig.~\ref{fig_5}, which shows raw (left column) and scaled (right column) numerical data for different values of $\tau_Q$ as a function of the shifted time $(t-t_c)$. The raw data demonstrate a strikingly different behaviour for different quench times. Nonetheless, plotting the same data scaled according to the law, Eq.~(\ref{eq:scalable}), reveal great similarity, particularly for the three slowest quenches (bottom three panels). In other words, for long enough $\tau_Q$, the scaled spectral functions collapse to a common scaling function thus confirming the KZ scaling hypothesis. In the same panels we also attempt a test of the scaling law in its stronger form of Eq.~(\ref{eq:strongscalable}). The added grey parabolas satisfy $u=(t-t_c)/\hat t-\hat\xi^2k^2=1$. According to Eq. (\ref{eq:strongscalable}) the spectral functions should be constant along these lines, which appears to be the case here, up to statistical fluctuations. 

\begin{figure}[t!]
\centering
\includegraphics[width=1\linewidth]{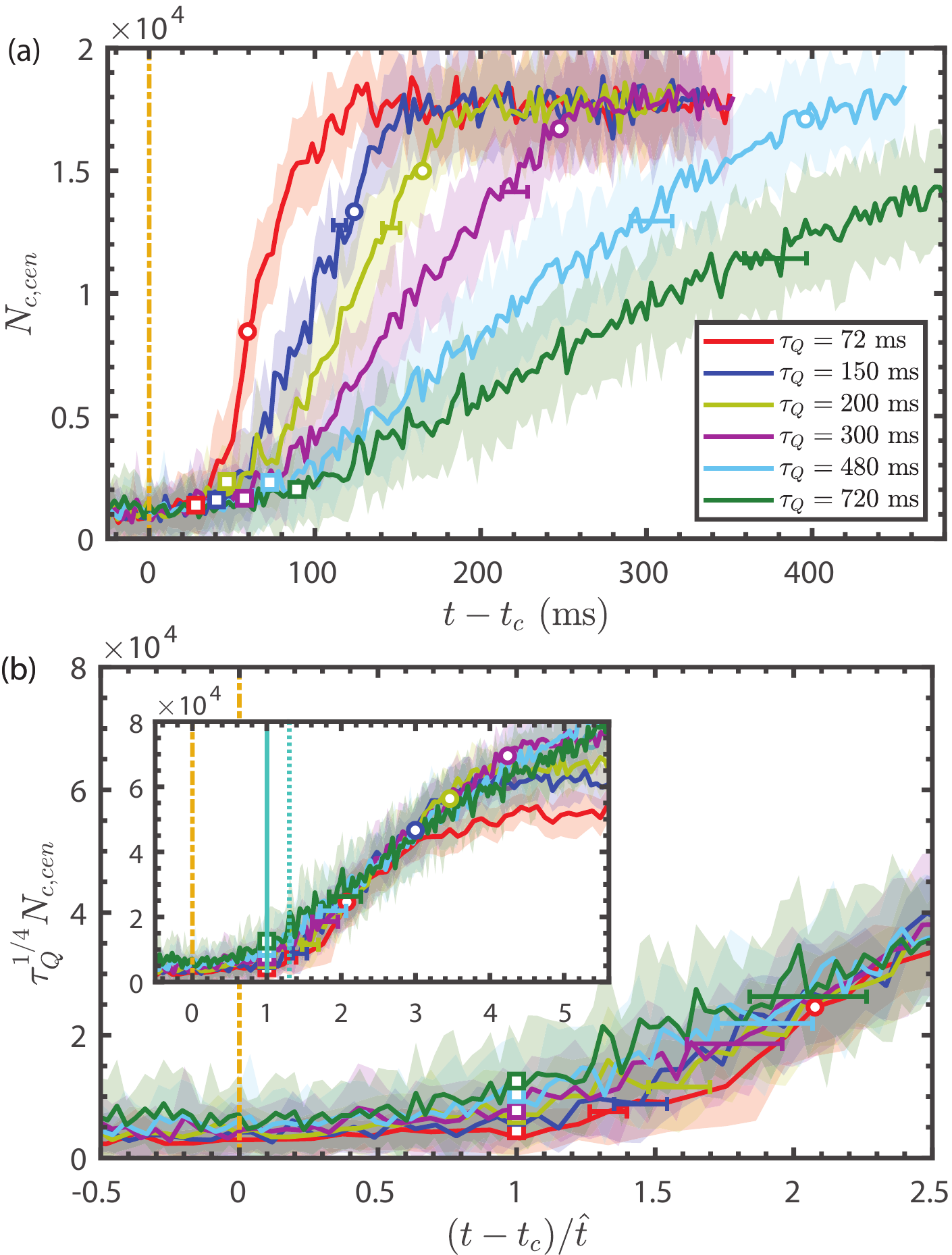}
\caption{
(a) Evolution of the central particle number Eq.~(\ref{Ncen}) in the SPGPE simulations as a function $(t-t_c)$; (b) same curves but rescaled according to the KZ scaling predicted by the linearized theory. Vertical lines and hollow points have the same meaning as in Fig.~\ref{fig_4}.
}
\label{fig_6}
\end{figure}

Nonlinear effects becomes relevant when the density is large. In Fig.~\ref{fig_6} we plot the time evolution of the number of atoms in the classical field near the center of the trap, defined as
\begin{equation}
N_{c,\text{cen}}=\int_{V_{\rm cen}}d\mathbf{r}~ \langle |\psi(\mathbf{r})|^2 \rangle \ , 
\label{Ncen}
\end{equation}
where, upon accounting for the system anisotropy, $V_{\rm cen}$ has been chosen as the ellipsoid around the center within half harmonic lengths in all directions. The results for different ramps are shown in panel (a) as a function of $t-t_c$, while in panel (b) the same curves are plotted according to the scaling law Eq.~(\ref{scalingdensity}) predicted by the linearized theory. Again, the curves nicely collapse onto each other in the same early-time regime $(t-t_c) \lesssim 2 \hat t$, where the spectral function also collapses, while for larger times, the scaling is less effective.

\begin{figure}[t!]
\centering
\includegraphics[width=1\linewidth]{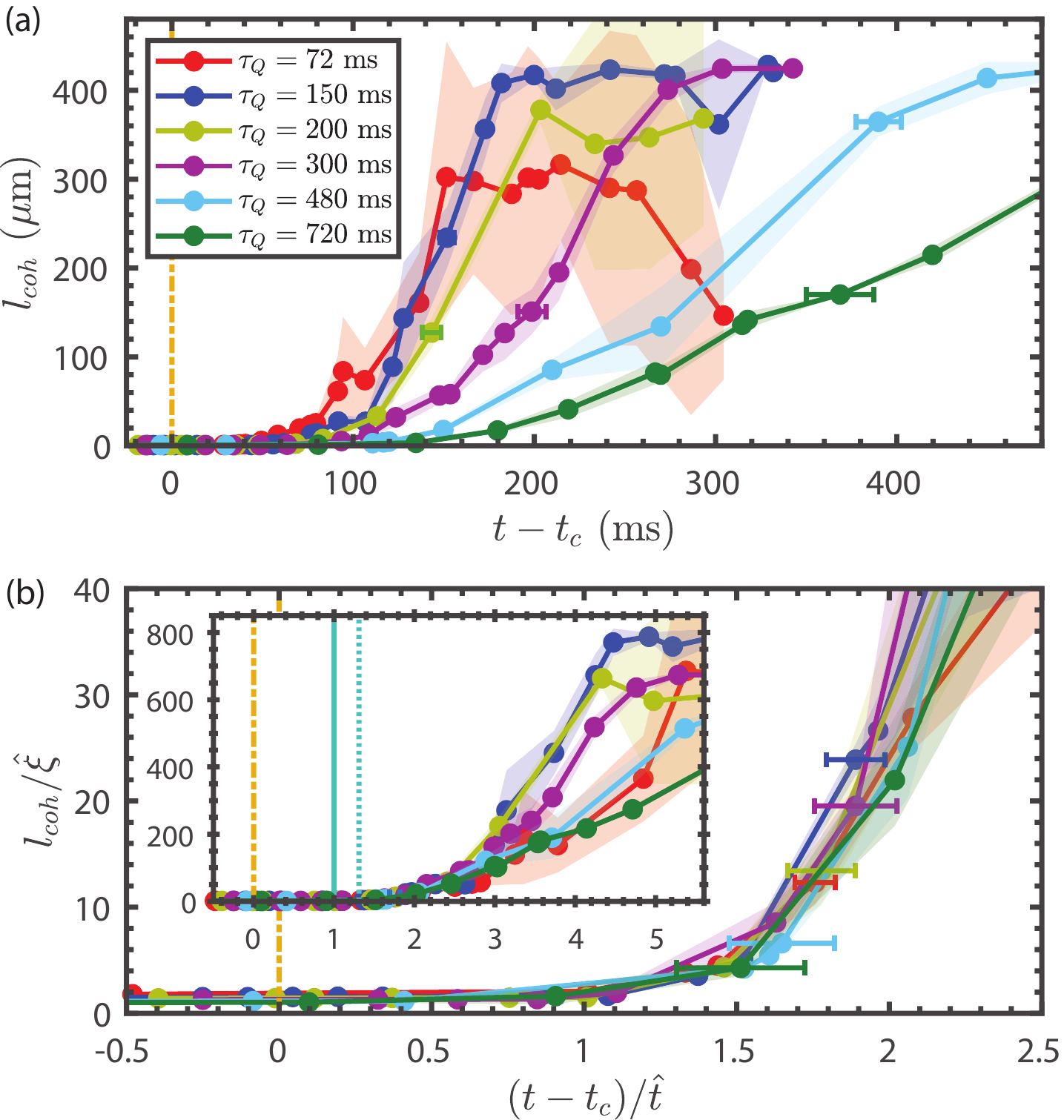}
\caption{
Evolution of the longitudinal correlation length defined by Eq.~(\ref{lcoh}) and calculated in the SPGPE simulations for different quench times: (a) raw data as a function of $(t-t_c)$; (b) same curves rescaled according to the KZ scaling predicted by the linearized theory. The light blue vertical solid and dotted lines in the inset respectively mark the positions of $\hat{t}$ and $1.3\hat{t}$.
}
\label{fig_7}
\end{figure}

Finally, in Fig.~\ref{fig_7}, we show that the longitudinal correlation length growth defined by Eq.~(\ref{lcoh}) also collapses onto a single scaling function by applying the same KZ rescaling previously used for the spectral function. 
All quenches exhibit an initial growth at some delayed time after $(t-t_c)=0$, with faster quenches displaying a faster initial growth. In the case of the slower quenches, $l_{coh}$ grows smoothly to the final value of around $430\;\mu$m. However, the 
three fastest quenches ($72$, $150$ and $200$~ms) reveal evident fluctuations in the value of $l_{coh}$ during its growth. These have been previously identified~\cite{KZexp-w} as being due to the persistence/dynamics of defects (vortices) within the region $|x|\leq 27.2\;\mu$m over which this correlation function is evaluated. 
This is more pronounced for the very fast quench [ $\tau_Q=72$ ms (red)], for which the cooling ramp terminates at $t-t_c\approx2\hat{t}$ (dashed black vertical lines in Fig.~\ref{fig_5}(a)(i)-(ii)), thereafter exciting higher momentum modes. 
Nonetheless, within $(t-t_c) \lesssim 2 \hat t$, and after rescaling, $l_{coh}$ reveals excellent collapse for all curves, as evident from Fig.~\ref{fig_7}(b).
%

\section{Late time dynamics}

Up to now, we have accounted for the early-time dynamical phase transition crossing within SPGPE, interpreting the result in the context of the homogeneous KZ mechanism and the linearized theory.
In this section we examine the extent to which the late-time dynamics of the nonlinear SPGPE -- based on our quench protocol of fixed initial and final states, and different quench duration $\tau_Q$ -- are also collapsible onto a single curve in a way dictated by the KZ mechanism.

\begin{figure}[t!]
\centering
\includegraphics[width=1\linewidth]{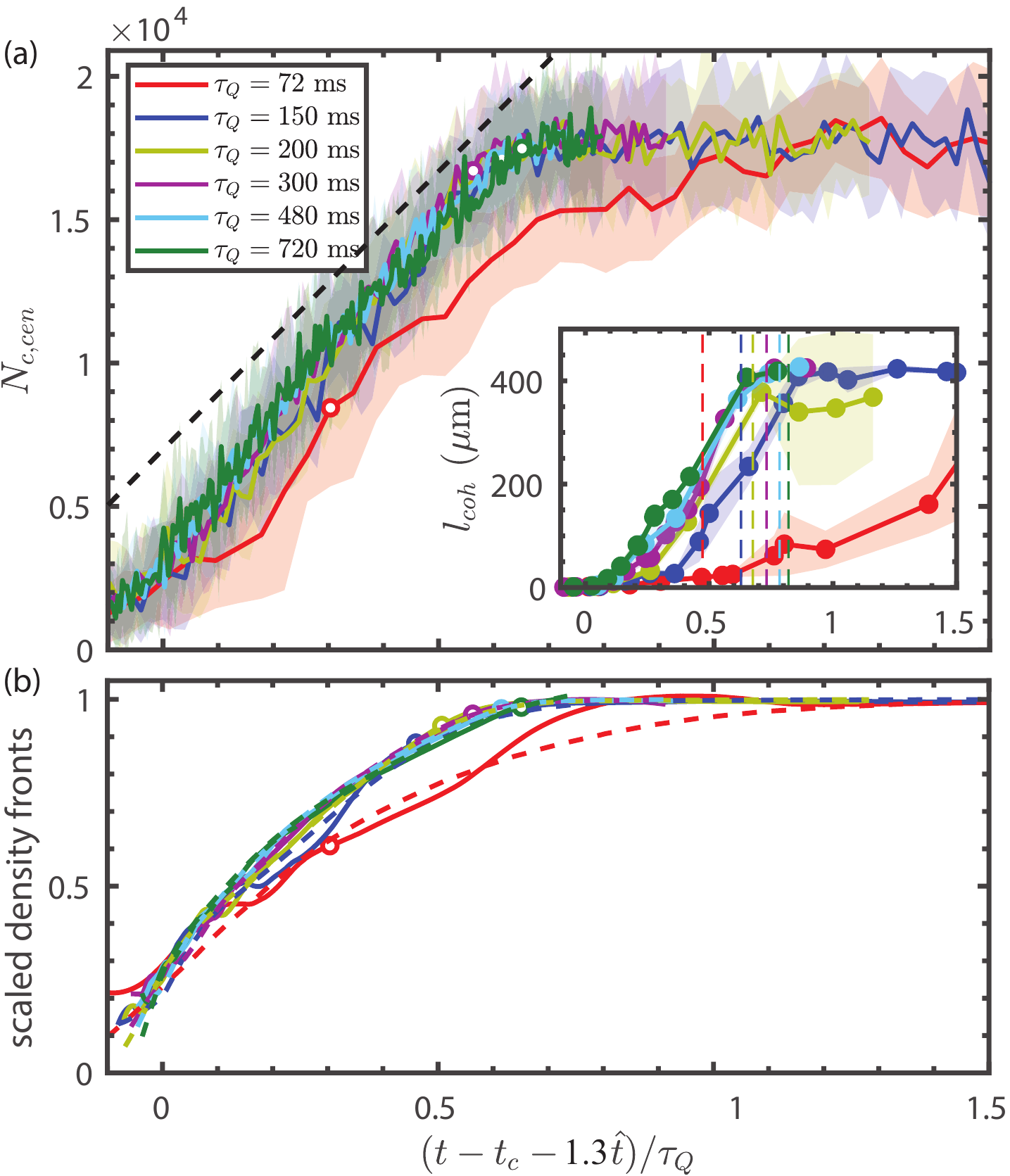}
\caption{
Evolution of (a) the central particle number and (b) scaled density wave fronts plotted in terms of $(t-t_c-1.3\hat{t})/\tau_Q$.
(a) Curves corresponding to different values of $\tau_Q$ collapse onto a single curve, with the exception of the fastest quench $\tau_Q=72$ms (red). The slope of this curve matches the slope of $\mu_f/g$, according to Eq.~(\ref{eqcrude}), which is portrayed by the black dashed line. Hollow circles mark the ends of linear ramps, while faint bands indicate the error bar in the determination of $N_{c,\,cen}$.
Inset: Evolution of $l_{coh}$ with the same time axis clearly demonstrating that coherence growth does not follow the same timescale, except for very slow ramps. Vertical dashed lines label the end of the different ramps. 
(b) Evolution of the position of the density wave fronts (see Appendix C) showing both longitudinal (solid curves) and transverse (dashed curves), scaled according to the corresponding final equilibrium spatial extents ($\sim114\mu$m and $\sim 11\;\mu$m for longitudinal and transverse directions). 
}
\label{fig_8}
\end{figure}

Firstly, we examine the late-time evolution of the central particle number, defined by Eq.~(\ref{Ncen}), in Fig.~\ref{fig_8}(a). While the raw density evolution curves corresponding to different quench rates differ widely [see earlier Fig.~(\ref{fig_6})(a)], when plotted as a function of $(t-t_c-\alpha \hat t)/\tau_Q$ as suggested by Eq.~(\ref{eqcrude}), instead of $(t-t_c)/\hat t$, the different curves collapse nicely for the non-universal constant $\alpha=1.3$. There is only one notable, but not unexpected, outlier: the fastest quench with $\tau_Q=72$ms. This quench has $\hat t$ close to the end of the linear ramp, hence its late time evolution is largely after the end of the linear ramp. 

A similar behaviour is also found for the evolution of the position of the density wave fronts in Fig.~\ref{fig_8}(b), determined by tracing a near constant value of the classical-field density $|\psi(\mathbf{r})|^2$, arbitrarily chosen here in the range $[16,20]~\mu\textrm{m}^{-3}$ to ensure relatively smooth curves (more details on this are given in Appendix~\ref{Appendix-C}).
Again, all data nicely collapse onto a single curve along both directions once the ellipsoidal growth mimicking the underlying anisotropic harmonic confinement is accounted for, consistent with the arguments exposed earlier in Sec.~\ref{homogeneous}, namely that the growth always occurs along an ellipsoid. 
The slightly different dynamical behaviour of the wave front for the $72$~ms ramp at intermediate times can be understood by the fact that this particular fast ramp terminates at $(t-t_c-1.3 \hat t)/\tau_Q \sim 0.3$, as marked by the hollow red circle.

\section{Discussion and conclusions}

We performed a detailed analysis of the early-stage quenched symmetry-breaking dynamics of an elongated harmonically trapped three-dimensional ultracold atomic gas evaporatively cooled from above the Bose-Einstein condensation phase transition temperature at variable rates. Our study was conducted by means of the stochastic projected Gross-Pitaevskii equation for parameters corresponding to a recent experiment, and cast in the language of the Kibble-Zurek mechanism.

Schematically, as the quenched system approaches the critical point from above, it enters a regime where it cannot follow the adiabatic evolution of the equilibrium state, due to the quench proceeding faster than the characteristic diverging relaxation time of the corresponding equilibrium system. Adiabaticity is resumed at a certain time around $+\hat t$ (actually we find a short delay prefactor of $\sim\mathcal{O}(1)$ compared to the standard Kibble Zureck prediction) and the overall process leads to the spontaneous emergence of defects (in this case vortices), with some of those gradually becoming embedded in the growing condensate. Although the system is still evolving during its quenched evolution within the critical region -- rather than remaining frozen in the `impulse' limit of the `cartoon' KZ version -- such evolution still exhibits scaling properties predicted by the KZ mechanism. 

In order to properly characterize the scaling laws for the observables in our SPGPE, we needed to first extract the equilibrium critical temperature of the interacting system numerically. Identification of the equilibrium critical point is crucial to correctly apply the KZ model to a shifted evolution time after the time $t_c$ when the system crosses the corresponding equilibrium critical point. 

Then we used the analytical predictions based on the linearized form of the stochastic Gross-Pitaevskii equation and KZ ordering considerations. Such predictions were found to be valid, allowing quantities like spectral functions, correlation lengths and density growth to collapse onto unique curves for all different quench times probed here and performed in the experiments motivating this work. 

At later times, the growth of the $\mathbf{k=0}$ mode and of the (ellipsoidal) density wave front proceed on similar timescales. However, the presence of highly-excited $\mathbf{k}$-modes associated with the existence of defects in the growing condensate -- which are more pronounced for the fastest quenches -- implies that the phase-ordering process and coherence growth depend on $\tau_Q$ and system geometry/inhomogeneity in a more complicated manner. This highlights the important nature of the decoupling of density and coherence degrees of freedom \cite{KZexp-w}. Although phase ordering for homogeneous systems is an established topic with known scaling laws, the presence of inhomogeneity and anisotropy introduces finite-size effects from the early stages of the evolution, making a collapse of the late-time dynamics particularly tricky even in numerical simulations.

Our work fills the gap between the experimentally observed long time evolution of a temperature quenched condensate and the Kibble-Zurek dynamics at earlier times, near the transition. The study of the early time dynamics is relevant for at least three reasons: i) it is needed to prove that there is an overall consistency in the interpretation of the SPGPE simulations over the whole range of timescales, including the effects of the KZ mechanism which can be understood in terms of a linearized theory and can be related to the later time evolution of the condensate; ii) it clarifies the role of different time and spatial scales in the quench, thus helping to place the homogeneous vs. inhomogeneous KZ mechanism in the proper context of realistic trapped condensates; iii) to our knowledge, new experiments are already planned to observe the early time dynamics during a temperature quench and this work is also meant to serve as a guide for the choice of the appropriate observables and parameters.


\acknowledgements
We would like to thank G. Ferrari and G. Lamporesi for numerous discussions and for providing us with unpublished experimental data. We also acknowledge early discussions with L. Cugliandolo and R. Smith. 
Financial support was provided by the Quantera ERA-NET cofund project NAQUAS through the Engineering and Physical Science Research Council, Grant No. EP/R043434/1 (I-K.L. and N.P.P.), the National Science Centre (NCN), Grant No.~2017/25/Z/ST2/03028 (J.D.), 
and the Consiglio Nazionale delle Ricerche (F.D.).
This work was also supported by Provincia Autonoma di Trento.
S.-C. Gou acknowledges the financial support from the Taiwan MOST 103-2112-M-018-002-MY3 grant.

Data supporting this publication is openly available under an 'Open Data Commons Open Database License' on the data.ncl.ac.uk site~\cite{data_open}.

\appendix
\section*{Appendix}

\section{Determination of Equilibrium Critical Point}
\label{Appendix-A}

\begin{figure}[t!]
\centering
\includegraphics[width=1\linewidth]{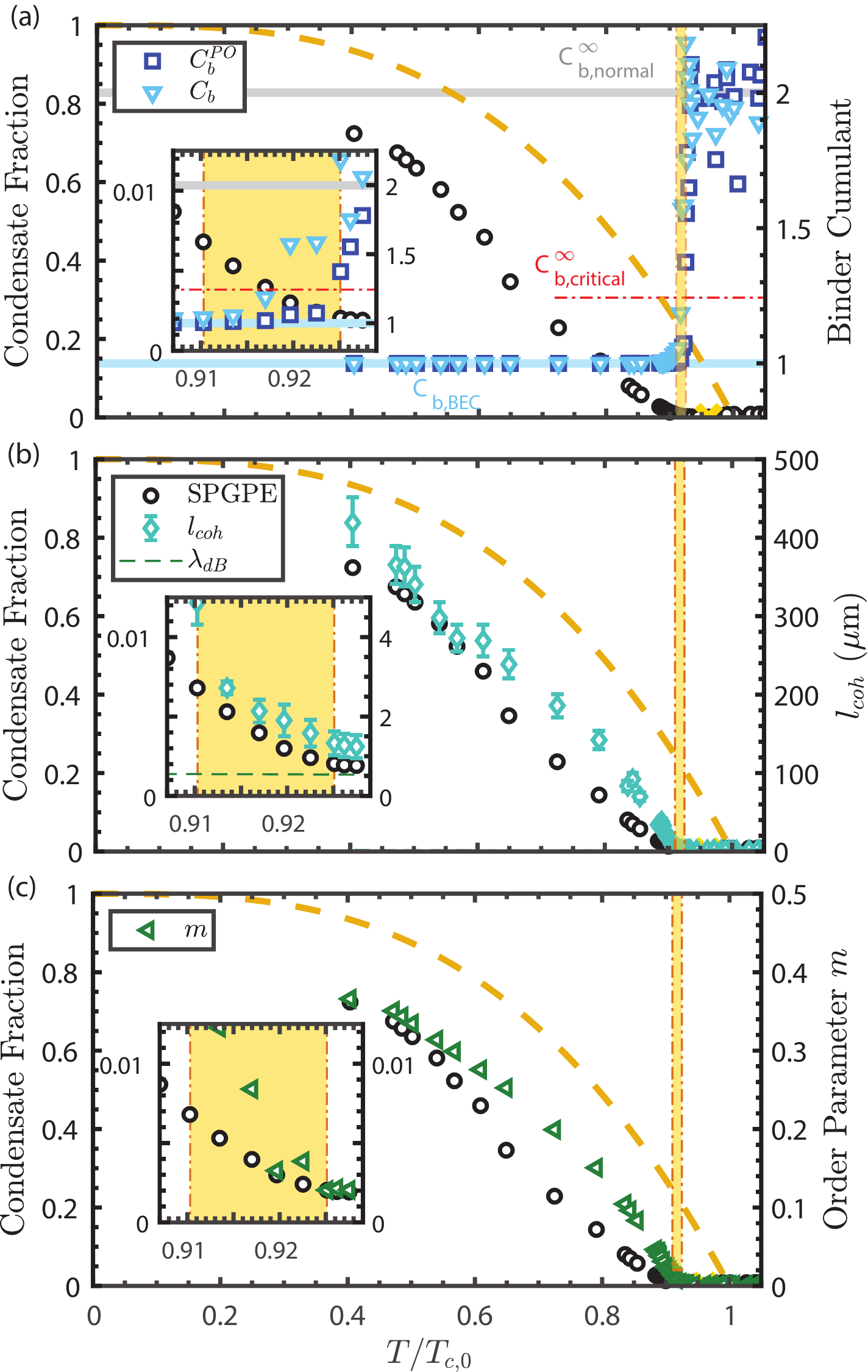}
\caption{
Numerical identification of the critical point from SPGPE equilibrium simulations.
We compare our findings based on (a) the Binder cumulant, (b) the correlation length, and (c) the order parameter $m$. The critical point is found to lie in the yellow band, during which region the Binder cumulant rapidly decreases from $2$ to $1$, by lowering $T$, while the correlation length and order parameter start increasing. For comparison, all plots also depict the corresponding SPGPE condensate fraction already shown in Fig.~\ref{fig_2}(a).
Insets show zoomed in versions of the main plots.
Horizontal lines in (a) indicate limiting values of $C_{b}$, and corresponding homogeneous critical value. 
Horizontal dashed line in inset to (b) indicates the value of the thermal de Broglie wavelength $\lambda_{dB}$ whose value in the critical region is $\sim 0.55 \mu$m. 
}
\label{fig_critical_probe}
\end{figure}

The numerical identification of the critical point from SPGPE equilibrium simulations is performed by using three quantities: the Binder cumulant, the correlation length and the order parameter $m$.

Two closely-related definitions of the Binder cumulant appear in the literature~\cite{Campostrini2001,Davis2003,Bezett2009,wright_proukakis_11,Kobayashi_16b,Fabrizio2018,comaron2019quench}. 
In the first definition, appropriate for a homogeneous system, it is defined in terms of the full classical-field $\psi$ via~\cite{Kobayashi_16b,Fabrizio2018,comaron2019quench}
\begin{equation}
C_b\equiv\frac{\langle |A|^4\rangle}{\langle|A|^2\rangle^2}\ , \hspace{0.3cm} \mathrm{with } \hspace{0.3cm} A=\int d\mathbf{r}\, \psi(\mathbf{r}) \;.
\end{equation}
The second definition has been implemented in the context of the trapped Bose gas, and extracts a similar information but from the condensate mode only~\cite{Bezett2009}: 
\begin{equation}
C_b^{\rm PO}=\frac{\langle |A_{\rm con}|^4\rangle}{\langle |A_{\rm con}|^2\rangle^2}, \hspace{0.3cm} \mathrm{with } \hspace{0.3cm} A_{\rm con}=\int d\mathbf{r}\, \psi_{0}^\ast(\mathbf{r})\psi(\mathbf{r}) \;.
\end{equation}
In both cases, one expects a sharp jump from the value $1$, below $T_c$, to the value $2$, above. The critical value of the Binder cumulant at the transition in the thermodynamic limit is $C_\mathrm{b,\,critical}^{\infty}\sim 1.2430$~\cite{Campostrini2001}, while for trapped Bose gases it is expected to be smaller than $C_\mathrm{b,\,critical}^\infty$ and affected by finite-size effects~\cite{Bezett2009}. Fig.~\ref{fig_critical_probe}(a) shows our numerical results based on both definitions, with their results convincingly overlapping with each other. The jump from $1$ to $2$ is clearly visible and
the critical value $C_\mathrm{b,\,critical}^\infty$ in found in the range $T/T_{c,0}\in(0.91,0.93)$, corresponding to $T\in(438,453)$~nK.

In the critical region, the correlation length is also expected to diverge as $|1-(T/T_c)|^{-\nu}$~\cite{Donner2007,Bezett2009}.
Based on our chosen extraction method for the correlation length, $l_{coh}$, defined by Eq.~(\ref{lcoh}), we indeed find $l_{coh}$ 
starts increasing rapidly in the above probed region, as evident from Fig.~\ref{fig_critical_probe}(b). 
However, the inhomogeneous finite-size nature of the system, our chosen definition of an {\em integrated} coherence length, and our numerical accuracy do not allow for the identification of a sharp critical point, and thus cannot facilitate an accurate determination of the static critical exponent $\nu$.


Finally, we also follow Refs.~\cite{Kobayashi_16b,Fabrizio2018,comaron2019quench} and investigate the behaviour of the order parameter $m$ defined within our computational volume $V$ by
\begin{equation}
m\equiv\frac{1}{\sqrt{V}}\frac{\left\langle\left|\int d\mathbf{r}\, \psi(\mathbf{r})\right|\right\rangle}{\sqrt{\left\langle\int d\mathbf{r}\, \left|\psi(\mathbf{r})\right|^2\right\rangle}}\ .
\end{equation}
This quantity, plotted in Fig.~\ref{fig_critical_probe}(c), is expected to be $m\sim 0$ above the phase transition and $m=1$ for a pure condensate at $T=0$~\cite{Kobayashi_16b,Fabrizio2018}. Again we see that $m$ starts increasing within the same critical region of the Binder cumulant and the correlation length.

The verification that both $l_{coh}$ and $m$ start increasing within the critical regime identified by the Binder cumulant, and the fact that this also coincides with the region when the condensate fraction decreases to zero, provide strong evidence for the consistency of the identification of our critical regime.

The vertical yellow area in Fig.~\ref{fig_critical_probe} highlights the above determined range $T/T_{c,0}\in(0.91,0.93)$. In our system, the corresponding critical chemical potential is $\mu_c=(4.13\pm0.55)\hbar\omega_\perp$ and the critical time $t_c$ in our quench protocol is 
\be 
t_c=\tau_Q\frac{\mu_c}{\mu_f}\sim(0.188\pm0.026)\tau_Q \ . 
\ee 

\begin{figure}[t!]
\centering
\includegraphics[width=1\linewidth]{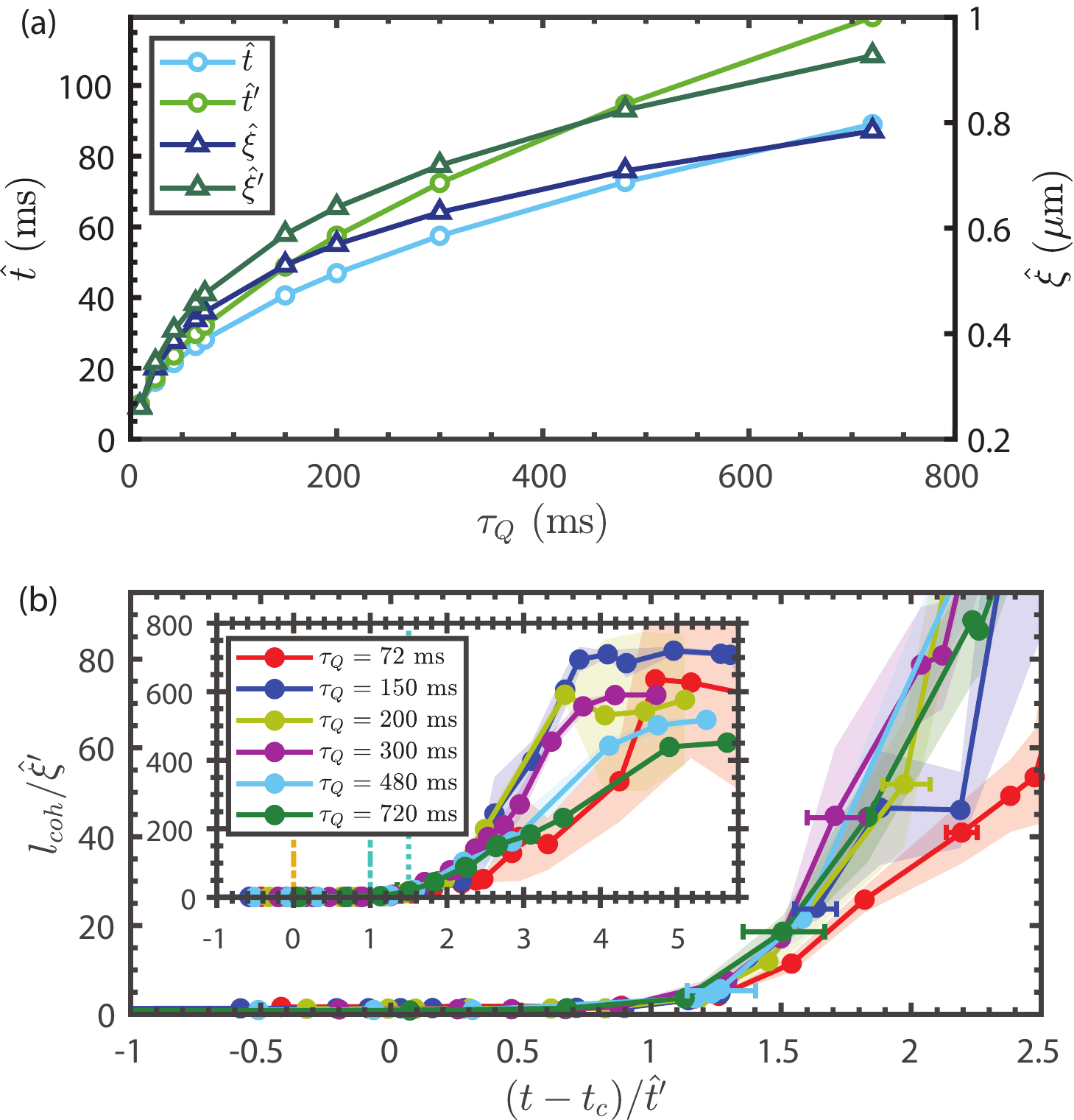}
\caption{(a) Quantities $\hat t^\prime$ and $\hat\xi^\prime$ given in Eqs.~(\ref{hattnu}) as a function of $\tau_Q$, compared to $\hat t$ and $\hat\xi$ used in the main text. The former correspond to the exact equilibrium critical exponent $\nu=0.67$; the latter to mean-field value $\nu=0.5$. (b)
Longitudinal correlation length, as in Fig.~\ref{fig_7}, but rescaled by using $(t-t_c)/\hat t^\prime$ and $\hat\xi^\prime$.
}
\label{fig_appendix_KZ_scaling}
\end{figure}

\begin{figure*}[t!]
\centering
\includegraphics[width=1\linewidth]{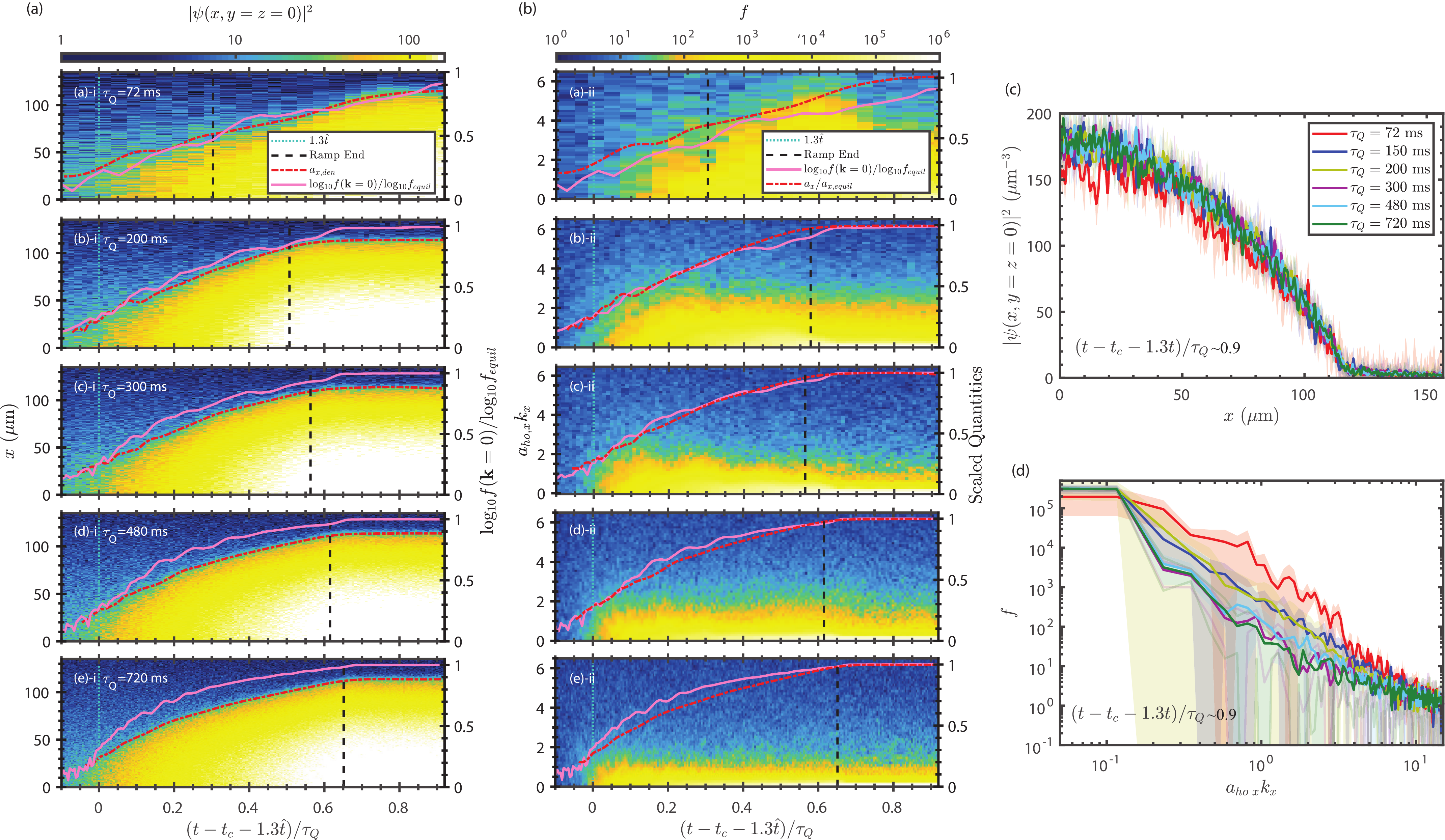}
\caption{Column (a): Density distribution of atoms in the classical field for different quench times, plotted as a function of $(t-t_c-1.3\hat{t})/\tau_Q$. Column (b) spectral function (\ref{fs}) for the same cases. In both columns, the dashed red line is the numerically-traced position of the density wave front for the density within $[16,20]$~$\mu\textrm{m}^{-3}$,
while the pink line is spectral function $f$ of the $\mathbf{k=0}$ mode. Panels (c) and (d) show the profiles of the density distribution and the spectral function, respectively, at the time $(t-t_c-1.3\hat{t})/\tau_Q\approx0.9$, revealing the extent of excitation still present for the faster quenches.
}
\label{fig_den_nk}
\end{figure*}

\section{Mean field vs. exact critical behaviour}
\label{Appendix-B}

In Section \ref{linearizedSPGPE} we introduced the linearized SPGPE approach with the Gaussian approximation. Within this approach, the equilibrium critical exponent should coincide with the mean-field exponent $\nu=1/2$, leading to the scaling law $\hat{t}\propto\tau_Q^{1/2}$. 

A natural question arises about whether the scalings/collapses presented in the main text would be significantly affected upon using instead the exact value of $\nu=0.67$. This would in fact lead to the slightly amended scalings:
\begin{equation}
\hat t^\prime = \tau_0^{0.43} \tau_Q^{0.57} \ \ \ {\rm and } \ \
\hat \xi^\prime = \xi_0
\left(\tau_Q/\tau_0\right)^{0.29}\ ,
\label{hattnu}
\end{equation}
with $\tau_0=\hbar/(\gamma\mu_f)$ and $\xi_0=(2M\mu_f/\hbar^2)^{-1/2}$.
The differences between the above expressions and the corresponding mean-field values of $\hat{t}$ and $\hat{\xi}$ are shown in Fig.~\ref{fig_appendix_KZ_scaling}(a). 
They are not significant in the considered regime of parameters. 
In panel (b) of the same figure, we present the correlation length and its rescaling with $\hat{t}^\prime$ and $\hat{\xi}^\prime$. Similarly to Fig.~\ref{fig_7}, except for the fastest quench with $\tau_Q=72$ms, the curves for different quench times $\tau_Q$ collapse onto a single curve for short times. Overall, this analysis suggests that our dynamical results cannot accurately distinguish between the two values of $\nu$. The KZ exponents they imply do not differ enough.

\section{Density distributions and spectral function}
\label{Appendix-C}

In the main text, we have identified the shifted time $(t-t_c-1.3\hat t)$ as the time when the system exits the KZ self-similar regime where evolutions corresponding to different quench times $\tau_Q$ collapse onto one another. At later times, the dominant timescale governing the system is the quench time, which determines the rate at which the system is ramped to its low-$T$ state (with the exception of the fastest ramp, $\tau_Q=72$~ms, which ends while the system is still well within the self-similar regime).

As a complement to our previous analysis, here we further investigate the behavior of the density and the spectral function of the gas during a quench. Fig.~\ref{fig_den_nk}(a)-(b) compares the evolution of density wavefront and $\mathbf{k=0}$ mode alongside the evolution of the full density distribution and spectral function. Specifically, Fig.~\ref{fig_den_nk}(a) shows how the axial system density grows as a function of time. The front of the growing density area has been traced for the lowest value of the density which allows a relatively smooth curve, and is shown by the dashed red line. The corresponding wave fronts for different $\tau_Q$ have already been discussed in Fig.~\ref{fig_8}(b), where they were shown not only to collapse on top of each other, but also on top of the corresponding transversal evolution wave front, once the system geometry/anisotropy were appropriately accounted for.
In Fig.~\ref{fig_den_nk}(c)-(d) we also plot the long term evolution of the the density distribution and the spectral function at time $(t-t_c-1.3\hat t)/\tau_Q\approx0.9$

We can thus draw various conclusions already briefly commented upon in the main text:
(i) From Fig.~\ref{fig_den_nk}(a)-(b), we see that the growth of the density wavefronts (dashed red lines) overlaps almost perfectly with that of the $\mathbf{k=0}$ modes (solid pink lines). This suggests that growth on this timescale is driven by the $\mathbf{k=0}$ mode, consistent with bosonic amplification. However, (ii) although density and $\mathbf{k=0}$ mode grow in parallel, the spectrum of higher excited modes looks very different on such scaled time [Fig.~\ref{fig_den_nk}(b)]. For comparison, the instantaneous spectral function at time $(t-t_c-1.3\hat t)/\tau_Q\approx0.9$ (when densities have mostly saturated) is plotted in Fig.~\ref{fig_den_nk}(d).
We thus see that although low momentum modes are mostly excited for slow quenches, whose late-time momentum distribution is consistent with the Bose-Einstein distribution, faster quenches generate more modes with higher $\mathbf{k}$, with the highest excited modes subsequently relaxing only gradually, and on a much longer timescale.
Importantly, faster ramps are still in a far-from-equilibrium state at $(t-t_c-1.3\hat t)/\tau_Q\approx0.9$, even though both density wavefronts and $\mathbf{k=0}$ mode occupations are close to saturating at such a time. This offers a clear perspective of the previously inferred decoupling between density and momentum/coherence relaxation.



\bibliographystyle{apsrev4-1}

\begin{thebibliography}{142}%
\makeatletter
\providecommand \@ifxundefined [1]{%
 \@ifx{#1\undefined}
}%
\providecommand \@ifnum [1]{%
 \ifnum #1\expandafter \@firstoftwo
 \else \expandafter \@secondoftwo
 \fi
}%
\providecommand \@ifx [1]{%
 \ifx #1\expandafter \@firstoftwo
 \else \expandafter \@secondoftwo
 \fi
}%
\providecommand \natexlab [1]{#1}%
\providecommand \enquote  [1]{``#1''}%
\providecommand \bibnamefont  [1]{#1}%
\providecommand \bibfnamefont [1]{#1}%
\providecommand \citenamefont [1]{#1}%
\providecommand \href@noop [0]{\@secondoftwo}%
\providecommand \href [0]{\begingroup \@sanitize@url \@href}%
\providecommand \@href[1]{\@@startlink{#1}\@@href}%
\providecommand \@@href[1]{\endgroup#1\@@endlink}%
\providecommand \@sanitize@url [0]{\catcode `\\12\catcode `\$12\catcode
  `\&12\catcode `\#12\catcode `\^12\catcode `\_12\catcode `\%12\relax}%
\providecommand \@@startlink[1]{}%
\providecommand \@@endlink[0]{}%
\providecommand \url  [0]{\begingroup\@sanitize@url \@url }%
\providecommand \@url [1]{\endgroup\@href {#1}{\urlprefix }}%
\providecommand \urlprefix  [0]{URL }%
\providecommand \Eprint [0]{\href }%
\providecommand \doibase [0]{http://dx.doi.org/}%
\providecommand \selectlanguage [0]{\@gobble}%
\providecommand \bibinfo  [0]{\@secondoftwo}%
\providecommand \bibfield  [0]{\@secondoftwo}%
\providecommand \translation [1]{[#1]}%
\providecommand \BibitemOpen [0]{}%
\providecommand \bibitemStop [0]{}%
\providecommand \bibitemNoStop [0]{.\EOS\space}%
\providecommand \EOS [0]{\spacefactor3000\relax}%
\providecommand \BibitemShut  [1]{\csname bibitem#1\endcsname}%
\let\auto@bib@innerbib\@empty
\bibitem [{\citenamefont {Kibble}(1976)}]{K-a}%
  \BibitemOpen
  \bibfield  {author} {\bibinfo {author} {\bibfnamefont {T.~W.~B.}\
  \bibnamefont {Kibble}},\ }\href {\doibase 10.1088/0305-4470/9/8/029}
  {\bibfield  {journal} {\bibinfo  {journal} {J. Phys.}\ }\textbf {\bibinfo
  {volume} {A9}},\ \bibinfo {pages} {1387} (\bibinfo {year}
  {1976})}\BibitemShut {NoStop}%
\bibitem [{\citenamefont {Kibble}(1980)}]{K-b}%
  \BibitemOpen
  \bibfield  {author} {\bibinfo {author} {\bibfnamefont {T.~W.~B.}\
  \bibnamefont {Kibble}},\ }\href {\doibase
  https://doi.org/10.1016/0370-1573(80)90091-5} {\bibfield  {journal} {\bibinfo
   {journal} {Physics Reports}\ }\textbf {\bibinfo {volume} {67}},\ \bibinfo
  {pages} {183 } (\bibinfo {year} {1980})}\BibitemShut {NoStop}%
\bibitem [{\citenamefont {Kibble}(2007)}]{K-c}%
  \BibitemOpen
  \bibfield  {author} {\bibinfo {author} {\bibfnamefont {T.~W.~B.}\
  \bibnamefont {Kibble}},\ }\href {\doibase 10.1063/1.2784684} {\bibfield
  {journal} {\bibinfo  {journal} {Physics Today}\ }\textbf {\bibinfo {volume}
  {60}},\ \bibinfo {pages} {47} (\bibinfo {year} {2007})}\BibitemShut {NoStop}%
\bibitem [{\citenamefont {Zurek}(1985)}]{Z-a}%
  \BibitemOpen
  \bibfield  {author} {\bibinfo {author} {\bibfnamefont {W.~H.}\ \bibnamefont
  {Zurek}},\ }\href {\doibase 10.1038/317505a0} {\bibfield  {journal} {\bibinfo
   {journal} {Nature}\ }\textbf {\bibinfo {volume} {317}},\ \bibinfo {pages}
  {505} (\bibinfo {year} {1985})}\BibitemShut {NoStop}%
\bibitem [{\citenamefont {Zurek}(1993)}]{Z-b}%
  \BibitemOpen
  \bibfield  {author} {\bibinfo {author} {\bibfnamefont {W.~H.}\ \bibnamefont
  {Zurek}},\ }\href {https://www.actaphys.uj.edu.pl/R/24/7/1301/pdf} {\bibfield
   {journal} {\bibinfo  {journal} {Acta Phys. Polon.}\ }\textbf {\bibinfo
  {volume} {B24}},\ \bibinfo {pages} {1301} (\bibinfo {year}
  {1993})}\BibitemShut {NoStop}%
\bibitem [{\citenamefont {Zurek}(1996)}]{Z-c}%
  \BibitemOpen
  \bibfield  {author} {\bibinfo {author} {\bibfnamefont {W.~H.}\ \bibnamefont
  {Zurek}},\ }\href {\doibase https://doi.org/10.1016/S0370-1573(96)00009-9}
  {\bibfield  {journal} {\bibinfo  {journal} {Physics Reports}\ }\textbf
  {\bibinfo {volume} {276}},\ \bibinfo {pages} {177 } (\bibinfo {year}
  {1996})}\BibitemShut {NoStop}%
\bibitem [{\citenamefont {del Campo}\ and\ \citenamefont {Zurek}(2014)}]{Z-d}%
  \BibitemOpen
  \bibfield  {author} {\bibinfo {author} {\bibfnamefont {A.}~\bibnamefont {del
  Campo}}\ and\ \bibinfo {author} {\bibfnamefont {W.~H.}\ \bibnamefont
  {Zurek}},\ }\href {\doibase 10.1142/s0217751x1430018x} {\bibfield  {journal}
  {\bibinfo  {journal} {Int. J. Mod. Phys. A}\ }\textbf {\bibinfo {volume}
  {29}},\ \bibinfo {pages} {1430018} (\bibinfo {year} {2014})}\BibitemShut
  {NoStop}%
\bibitem [{\citenamefont {Laguna}\ and\ \citenamefont {Zurek}(1997)}]{KZnum-a}%
  \BibitemOpen
  \bibfield  {author} {\bibinfo {author} {\bibfnamefont {P.}~\bibnamefont
  {Laguna}}\ and\ \bibinfo {author} {\bibfnamefont {W.~H.}\ \bibnamefont
  {Zurek}},\ }\href {\doibase 10.1103/PhysRevLett.78.2519} {\bibfield
  {journal} {\bibinfo  {journal} {Phys. Rev. Lett.}\ }\textbf {\bibinfo
  {volume} {78}},\ \bibinfo {pages} {2519} (\bibinfo {year}
  {1997})}\BibitemShut {NoStop}%
\bibitem [{\citenamefont {Yates}\ and\ \citenamefont {Zurek}(1998)}]{KZnum-b}%
  \BibitemOpen
  \bibfield  {author} {\bibinfo {author} {\bibfnamefont {A.}~\bibnamefont
  {Yates}}\ and\ \bibinfo {author} {\bibfnamefont {W.~H.}\ \bibnamefont
  {Zurek}},\ }\href {\doibase 10.1103/PhysRevLett.80.5477} {\bibfield
  {journal} {\bibinfo  {journal} {Phys. Rev. Lett.}\ }\textbf {\bibinfo
  {volume} {80}},\ \bibinfo {pages} {5477} (\bibinfo {year}
  {1998})}\BibitemShut {NoStop}%
\bibitem [{\citenamefont {Dziarmaga}\ \emph {et~al.}(1999)\citenamefont
  {Dziarmaga}, \citenamefont {Laguna},\ and\ \citenamefont {Zurek}}]{KZnum-c}%
  \BibitemOpen
  \bibfield  {author} {\bibinfo {author} {\bibfnamefont {J.}~\bibnamefont
  {Dziarmaga}}, \bibinfo {author} {\bibfnamefont {P.}~\bibnamefont {Laguna}}, \
  and\ \bibinfo {author} {\bibfnamefont {W.~H.}\ \bibnamefont {Zurek}},\ }\href
  {\doibase 10.1103/PhysRevLett.82.4749} {\bibfield  {journal} {\bibinfo
  {journal} {Phys. Rev. Lett.}\ }\textbf {\bibinfo {volume} {82}},\ \bibinfo
  {pages} {4749} (\bibinfo {year} {1999})}\BibitemShut {NoStop}%
\bibitem [{\citenamefont {Antunes}\ \emph {et~al.}(1999)\citenamefont
  {Antunes}, \citenamefont {Bettencourt},\ and\ \citenamefont
  {Zurek}}]{KZnum-d}%
  \BibitemOpen
  \bibfield  {author} {\bibinfo {author} {\bibfnamefont {N.~D.}\ \bibnamefont
  {Antunes}}, \bibinfo {author} {\bibfnamefont {L.~M.~A.}\ \bibnamefont
  {Bettencourt}}, \ and\ \bibinfo {author} {\bibfnamefont {W.~H.}\ \bibnamefont
  {Zurek}},\ }\href {\doibase 10.1103/PhysRevLett.82.2824} {\bibfield
  {journal} {\bibinfo  {journal} {Phys. Rev. Lett.}\ }\textbf {\bibinfo
  {volume} {82}},\ \bibinfo {pages} {2824} (\bibinfo {year}
  {1999})}\BibitemShut {NoStop}%
\bibitem [{\citenamefont {Bettencourt}\ \emph {et~al.}(2000)\citenamefont
  {Bettencourt}, \citenamefont {Antunes},\ and\ \citenamefont
  {Zurek}}]{KZnum-e}%
  \BibitemOpen
  \bibfield  {author} {\bibinfo {author} {\bibfnamefont {L.~M.~A.}\
  \bibnamefont {Bettencourt}}, \bibinfo {author} {\bibfnamefont {N.~D.}\
  \bibnamefont {Antunes}}, \ and\ \bibinfo {author} {\bibfnamefont {W.~H.}\
  \bibnamefont {Zurek}},\ }\href {\doibase 10.1103/PhysRevD.62.065005}
  {\bibfield  {journal} {\bibinfo  {journal} {Phys. Rev. D}\ }\textbf {\bibinfo
  {volume} {62}},\ \bibinfo {pages} {065005} (\bibinfo {year}
  {2000})}\BibitemShut {NoStop}%
\bibitem [{\citenamefont {Zurek}\ \emph {et~al.}(2000)\citenamefont {Zurek},
  \citenamefont {Bettencourt}, \citenamefont {Dziarmaga},\ and\ \citenamefont
  {Antunes}}]{KZnum-f}%
  \BibitemOpen
  \bibfield  {author} {\bibinfo {author} {\bibfnamefont {W.~H.}\ \bibnamefont
  {Zurek}}, \bibinfo {author} {\bibfnamefont {L.~M.~A.}\ \bibnamefont
  {Bettencourt}}, \bibinfo {author} {\bibfnamefont {J.}~\bibnamefont
  {Dziarmaga}}, \ and\ \bibinfo {author} {\bibfnamefont {N.~D.}\ \bibnamefont
  {Antunes}},\ }\href {https://www.actaphys.uj.edu.pl/R/31/12/2937/pdf}
  {\bibfield  {journal} {\bibinfo  {journal} {Acta Phys. Pol. B}\ }\textbf
  {\bibinfo {volume} {31}},\ \bibinfo {pages} {2937} (\bibinfo {year}
  {2000})}\BibitemShut {NoStop}%
\bibitem [{\citenamefont {Uhlmann}\ \emph {et~al.}(2007)\citenamefont
  {Uhlmann}, \citenamefont {Sch\"utzhold},\ and\ \citenamefont
  {Fischer}}]{KZnum-g}%
  \BibitemOpen
  \bibfield  {author} {\bibinfo {author} {\bibfnamefont {M.}~\bibnamefont
  {Uhlmann}}, \bibinfo {author} {\bibfnamefont {R.}~\bibnamefont
  {Sch\"utzhold}}, \ and\ \bibinfo {author} {\bibfnamefont {U.~R.}\
  \bibnamefont {Fischer}},\ }\href {\doibase 10.1103/PhysRevLett.99.120407}
  {\bibfield  {journal} {\bibinfo  {journal} {Phys. Rev. Lett.}\ }\textbf
  {\bibinfo {volume} {99}},\ \bibinfo {pages} {120407} (\bibinfo {year}
  {2007})}\BibitemShut {NoStop}%
\bibitem [{\citenamefont {Zurek}(2009)}]{inhomo_classical-c}%
  \BibitemOpen
  \bibfield  {author} {\bibinfo {author} {\bibfnamefont {W.~H.}\ \bibnamefont
  {Zurek}},\ }\href {\doibase 10.1103/PhysRevLett.102.105702} {\bibfield
  {journal} {\bibinfo  {journal} {Phys. Rev. Lett.}\ }\textbf {\bibinfo
  {volume} {102}},\ \bibinfo {pages} {105702} (\bibinfo {year}
  {2009})}\BibitemShut {NoStop}%
\bibitem [{\citenamefont {Damski}\ and\ \citenamefont
  {Zurek}(2010)}]{QKZteor-i}%
  \BibitemOpen
  \bibfield  {author} {\bibinfo {author} {\bibfnamefont {B.}~\bibnamefont
  {Damski}}\ and\ \bibinfo {author} {\bibfnamefont {W.~H.}\ \bibnamefont
  {Zurek}},\ }\href {\doibase 10.1103/PhysRevLett.104.160404} {\bibfield
  {journal} {\bibinfo  {journal} {Phys. Rev. Lett.}\ }\textbf {\bibinfo
  {volume} {104}},\ \bibinfo {pages} {160404} (\bibinfo {year}
  {2010})}\BibitemShut {NoStop}%
\bibitem [{\citenamefont {Uhlmann}\ \emph
  {et~al.}(2010{\natexlab{a}})\citenamefont {Uhlmann}, \citenamefont
  {Sch\"utzhold},\ and\ \citenamefont {Fischer}}]{KZnum-h}%
  \BibitemOpen
  \bibfield  {author} {\bibinfo {author} {\bibfnamefont {M.}~\bibnamefont
  {Uhlmann}}, \bibinfo {author} {\bibfnamefont {R.}~\bibnamefont
  {Sch\"utzhold}}, \ and\ \bibinfo {author} {\bibfnamefont {U.~R.}\
  \bibnamefont {Fischer}},\ }\href {\doibase 10.1103/PhysRevD.81.025017}
  {\bibfield  {journal} {\bibinfo  {journal} {Phys. Rev. D}\ }\textbf {\bibinfo
  {volume} {81}},\ \bibinfo {pages} {025017} (\bibinfo {year}
  {2010}{\natexlab{a}})}\BibitemShut {NoStop}%
\bibitem [{\citenamefont {Uhlmann}\ \emph
  {et~al.}(2010{\natexlab{b}})\citenamefont {Uhlmann}, \citenamefont
  {Schützhold},\ and\ \citenamefont {Fischer}}]{KZnum-i}%
  \BibitemOpen
  \bibfield  {author} {\bibinfo {author} {\bibfnamefont {M.}~\bibnamefont
  {Uhlmann}}, \bibinfo {author} {\bibfnamefont {R.}~\bibnamefont
  {Schützhold}}, \ and\ \bibinfo {author} {\bibfnamefont {U.~R.}\ \bibnamefont
  {Fischer}},\ }\href {\doibase 10.1088/1367-2630/12/9/095020} {\bibfield
  {journal} {\bibinfo  {journal} {New J. Phys}\ }\textbf {\bibinfo {volume}
  {12}},\ \bibinfo {pages} {095020} (\bibinfo {year}
  {2010}{\natexlab{b}})}\BibitemShut {NoStop}%
\bibitem [{\citenamefont {Witkowska}\ \emph {et~al.}(2011)\citenamefont
  {Witkowska}, \citenamefont {Deuar}, \citenamefont {Gajda},\ and\
  \citenamefont {Rz\k{a}\ifmmode~\dot{z}\else \.{z}\fi{}ewski}}]{KZnum-j}%
  \BibitemOpen
  \bibfield  {author} {\bibinfo {author} {\bibfnamefont {E.}~\bibnamefont
  {Witkowska}}, \bibinfo {author} {\bibfnamefont {P.}~\bibnamefont {Deuar}},
  \bibinfo {author} {\bibfnamefont {M.}~\bibnamefont {Gajda}}, \ and\ \bibinfo
  {author} {\bibfnamefont {K.}~\bibnamefont {Rz\k{a}\ifmmode~\dot{z}\else
  \.{z}\fi{}ewski}},\ }\href {\doibase 10.1103/PhysRevLett.106.135301}
  {\bibfield  {journal} {\bibinfo  {journal} {Phys. Rev. Lett.}\ }\textbf
  {\bibinfo {volume} {106}},\ \bibinfo {pages} {135301} (\bibinfo {year}
  {2011})}\BibitemShut {NoStop}%
\bibitem [{\citenamefont {Das}\ \emph {et~al.}(2012)\citenamefont {Das},
  \citenamefont {Sabbatini},\ and\ \citenamefont {Zurek}}]{KZnum-k}%
  \BibitemOpen
  \bibfield  {author} {\bibinfo {author} {\bibfnamefont {A.}~\bibnamefont
  {Das}}, \bibinfo {author} {\bibfnamefont {J.}~\bibnamefont {Sabbatini}}, \
  and\ \bibinfo {author} {\bibfnamefont {W.~H.}\ \bibnamefont {Zurek}},\ }\href
  {\doibase 10.1038/srep00352} {\bibfield  {journal} {\bibinfo  {journal} {Sci.
  Rep.}\ }\textbf {\bibinfo {volume} {2}} (\bibinfo {year} {2012}),\
  10.1038/srep00352}\BibitemShut {NoStop}%
\bibitem [{\citenamefont {Sonner}\ \emph {et~al.}(2015)\citenamefont {Sonner},
  \citenamefont {del Campo},\ and\ \citenamefont {Zurek}}]{KZnum-l}%
  \BibitemOpen
  \bibfield  {author} {\bibinfo {author} {\bibfnamefont {J.}~\bibnamefont
  {Sonner}}, \bibinfo {author} {\bibfnamefont {A.}~\bibnamefont {del Campo}}, \
  and\ \bibinfo {author} {\bibfnamefont {W.~H.}\ \bibnamefont {Zurek}},\ }\href
  {\doibase 10.1038/ncomms8406} {\bibfield  {journal} {\bibinfo  {journal}
  {Nat. Comm.}\ }\textbf {\bibinfo {volume} {6}},\ \bibinfo {pages} {7406}
  (\bibinfo {year} {2015})}\BibitemShut {NoStop}%
\bibitem [{\citenamefont {Chesler}\ \emph {et~al.}(2015)\citenamefont
  {Chesler}, \citenamefont {Garc\'{\i}a-Garc\'{\i}a},\ and\ \citenamefont
  {Liu}}]{KZnum-m}%
  \BibitemOpen
  \bibfield  {author} {\bibinfo {author} {\bibfnamefont {P.~M.}\ \bibnamefont
  {Chesler}}, \bibinfo {author} {\bibfnamefont {A.~M.}\ \bibnamefont
  {Garc\'{\i}a-Garc\'{\i}a}}, \ and\ \bibinfo {author} {\bibfnamefont
  {H.}~\bibnamefont {Liu}},\ }\href {\doibase 10.1103/PhysRevX.5.021015}
  {\bibfield  {journal} {\bibinfo  {journal} {Phys. Rev. X}\ }\textbf {\bibinfo
  {volume} {5}},\ \bibinfo {pages} {021015} (\bibinfo {year}
  {2015})}\BibitemShut {NoStop}%
\bibitem [{\citenamefont {Liu}\ \emph {et~al.}(2018)\citenamefont {Liu},
  \citenamefont {Donadello}, \citenamefont {Lamporesi}, \citenamefont
  {Ferrari}, \citenamefont {Gou}, \citenamefont {Dalfovo},\ and\ \citenamefont
  {Proukakis}}]{KZexp-w}%
  \BibitemOpen
  \bibfield  {author} {\bibinfo {author} {\bibfnamefont {I.-K.}\ \bibnamefont
  {Liu}}, \bibinfo {author} {\bibfnamefont {S.}~\bibnamefont {Donadello}},
  \bibinfo {author} {\bibfnamefont {G.}~\bibnamefont {Lamporesi}}, \bibinfo
  {author} {\bibfnamefont {G.}~\bibnamefont {Ferrari}}, \bibinfo {author}
  {\bibfnamefont {S.-C.}\ \bibnamefont {Gou}}, \bibinfo {author} {\bibfnamefont
  {F.}~\bibnamefont {Dalfovo}}, \ and\ \bibinfo {author} {\bibfnamefont
  {N.~P.}\ \bibnamefont {Proukakis}},\ }\href {\doibase
  10.1038/s42005-018-0023-6} {\bibfield  {journal} {\bibinfo  {journal}
  {Commun. Phys.}\ }\textbf {\bibinfo {volume} {1}},\ \bibinfo {pages} {24}
  (\bibinfo {year} {2018})}\BibitemShut {NoStop}%
\bibitem [{\citenamefont {Su}\ \emph {et~al.}(2013)\citenamefont {Su},
  \citenamefont {Gou}, \citenamefont {Bradley}, \citenamefont {Fialko},\ and\
  \citenamefont {Brand}}]{Su13}%
  \BibitemOpen
  \bibfield  {author} {\bibinfo {author} {\bibfnamefont {S.-W.}\ \bibnamefont
  {Su}}, \bibinfo {author} {\bibfnamefont {S.-C.}\ \bibnamefont {Gou}},
  \bibinfo {author} {\bibfnamefont {A.}~\bibnamefont {Bradley}}, \bibinfo
  {author} {\bibfnamefont {O.}~\bibnamefont {Fialko}}, \ and\ \bibinfo {author}
  {\bibfnamefont {J.}~\bibnamefont {Brand}},\ }\href {\doibase
  10.1103/PhysRevLett.110.215302} {\bibfield  {journal} {\bibinfo  {journal}
  {Phys. Rev. Lett.}\ }\textbf {\bibinfo {volume} {110}},\ \bibinfo {pages}
  {215302} (\bibinfo {year} {2013})}\BibitemShut {NoStop}%
\bibitem [{\citenamefont {McDonald}\ and\ \citenamefont
  {Bradley}(2015)}]{McDonald2015}%
  \BibitemOpen
  \bibfield  {author} {\bibinfo {author} {\bibfnamefont {R.~G.}\ \bibnamefont
  {McDonald}}\ and\ \bibinfo {author} {\bibfnamefont {A.~S.}\ \bibnamefont
  {Bradley}},\ }\href {\doibase 10.1103/PhysRevA.92.033616} {\bibfield
  {journal} {\bibinfo  {journal} {Phys. Rev. A}\ }\textbf {\bibinfo {volume}
  {92}},\ \bibinfo {pages} {033616} (\bibinfo {year} {2015})}\BibitemShut
  {NoStop}%
\bibitem [{\citenamefont {Liu}\ \emph {et~al.}(2016)\citenamefont {Liu},
  \citenamefont {Pattinson}, \citenamefont {Billam}, \citenamefont {Gardiner},
  \citenamefont {Cornish}, \citenamefont {Huang}, \citenamefont {Lin},
  \citenamefont {Gou}, \citenamefont {Parker},\ and\ \citenamefont
  {Proukakis}}]{Liu_2016}%
  \BibitemOpen
  \bibfield  {author} {\bibinfo {author} {\bibfnamefont {I.-K.}\ \bibnamefont
  {Liu}}, \bibinfo {author} {\bibfnamefont {R.~W.}\ \bibnamefont {Pattinson}},
  \bibinfo {author} {\bibfnamefont {T.~P.}\ \bibnamefont {Billam}}, \bibinfo
  {author} {\bibfnamefont {S.~A.}\ \bibnamefont {Gardiner}}, \bibinfo {author}
  {\bibfnamefont {S.~L.}\ \bibnamefont {Cornish}}, \bibinfo {author}
  {\bibfnamefont {T.-M.}\ \bibnamefont {Huang}}, \bibinfo {author}
  {\bibfnamefont {W.-W.}\ \bibnamefont {Lin}}, \bibinfo {author} {\bibfnamefont
  {S.-C.}\ \bibnamefont {Gou}}, \bibinfo {author} {\bibfnamefont {N.~G.}\
  \bibnamefont {Parker}}, \ and\ \bibinfo {author} {\bibfnamefont {N.~P.}\
  \bibnamefont {Proukakis}},\ }\href {\doibase 10.1103/PhysRevA.93.023628}
  {\bibfield  {journal} {\bibinfo  {journal} {Phys. Rev. A}\ }\textbf {\bibinfo
  {volume} {93}},\ \bibinfo {pages} {023628} (\bibinfo {year}
  {2016})}\BibitemShut {NoStop}%
\bibitem [{\citenamefont {Bland}\ \emph {et~al.}(2020)\citenamefont {Bland},
  \citenamefont {Marolleau}, \citenamefont {Comaron}, \citenamefont {Malomed},\
  and\ \citenamefont {Proukakis}}]{bland_marolleau_20}%
  \BibitemOpen
  \bibfield  {author} {\bibinfo {author} {\bibfnamefont {T.}~\bibnamefont
  {Bland}}, \bibinfo {author} {\bibfnamefont {Q.}~\bibnamefont {Marolleau}},
  \bibinfo {author} {\bibfnamefont {P.}~\bibnamefont {Comaron}}, \bibinfo
  {author} {\bibfnamefont {B.}~\bibnamefont {Malomed}}, \ and\ \bibinfo
  {author} {\bibfnamefont {N.~P.}\ \bibnamefont {Proukakis}},\ }\href
  {http://iopscience.iop.org/10.1088/1361-6455/ab81e9} {\bibfield  {journal}
  {\bibinfo  {journal} {Journal of Physics B: Atomic, Molecular and Optical
  Physics}\ } (\bibinfo {year} {2020})}\BibitemShut {NoStop}%
\bibitem [{\citenamefont {Zamora}\ \emph {et~al.}(2020)\citenamefont {Zamora},
  \citenamefont {Dagvadorj}, \citenamefont {Comaron}, \citenamefont
  {Carusotto}, \citenamefont {Proukakis},\ and\ \citenamefont
  {Szymanska}}]{zamora_20}%
  \BibitemOpen
  \bibfield  {author} {\bibinfo {author} {\bibfnamefont {A.}~\bibnamefont
  {Zamora}}, \bibinfo {author} {\bibfnamefont {G.}~\bibnamefont {Dagvadorj}},
  \bibinfo {author} {\bibfnamefont {P.}~\bibnamefont {Comaron}}, \bibinfo
  {author} {\bibfnamefont {I.}~\bibnamefont {Carusotto}}, \bibinfo {author}
  {\bibfnamefont {N.~P.}\ \bibnamefont {Proukakis}}, \ and\ \bibinfo {author}
  {\bibfnamefont {M.~H.}\ \bibnamefont {Szymanska}},\ }\href@noop {} {\enquote
  {\bibinfo {title} {Kibble-zurek mechanism in driven-dissipative systems
  crossing a non-equilibrium phase transition},}\ } (\bibinfo {year} {2020}),\
  \Eprint {http://arxiv.org/abs/2004.00918} {arXiv:2004.00918
  [cond-mat.mes-hall]} \BibitemShut {NoStop}%
\bibitem [{\citenamefont {Chung}\ \emph {et~al.}(1991)\citenamefont {Chung},
  \citenamefont {Durrer}, \citenamefont {Turok},\ and\ \citenamefont
  {Yurke}}]{KZexp-a}%
  \BibitemOpen
  \bibfield  {author} {\bibinfo {author} {\bibfnamefont {I.}~\bibnamefont
  {Chung}}, \bibinfo {author} {\bibfnamefont {R.}~\bibnamefont {Durrer}},
  \bibinfo {author} {\bibfnamefont {N.}~\bibnamefont {Turok}}, \ and\ \bibinfo
  {author} {\bibfnamefont {B.}~\bibnamefont {Yurke}},\ }\href {\doibase
  10.1126/science.251.4999.1336} {\bibfield  {journal} {\bibinfo  {journal}
  {Science}\ }\textbf {\bibinfo {volume} {251}},\ \bibinfo {pages} {1336}
  (\bibinfo {year} {1991})}\BibitemShut {NoStop}%
\bibitem [{\citenamefont {Bowick}\ \emph {et~al.}(1994)\citenamefont {Bowick},
  \citenamefont {Chandar}, \citenamefont {Schiff},\ and\ \citenamefont
  {Srivastava}}]{KZexp-b}%
  \BibitemOpen
  \bibfield  {author} {\bibinfo {author} {\bibfnamefont {M.~J.}\ \bibnamefont
  {Bowick}}, \bibinfo {author} {\bibfnamefont {L.}~\bibnamefont {Chandar}},
  \bibinfo {author} {\bibfnamefont {E.~A.}\ \bibnamefont {Schiff}}, \ and\
  \bibinfo {author} {\bibfnamefont {A.~M.}\ \bibnamefont {Srivastava}},\ }\href
  {\doibase 10.1126/science.263.5149.943} {\bibfield  {journal} {\bibinfo
  {journal} {Science}\ }\textbf {\bibinfo {volume} {263}},\ \bibinfo {pages}
  {943} (\bibinfo {year} {1994})}\BibitemShut {NoStop}%
\bibitem [{\citenamefont {Ruutu}\ \emph {et~al.}(1996)\citenamefont {Ruutu},
  \citenamefont {Eltsov}, \citenamefont {Gill}, \citenamefont {Kibble},
  \citenamefont {Krusius}, \citenamefont {Makhlin}, \citenamefont
  {Pla{\c{c}}ais}, \citenamefont {Volovik},\ and\ \citenamefont
  {Xu}}]{KZexp-c}%
  \BibitemOpen
  \bibfield  {author} {\bibinfo {author} {\bibfnamefont {V.~M.~H.}\
  \bibnamefont {Ruutu}}, \bibinfo {author} {\bibfnamefont {V.~B.}\ \bibnamefont
  {Eltsov}}, \bibinfo {author} {\bibfnamefont {A.~J.}\ \bibnamefont {Gill}},
  \bibinfo {author} {\bibfnamefont {T.~W.~B.}\ \bibnamefont {Kibble}}, \bibinfo
  {author} {\bibfnamefont {M.}~\bibnamefont {Krusius}}, \bibinfo {author}
  {\bibfnamefont {Y.~G.}\ \bibnamefont {Makhlin}}, \bibinfo {author}
  {\bibfnamefont {B.}~\bibnamefont {Pla{\c{c}}ais}}, \bibinfo {author}
  {\bibfnamefont {G.~E.}\ \bibnamefont {Volovik}}, \ and\ \bibinfo {author}
  {\bibfnamefont {W.}~\bibnamefont {Xu}},\ }\href {\doibase 10.1038/382334a0}
  {\bibfield  {journal} {\bibinfo  {journal} {Nature}\ }\textbf {\bibinfo
  {volume} {382}},\ \bibinfo {pages} {334} (\bibinfo {year}
  {1996})}\BibitemShut {NoStop}%
\bibitem [{\citenamefont {B\"{a}uerle}\ \emph {et~al.}(1996)\citenamefont
  {B\"{a}uerle}, \citenamefont {Bunkov}, \citenamefont {Fisher}, \citenamefont
  {Godfrin},\ and\ \citenamefont {Pickett}}]{KZexp-d}%
  \BibitemOpen
  \bibfield  {author} {\bibinfo {author} {\bibfnamefont {C.}~\bibnamefont
  {B\"{a}uerle}}, \bibinfo {author} {\bibfnamefont {Y.~M.}\ \bibnamefont
  {Bunkov}}, \bibinfo {author} {\bibfnamefont {S.~N.}\ \bibnamefont {Fisher}},
  \bibinfo {author} {\bibfnamefont {H.}~\bibnamefont {Godfrin}}, \ and\
  \bibinfo {author} {\bibfnamefont {G.~R.}\ \bibnamefont {Pickett}},\ }\href
  {\doibase 10.1038/382332a0} {\bibfield  {journal} {\bibinfo  {journal}
  {Nature}\ }\textbf {\bibinfo {volume} {382}},\ \bibinfo {pages} {332}
  (\bibinfo {year} {1996})}\BibitemShut {NoStop}%
\bibitem [{\citenamefont {Carmi}\ \emph {et~al.}(2000)\citenamefont {Carmi},
  \citenamefont {Polturak},\ and\ \citenamefont {Koren}}]{KZexp-e}%
  \BibitemOpen
  \bibfield  {author} {\bibinfo {author} {\bibfnamefont {R.}~\bibnamefont
  {Carmi}}, \bibinfo {author} {\bibfnamefont {E.}~\bibnamefont {Polturak}}, \
  and\ \bibinfo {author} {\bibfnamefont {G.}~\bibnamefont {Koren}},\ }\href
  {\doibase 10.1103/PhysRevLett.84.4966} {\bibfield  {journal} {\bibinfo
  {journal} {Phys. Rev. Lett.}\ }\textbf {\bibinfo {volume} {84}},\ \bibinfo
  {pages} {4966} (\bibinfo {year} {2000})}\BibitemShut {NoStop}%
\bibitem [{\citenamefont {Monaco}\ \emph {et~al.}(2002)\citenamefont {Monaco},
  \citenamefont {Mygind},\ and\ \citenamefont {Rivers}}]{KZexp-f}%
  \BibitemOpen
  \bibfield  {author} {\bibinfo {author} {\bibfnamefont {R.}~\bibnamefont
  {Monaco}}, \bibinfo {author} {\bibfnamefont {J.}~\bibnamefont {Mygind}}, \
  and\ \bibinfo {author} {\bibfnamefont {R.~J.}\ \bibnamefont {Rivers}},\
  }\href {\doibase 10.1103/PhysRevLett.89.080603} {\bibfield  {journal}
  {\bibinfo  {journal} {Phys. Rev. Lett.}\ }\textbf {\bibinfo {volume} {89}},\
  \bibinfo {pages} {080603} (\bibinfo {year} {2002})}\BibitemShut {NoStop}%
\bibitem [{\citenamefont {Maniv}\ \emph {et~al.}(2003)\citenamefont {Maniv},
  \citenamefont {Polturak},\ and\ \citenamefont {Koren}}]{KZexp-g}%
  \BibitemOpen
  \bibfield  {author} {\bibinfo {author} {\bibfnamefont {A.}~\bibnamefont
  {Maniv}}, \bibinfo {author} {\bibfnamefont {E.}~\bibnamefont {Polturak}}, \
  and\ \bibinfo {author} {\bibfnamefont {G.}~\bibnamefont {Koren}},\ }\href
  {\doibase 10.1103/PhysRevLett.91.197001} {\bibfield  {journal} {\bibinfo
  {journal} {Phys. Rev. Lett.}\ }\textbf {\bibinfo {volume} {91}},\ \bibinfo
  {pages} {197001} (\bibinfo {year} {2003})}\BibitemShut {NoStop}%
\bibitem [{\citenamefont {Monaco}\ \emph {et~al.}(2009)\citenamefont {Monaco},
  \citenamefont {Mygind}, \citenamefont {Rivers},\ and\ \citenamefont
  {Koshelets}}]{KZexp-i}%
  \BibitemOpen
  \bibfield  {author} {\bibinfo {author} {\bibfnamefont {R.}~\bibnamefont
  {Monaco}}, \bibinfo {author} {\bibfnamefont {J.}~\bibnamefont {Mygind}},
  \bibinfo {author} {\bibfnamefont {R.~J.}\ \bibnamefont {Rivers}}, \ and\
  \bibinfo {author} {\bibfnamefont {V.~P.}\ \bibnamefont {Koshelets}},\ }\href
  {\doibase 10.1103/PhysRevB.80.180501} {\bibfield  {journal} {\bibinfo
  {journal} {Phys. Rev. B}\ }\textbf {\bibinfo {volume} {80}},\ \bibinfo
  {pages} {180501} (\bibinfo {year} {2009})}\BibitemShut {NoStop}%
\bibitem [{\citenamefont {Golubchik}\ \emph {et~al.}(2010)\citenamefont
  {Golubchik}, \citenamefont {Polturak},\ and\ \citenamefont
  {Koren}}]{KZexp-j}%
  \BibitemOpen
  \bibfield  {author} {\bibinfo {author} {\bibfnamefont {D.}~\bibnamefont
  {Golubchik}}, \bibinfo {author} {\bibfnamefont {E.}~\bibnamefont {Polturak}},
  \ and\ \bibinfo {author} {\bibfnamefont {G.}~\bibnamefont {Koren}},\ }\href
  {\doibase 10.1103/PhysRevLett.104.247002} {\bibfield  {journal} {\bibinfo
  {journal} {Phys. Rev. Lett.}\ }\textbf {\bibinfo {volume} {104}},\ \bibinfo
  {pages} {247002} (\bibinfo {year} {2010})}\BibitemShut {NoStop}%
\bibitem [{\citenamefont {Chiara}\ \emph {et~al.}(2010)\citenamefont {Chiara},
  \citenamefont {del Campo}, \citenamefont {Morigi}, \citenamefont {Plenio},\
  and\ \citenamefont {Retzker}}]{KZexp-k}%
  \BibitemOpen
  \bibfield  {author} {\bibinfo {author} {\bibfnamefont {G.~D.}\ \bibnamefont
  {Chiara}}, \bibinfo {author} {\bibfnamefont {A.}~\bibnamefont {del Campo}},
  \bibinfo {author} {\bibfnamefont {G.}~\bibnamefont {Morigi}}, \bibinfo
  {author} {\bibfnamefont {M.~B.}\ \bibnamefont {Plenio}}, \ and\ \bibinfo
  {author} {\bibfnamefont {A.}~\bibnamefont {Retzker}},\ }\href {\doibase
  10.1088/1367-2630/12/11/115003} {\bibfield  {journal} {\bibinfo  {journal}
  {New J. Phys.}\ }\textbf {\bibinfo {volume} {12}},\ \bibinfo {pages} {115003}
  (\bibinfo {year} {2010})}\BibitemShut {NoStop}%
\bibitem [{\citenamefont {Mielenz}\ \emph {et~al.}(2013)\citenamefont
  {Mielenz}, \citenamefont {Brox}, \citenamefont {Kahra}, \citenamefont
  {Leschhorn}, \citenamefont {Albert}, \citenamefont {Schaetz}, \citenamefont
  {Landa},\ and\ \citenamefont {Reznik}}]{KZexp-l}%
  \BibitemOpen
  \bibfield  {author} {\bibinfo {author} {\bibfnamefont {M.}~\bibnamefont
  {Mielenz}}, \bibinfo {author} {\bibfnamefont {J.}~\bibnamefont {Brox}},
  \bibinfo {author} {\bibfnamefont {S.}~\bibnamefont {Kahra}}, \bibinfo
  {author} {\bibfnamefont {G.}~\bibnamefont {Leschhorn}}, \bibinfo {author}
  {\bibfnamefont {M.}~\bibnamefont {Albert}}, \bibinfo {author} {\bibfnamefont
  {T.}~\bibnamefont {Schaetz}}, \bibinfo {author} {\bibfnamefont
  {H.}~\bibnamefont {Landa}}, \ and\ \bibinfo {author} {\bibfnamefont
  {B.}~\bibnamefont {Reznik}},\ }\href {\doibase
  10.1103/PhysRevLett.110.133004} {\bibfield  {journal} {\bibinfo  {journal}
  {Phys. Rev. Lett.}\ }\textbf {\bibinfo {volume} {110}},\ \bibinfo {pages}
  {133004} (\bibinfo {year} {2013})}\BibitemShut {NoStop}%
\bibitem [{\citenamefont {Ulm}\ \emph {et~al.}(2013)\citenamefont {Ulm},
  \citenamefont {Ro{\ss}nagel}, \citenamefont {Jacob}, \citenamefont
  {Deg\"{u}nther}, \citenamefont {Dawkins}, \citenamefont {Poschinger},
  \citenamefont {Nigmatullin}, \citenamefont {Retzker}, \citenamefont {Plenio},
  \citenamefont {Schmidt-Kaler},\ and\ \citenamefont {Singer}}]{KZexp-m}%
  \BibitemOpen
  \bibfield  {author} {\bibinfo {author} {\bibfnamefont {S.}~\bibnamefont
  {Ulm}}, \bibinfo {author} {\bibfnamefont {J.}~\bibnamefont {Ro{\ss}nagel}},
  \bibinfo {author} {\bibfnamefont {G.}~\bibnamefont {Jacob}}, \bibinfo
  {author} {\bibfnamefont {C.}~\bibnamefont {Deg\"{u}nther}}, \bibinfo {author}
  {\bibfnamefont {S.~T.}\ \bibnamefont {Dawkins}}, \bibinfo {author}
  {\bibfnamefont {U.~G.}\ \bibnamefont {Poschinger}}, \bibinfo {author}
  {\bibfnamefont {R.}~\bibnamefont {Nigmatullin}}, \bibinfo {author}
  {\bibfnamefont {A.}~\bibnamefont {Retzker}}, \bibinfo {author} {\bibfnamefont
  {M.~B.}\ \bibnamefont {Plenio}}, \bibinfo {author} {\bibfnamefont
  {F.}~\bibnamefont {Schmidt-Kaler}}, \ and\ \bibinfo {author} {\bibfnamefont
  {K.}~\bibnamefont {Singer}},\ }\href {\doibase 10.1038/ncomms3290} {\bibfield
   {journal} {\bibinfo  {journal} {Nat. Comm.}\ }\textbf {\bibinfo {volume}
  {4}},\ \bibinfo {pages} {2290} (\bibinfo {year} {2013})}\BibitemShut
  {NoStop}%
\bibitem [{\citenamefont {Pyka}\ \emph {et~al.}(2013)\citenamefont {Pyka},
  \citenamefont {Keller}, \citenamefont {Partner}, \citenamefont {Nigmatullin},
  \citenamefont {Burgermeister}, \citenamefont {Meier}, \citenamefont
  {Kuhlmann}, \citenamefont {Retzker}, \citenamefont {Plenio}, \citenamefont
  {Zurek}, \citenamefont {del Campo},\ and\ \citenamefont
  {Mehlst\"{a}ubler}}]{KZexp-n}%
  \BibitemOpen
  \bibfield  {author} {\bibinfo {author} {\bibfnamefont {K.}~\bibnamefont
  {Pyka}}, \bibinfo {author} {\bibfnamefont {J.}~\bibnamefont {Keller}},
  \bibinfo {author} {\bibfnamefont {H.~L.}\ \bibnamefont {Partner}}, \bibinfo
  {author} {\bibfnamefont {R.}~\bibnamefont {Nigmatullin}}, \bibinfo {author}
  {\bibfnamefont {T.}~\bibnamefont {Burgermeister}}, \bibinfo {author}
  {\bibfnamefont {D.~M.}\ \bibnamefont {Meier}}, \bibinfo {author}
  {\bibfnamefont {K.}~\bibnamefont {Kuhlmann}}, \bibinfo {author}
  {\bibfnamefont {A.}~\bibnamefont {Retzker}}, \bibinfo {author} {\bibfnamefont
  {M.~B.}\ \bibnamefont {Plenio}}, \bibinfo {author} {\bibfnamefont {W.~H.}\
  \bibnamefont {Zurek}}, \bibinfo {author} {\bibfnamefont {A.}~\bibnamefont
  {del Campo}}, \ and\ \bibinfo {author} {\bibfnamefont {T.~E.}\ \bibnamefont
  {Mehlst\"{a}ubler}},\ }\href {\doibase 10.1038/ncomms3291} {\bibfield
  {journal} {\bibinfo  {journal} {Nat. Comm.}\ }\textbf {\bibinfo {volume}
  {4}},\ \bibinfo {pages} {2291} (\bibinfo {year} {2013})}\BibitemShut
  {NoStop}%
\bibitem [{\citenamefont {Chae}\ \emph {et~al.}(2012)\citenamefont {Chae},
  \citenamefont {Lee}, \citenamefont {Horibe}, \citenamefont {Tanimura},
  \citenamefont {Mori}, \citenamefont {Gao}, \citenamefont {Carr},\ and\
  \citenamefont {Cheong}}]{KZexp-o}%
  \BibitemOpen
  \bibfield  {author} {\bibinfo {author} {\bibfnamefont {S.~C.}\ \bibnamefont
  {Chae}}, \bibinfo {author} {\bibfnamefont {N.}~\bibnamefont {Lee}}, \bibinfo
  {author} {\bibfnamefont {Y.}~\bibnamefont {Horibe}}, \bibinfo {author}
  {\bibfnamefont {M.}~\bibnamefont {Tanimura}}, \bibinfo {author}
  {\bibfnamefont {S.}~\bibnamefont {Mori}}, \bibinfo {author} {\bibfnamefont
  {B.}~\bibnamefont {Gao}}, \bibinfo {author} {\bibfnamefont {S.}~\bibnamefont
  {Carr}}, \ and\ \bibinfo {author} {\bibfnamefont {S.-W.}\ \bibnamefont
  {Cheong}},\ }\href {\doibase 10.1103/PhysRevLett.108.167603} {\bibfield
  {journal} {\bibinfo  {journal} {Phys. Rev. Lett.}\ }\textbf {\bibinfo
  {volume} {108}},\ \bibinfo {pages} {167603} (\bibinfo {year}
  {2012})}\BibitemShut {NoStop}%
\bibitem [{\citenamefont {Lin}\ \emph {et~al.}(2014)\citenamefont {Lin},
  \citenamefont {Wang}, \citenamefont {Kamiya}, \citenamefont {Chern},
  \citenamefont {Fan}, \citenamefont {Fan}, \citenamefont {Casas},
  \citenamefont {Liu}, \citenamefont {Kiryukhin}, \citenamefont {Zurek},
  \citenamefont {Batista},\ and\ \citenamefont {Cheong}}]{KZexp-p}%
  \BibitemOpen
  \bibfield  {author} {\bibinfo {author} {\bibfnamefont {S.-Z.}\ \bibnamefont
  {Lin}}, \bibinfo {author} {\bibfnamefont {X.}~\bibnamefont {Wang}}, \bibinfo
  {author} {\bibfnamefont {Y.}~\bibnamefont {Kamiya}}, \bibinfo {author}
  {\bibfnamefont {G.-W.}\ \bibnamefont {Chern}}, \bibinfo {author}
  {\bibfnamefont {F.}~\bibnamefont {Fan}}, \bibinfo {author} {\bibfnamefont
  {D.}~\bibnamefont {Fan}}, \bibinfo {author} {\bibfnamefont {B.}~\bibnamefont
  {Casas}}, \bibinfo {author} {\bibfnamefont {Y.}~\bibnamefont {Liu}}, \bibinfo
  {author} {\bibfnamefont {V.}~\bibnamefont {Kiryukhin}}, \bibinfo {author}
  {\bibfnamefont {W.~H.}\ \bibnamefont {Zurek}}, \bibinfo {author}
  {\bibfnamefont {C.~D.}\ \bibnamefont {Batista}}, \ and\ \bibinfo {author}
  {\bibfnamefont {S.-W.}\ \bibnamefont {Cheong}},\ }\href {\doibase
  10.1038/nphys3142} {\bibfield  {journal} {\bibinfo  {journal} {Nat. Phys.}\
  }\textbf {\bibinfo {volume} {10}},\ \bibinfo {pages} {970} (\bibinfo {year}
  {2014})}\BibitemShut {NoStop}%
\bibitem [{\citenamefont {Griffin}\ \emph {et~al.}(2012)\citenamefont
  {Griffin}, \citenamefont {Lilienblum}, \citenamefont {Delaney}, \citenamefont
  {Kumagai}, \citenamefont {Fiebig},\ and\ \citenamefont {Spaldin}}]{KZexp-q}%
  \BibitemOpen
  \bibfield  {author} {\bibinfo {author} {\bibfnamefont {S.~M.}\ \bibnamefont
  {Griffin}}, \bibinfo {author} {\bibfnamefont {M.}~\bibnamefont {Lilienblum}},
  \bibinfo {author} {\bibfnamefont {K.~T.}\ \bibnamefont {Delaney}}, \bibinfo
  {author} {\bibfnamefont {Y.}~\bibnamefont {Kumagai}}, \bibinfo {author}
  {\bibfnamefont {M.}~\bibnamefont {Fiebig}}, \ and\ \bibinfo {author}
  {\bibfnamefont {N.~A.}\ \bibnamefont {Spaldin}},\ }\href {\doibase
  10.1103/PhysRevX.2.041022} {\bibfield  {journal} {\bibinfo  {journal} {Phys.
  Rev. X}\ }\textbf {\bibinfo {volume} {2}},\ \bibinfo {pages} {041022}
  (\bibinfo {year} {2012})}\BibitemShut {NoStop}%
\bibitem [{\citenamefont {Deutschl\"{a}nder}\ \emph {et~al.}(2015)\citenamefont
  {Deutschl\"{a}nder}, \citenamefont {Dillmann}, \citenamefont {Maret},\ and\
  \citenamefont {Keim}}]{KZexp-s}%
  \BibitemOpen
  \bibfield  {author} {\bibinfo {author} {\bibfnamefont {S.}~\bibnamefont
  {Deutschl\"{a}nder}}, \bibinfo {author} {\bibfnamefont {P.}~\bibnamefont
  {Dillmann}}, \bibinfo {author} {\bibfnamefont {G.}~\bibnamefont {Maret}}, \
  and\ \bibinfo {author} {\bibfnamefont {P.}~\bibnamefont {Keim}},\ }\href
  {\doibase 10.1073/pnas.1500763112} {\bibfield  {journal} {\bibinfo  {journal}
  {Proc. Natl. Acad. Sci. U.S.A.}\ }\textbf {\bibinfo {volume} {112}},\
  \bibinfo {pages} {6925} (\bibinfo {year} {2015})}\BibitemShut {NoStop}%
\bibitem [{\citenamefont {Chomaz}\ \emph {et~al.}(2015)\citenamefont {Chomaz},
  \citenamefont {Corman}, \citenamefont {Bienaim{\'{e}}}, \citenamefont
  {Desbuquois}, \citenamefont {Weitenberg}, \citenamefont {Nascimb{\`{e}}ne},
  \citenamefont {Beugnon},\ and\ \citenamefont {Dalibard}}]{KZexp-t}%
  \BibitemOpen
  \bibfield  {author} {\bibinfo {author} {\bibfnamefont {L.}~\bibnamefont
  {Chomaz}}, \bibinfo {author} {\bibfnamefont {L.}~\bibnamefont {Corman}},
  \bibinfo {author} {\bibfnamefont {T.}~\bibnamefont {Bienaim{\'{e}}}},
  \bibinfo {author} {\bibfnamefont {R.}~\bibnamefont {Desbuquois}}, \bibinfo
  {author} {\bibfnamefont {C.}~\bibnamefont {Weitenberg}}, \bibinfo {author}
  {\bibfnamefont {S.}~\bibnamefont {Nascimb{\`{e}}ne}}, \bibinfo {author}
  {\bibfnamefont {J.}~\bibnamefont {Beugnon}}, \ and\ \bibinfo {author}
  {\bibfnamefont {J.}~\bibnamefont {Dalibard}},\ }\href {\doibase
  10.1038/ncomms7162} {\bibfield  {journal} {\bibinfo  {journal} {Nature
  Communications}\ }\textbf {\bibinfo {volume} {6}},\ \bibinfo {pages} {6162}
  (\bibinfo {year} {2015})}\BibitemShut {NoStop}%
\bibitem [{\citenamefont {Rysti}\ \emph {et~al.}(2019)\citenamefont {Rysti},
  \citenamefont {Autti}, \citenamefont {Volovik},\ and\ \citenamefont
  {Eltsov}}]{KZexp-x}%
  \BibitemOpen
  \bibfield  {author} {\bibinfo {author} {\bibfnamefont {J.}~\bibnamefont
  {Rysti}}, \bibinfo {author} {\bibfnamefont {S.}~\bibnamefont {Autti}},
  \bibinfo {author} {\bibfnamefont {G.}~\bibnamefont {Volovik}}, \ and\
  \bibinfo {author} {\bibfnamefont {V.}~\bibnamefont {Eltsov}},\ }\href@noop {}
  {\enquote {\bibinfo {title} {Kibble-zurek creation of half-quantum vortices
  under symmetry violating bias},}\ } (\bibinfo {year} {2019}),\ \Eprint
  {http://arxiv.org/abs/arXiv:1906.11453} {arXiv:1906.11453} \BibitemShut
  {NoStop}%
\bibitem [{\citenamefont {Sadler}\ \emph {et~al.}(2006)\citenamefont {Sadler},
  \citenamefont {Higbie}, \citenamefont {Leslie}, \citenamefont
  {Vengalattore},\ and\ \citenamefont {Stamper-Kurn}}]{KZexp-gg}%
  \BibitemOpen
  \bibfield  {author} {\bibinfo {author} {\bibfnamefont {L.~E.}\ \bibnamefont
  {Sadler}}, \bibinfo {author} {\bibfnamefont {J.~M.}\ \bibnamefont {Higbie}},
  \bibinfo {author} {\bibfnamefont {S.~R.}\ \bibnamefont {Leslie}}, \bibinfo
  {author} {\bibfnamefont {M.}~\bibnamefont {Vengalattore}}, \ and\ \bibinfo
  {author} {\bibfnamefont {D.~M.}\ \bibnamefont {Stamper-Kurn}},\ }\href
  {\doibase 10.1038/nature05094} {\bibfield  {journal} {\bibinfo  {journal}
  {Nature}\ }\textbf {\bibinfo {volume} {443}},\ \bibinfo {pages} {312}
  (\bibinfo {year} {2006})}\BibitemShut {NoStop}%
\bibitem [{\citenamefont {Weiler}\ \emph {et~al.}(2008)\citenamefont {Weiler},
  \citenamefont {Neely}, \citenamefont {Scherer}, \citenamefont {Bradley},
  \citenamefont {Davis},\ and\ \citenamefont {Anderson}}]{KZexp-h}%
  \BibitemOpen
  \bibfield  {author} {\bibinfo {author} {\bibfnamefont {C.~N.}\ \bibnamefont
  {Weiler}}, \bibinfo {author} {\bibfnamefont {T.~W.}\ \bibnamefont {Neely}},
  \bibinfo {author} {\bibfnamefont {D.~R.}\ \bibnamefont {Scherer}}, \bibinfo
  {author} {\bibfnamefont {A.~S.}\ \bibnamefont {Bradley}}, \bibinfo {author}
  {\bibfnamefont {M.~J.}\ \bibnamefont {Davis}}, \ and\ \bibinfo {author}
  {\bibfnamefont {B.~P.}\ \bibnamefont {Anderson}},\ }\href {\doibase
  10.1038/nature07334} {\bibfield  {journal} {\bibinfo  {journal} {Nature}\
  }\textbf {\bibinfo {volume} {455}},\ \bibinfo {pages} {948} (\bibinfo {year}
  {2008})}\BibitemShut {NoStop}%
\bibitem [{\citenamefont {Lamporesi}\ \emph {et~al.}(2013)\citenamefont
  {Lamporesi}, \citenamefont {Donadello}, \citenamefont {Serafini},
  \citenamefont {Dalfovo},\ and\ \citenamefont
  {Ferrari}}]{lamporesi2013spontaneous}%
  \BibitemOpen
  \bibfield  {author} {\bibinfo {author} {\bibfnamefont {G.}~\bibnamefont
  {Lamporesi}}, \bibinfo {author} {\bibfnamefont {S.}~\bibnamefont
  {Donadello}}, \bibinfo {author} {\bibfnamefont {S.}~\bibnamefont {Serafini}},
  \bibinfo {author} {\bibfnamefont {F.}~\bibnamefont {Dalfovo}}, \ and\
  \bibinfo {author} {\bibfnamefont {G.}~\bibnamefont {Ferrari}},\ }\href@noop
  {} {\bibfield  {journal} {\bibinfo  {journal} {Nat. Phys.}\ }\textbf
  {\bibinfo {volume} {9}},\ \bibinfo {pages} {656} (\bibinfo {year}
  {2013})}\BibitemShut {NoStop}%
\bibitem [{\citenamefont {Donadello}\ \emph {et~al.}(2014)\citenamefont
  {Donadello}, \citenamefont {Serafini}, \citenamefont {Tylutki}, \citenamefont
  {Pitaevskii}, \citenamefont {Dalfovo}, \citenamefont {Lamporesi},\ and\
  \citenamefont {Ferrari}}]{KZexp-r}%
  \BibitemOpen
  \bibfield  {author} {\bibinfo {author} {\bibfnamefont {S.}~\bibnamefont
  {Donadello}}, \bibinfo {author} {\bibfnamefont {S.}~\bibnamefont {Serafini}},
  \bibinfo {author} {\bibfnamefont {M.}~\bibnamefont {Tylutki}}, \bibinfo
  {author} {\bibfnamefont {L.~P.}\ \bibnamefont {Pitaevskii}}, \bibinfo
  {author} {\bibfnamefont {F.}~\bibnamefont {Dalfovo}}, \bibinfo {author}
  {\bibfnamefont {G.}~\bibnamefont {Lamporesi}}, \ and\ \bibinfo {author}
  {\bibfnamefont {G.}~\bibnamefont {Ferrari}},\ }\href {\doibase
  10.1103/PhysRevLett.113.065302} {\bibfield  {journal} {\bibinfo  {journal}
  {Phys. Rev. Lett.}\ }\textbf {\bibinfo {volume} {113}},\ \bibinfo {pages}
  {065302} (\bibinfo {year} {2014})}\BibitemShut {NoStop}%
\bibitem [{\citenamefont {Corman}\ \emph {et~al.}(2014)\citenamefont {Corman},
  \citenamefont {Chomaz}, \citenamefont {Bienaime}, \citenamefont {Desbuquois},
  \citenamefont {Weitenberg}, \citenamefont {Nascimbene}, \citenamefont
  {Dalibard},\ and\ \citenamefont {Beugnon}}]{dalibard-ring-KZ-2014}%
  \BibitemOpen
  \bibfield  {author} {\bibinfo {author} {\bibfnamefont {L.}~\bibnamefont
  {Corman}}, \bibinfo {author} {\bibfnamefont {L.}~\bibnamefont {Chomaz}},
  \bibinfo {author} {\bibfnamefont {T.}~\bibnamefont {Bienaime}}, \bibinfo
  {author} {\bibfnamefont {R.}~\bibnamefont {Desbuquois}}, \bibinfo {author}
  {\bibfnamefont {C.}~\bibnamefont {Weitenberg}}, \bibinfo {author}
  {\bibfnamefont {S.}~\bibnamefont {Nascimbene}}, \bibinfo {author}
  {\bibfnamefont {J.}~\bibnamefont {Dalibard}}, \ and\ \bibinfo {author}
  {\bibfnamefont {J.}~\bibnamefont {Beugnon}},\ }\href@noop {} {\bibfield
  {journal} {\bibinfo  {journal} {Phys. Rev. Lett.}\ }\textbf {\bibinfo
  {volume} {113}} (\bibinfo {year} {2014})}\BibitemShut {NoStop}%
\bibitem [{\citenamefont {Navon}\ \emph {et~al.}(2015)\citenamefont {Navon},
  \citenamefont {Gaunt}, \citenamefont {Smith},\ and\ \citenamefont
  {Hadzibabic}}]{KZexp-v}%
  \BibitemOpen
  \bibfield  {author} {\bibinfo {author} {\bibfnamefont {N.}~\bibnamefont
  {Navon}}, \bibinfo {author} {\bibfnamefont {A.~L.}\ \bibnamefont {Gaunt}},
  \bibinfo {author} {\bibfnamefont {R.~P.}\ \bibnamefont {Smith}}, \ and\
  \bibinfo {author} {\bibfnamefont {Z.}~\bibnamefont {Hadzibabic}},\ }\href
  {\doibase 10.1126/science.1258676} {\bibfield  {journal} {\bibinfo  {journal}
  {Science}\ }\textbf {\bibinfo {volume} {347}},\ \bibinfo {pages} {167}
  (\bibinfo {year} {2015})}\BibitemShut {NoStop}%
\bibitem [{\citenamefont {Yukalov}\ \emph {et~al.}(2015)\citenamefont
  {Yukalov}, \citenamefont {Novikov},\ and\ \citenamefont {Bagnato}}]{KZexp-u}%
  \BibitemOpen
  \bibfield  {author} {\bibinfo {author} {\bibfnamefont {V.}~\bibnamefont
  {Yukalov}}, \bibinfo {author} {\bibfnamefont {A.}~\bibnamefont {Novikov}}, \
  and\ \bibinfo {author} {\bibfnamefont {V.}~\bibnamefont {Bagnato}},\ }\href
  {\doibase 10.1016/j.physleta.2015.02.033} {\bibfield  {journal} {\bibinfo
  {journal} {Phys. Lett. A}\ }\textbf {\bibinfo {volume} {379}},\ \bibinfo
  {pages} {1366} (\bibinfo {year} {2015})}\BibitemShut {NoStop}%
\bibitem [{\citenamefont {Donadello}\ \emph {et~al.}(2016)\citenamefont
  {Donadello}, \citenamefont {Serafini}, \citenamefont {Bienaim{\'e}},
  \citenamefont {Dalfovo}, \citenamefont {Lamporesi},\ and\ \citenamefont
  {Ferrari}}]{donadello2016creation}%
  \BibitemOpen
  \bibfield  {author} {\bibinfo {author} {\bibfnamefont {S.}~\bibnamefont
  {Donadello}}, \bibinfo {author} {\bibfnamefont {S.}~\bibnamefont {Serafini}},
  \bibinfo {author} {\bibfnamefont {T.}~\bibnamefont {Bienaim{\'e}}}, \bibinfo
  {author} {\bibfnamefont {F.}~\bibnamefont {Dalfovo}}, \bibinfo {author}
  {\bibfnamefont {G.}~\bibnamefont {Lamporesi}}, \ and\ \bibinfo {author}
  {\bibfnamefont {G.}~\bibnamefont {Ferrari}},\ }\href@noop {} {\bibfield
  {journal} {\bibinfo  {journal} {Phys. Rev. A}\ }\textbf {\bibinfo {volume}
  {94}},\ \bibinfo {pages} {023628} (\bibinfo {year} {2016})}\BibitemShut
  {NoStop}%
\bibitem [{\citenamefont {Ko}\ \emph {et~al.}(2019)\citenamefont {Ko},
  \citenamefont {W.},\ and\ \citenamefont {Shin}}]{shin-KZ-Fermi-2019}%
  \BibitemOpen
  \bibfield  {author} {\bibinfo {author} {\bibfnamefont {B.}~\bibnamefont
  {Ko}}, \bibinfo {author} {\bibfnamefont {P.~J.}\ \bibnamefont {W.}}, \ and\
  \bibinfo {author} {\bibfnamefont {Y.}~\bibnamefont {Shin}},\ }\href {\doibase
  10.1038/s41567-019-0650-1} {\bibfield  {journal} {\bibinfo  {journal} {Nat.
  Phys.}\ } (\bibinfo {year} {2019}),\ 10.1038/s41567-019-0650-1}\BibitemShut
  {NoStop}%
\bibitem [{\citenamefont {Polkovnikov}(2005)}]{Polkovnikov2005}%
  \BibitemOpen
  \bibfield  {author} {\bibinfo {author} {\bibfnamefont {A.}~\bibnamefont
  {Polkovnikov}},\ }\href {\doibase 10.1103/PhysRevB.72.161201} {\bibfield
  {journal} {\bibinfo  {journal} {Phys. Rev. B}\ }\textbf {\bibinfo {volume}
  {72}},\ \bibinfo {pages} {161201} (\bibinfo {year} {2005})}\BibitemShut
  {NoStop}%
\bibitem [{\citenamefont {Damski}(2005)}]{QKZ1}%
  \BibitemOpen
  \bibfield  {author} {\bibinfo {author} {\bibfnamefont {B.}~\bibnamefont
  {Damski}},\ }\href {\doibase 10.1103/PhysRevLett.95.035701} {\bibfield
  {journal} {\bibinfo  {journal} {Phys. Rev. Lett.}\ }\textbf {\bibinfo
  {volume} {95}},\ \bibinfo {pages} {035701} (\bibinfo {year}
  {2005})}\BibitemShut {NoStop}%
\bibitem [{\citenamefont {Zurek}\ \emph {et~al.}(2005)\citenamefont {Zurek},
  \citenamefont {Dorner},\ and\ \citenamefont {Zoller}}]{QKZ2}%
  \BibitemOpen
  \bibfield  {author} {\bibinfo {author} {\bibfnamefont {W.~H.}\ \bibnamefont
  {Zurek}}, \bibinfo {author} {\bibfnamefont {U.}~\bibnamefont {Dorner}}, \
  and\ \bibinfo {author} {\bibfnamefont {P.}~\bibnamefont {Zoller}},\ }\href
  {\doibase 10.1103/PhysRevLett.95.105701} {\bibfield  {journal} {\bibinfo
  {journal} {Phys. Rev. Lett.}\ }\textbf {\bibinfo {volume} {95}},\ \bibinfo
  {pages} {105701} (\bibinfo {year} {2005})}\BibitemShut {NoStop}%
\bibitem [{\citenamefont {Dziarmaga}(2005)}]{d2005}%
  \BibitemOpen
  \bibfield  {author} {\bibinfo {author} {\bibfnamefont {J.}~\bibnamefont
  {Dziarmaga}},\ }\href {\doibase 10.1103/PhysRevLett.95.245701} {\bibfield
  {journal} {\bibinfo  {journal} {Phys. Rev. Lett.}\ }\textbf {\bibinfo
  {volume} {95}},\ \bibinfo {pages} {245701} (\bibinfo {year}
  {2005})}\BibitemShut {NoStop}%
\bibitem [{\citenamefont {Dziarmaga}(2010)}]{d2010-a}%
  \BibitemOpen
  \bibfield  {author} {\bibinfo {author} {\bibfnamefont {J.}~\bibnamefont
  {Dziarmaga}},\ }\href {\doibase 10.1080/00018732.2010.514702} {\bibfield
  {journal} {\bibinfo  {journal} {Adv. Phys.}\ }\textbf {\bibinfo {volume}
  {59}},\ \bibinfo {pages} {1063} (\bibinfo {year} {2010})}\BibitemShut
  {NoStop}%
\bibitem [{\citenamefont {Polkovnikov}\ \emph {et~al.}(2011)\citenamefont
  {Polkovnikov}, \citenamefont {Sengupta}, \citenamefont {Silva},\ and\
  \citenamefont {Vengalattore}}]{d2010-b}%
  \BibitemOpen
  \bibfield  {author} {\bibinfo {author} {\bibfnamefont {A.}~\bibnamefont
  {Polkovnikov}}, \bibinfo {author} {\bibfnamefont {K.}~\bibnamefont
  {Sengupta}}, \bibinfo {author} {\bibfnamefont {A.}~\bibnamefont {Silva}}, \
  and\ \bibinfo {author} {\bibfnamefont {M.}~\bibnamefont {Vengalattore}},\
  }\href {\doibase 10.1103/RevModPhys.83.863} {\bibfield  {journal} {\bibinfo
  {journal} {Rev. Mod. Phys.}\ }\textbf {\bibinfo {volume} {83}},\ \bibinfo
  {pages} {863} (\bibinfo {year} {2011})}\BibitemShut {NoStop}%
\bibitem [{\citenamefont {Sch\"utzhold}\ \emph {et~al.}(2006)\citenamefont
  {Sch\"utzhold}, \citenamefont {Uhlmann}, \citenamefont {Xu},\ and\
  \citenamefont {Fischer}}]{QKZteor-a}%
  \BibitemOpen
  \bibfield  {author} {\bibinfo {author} {\bibfnamefont {R.}~\bibnamefont
  {Sch\"utzhold}}, \bibinfo {author} {\bibfnamefont {M.}~\bibnamefont
  {Uhlmann}}, \bibinfo {author} {\bibfnamefont {Y.}~\bibnamefont {Xu}}, \ and\
  \bibinfo {author} {\bibfnamefont {U.~R.}\ \bibnamefont {Fischer}},\ }\href
  {\doibase 10.1103/PhysRevLett.97.200601} {\bibfield  {journal} {\bibinfo
  {journal} {Phys. Rev. Lett.}\ }\textbf {\bibinfo {volume} {97}},\ \bibinfo
  {pages} {200601} (\bibinfo {year} {2006})}\BibitemShut {NoStop}%
\bibitem [{\citenamefont {Saito}\ \emph {et~al.}(2007)\citenamefont {Saito},
  \citenamefont {Kawaguchi},\ and\ \citenamefont {Ueda}}]{QKZteor-b}%
  \BibitemOpen
  \bibfield  {author} {\bibinfo {author} {\bibfnamefont {H.}~\bibnamefont
  {Saito}}, \bibinfo {author} {\bibfnamefont {Y.}~\bibnamefont {Kawaguchi}}, \
  and\ \bibinfo {author} {\bibfnamefont {M.}~\bibnamefont {Ueda}},\ }\href
  {\doibase 10.1103/PhysRevA.76.043613} {\bibfield  {journal} {\bibinfo
  {journal} {Phys. Rev. A}\ }\textbf {\bibinfo {volume} {76}},\ \bibinfo
  {pages} {043613} (\bibinfo {year} {2007})}\BibitemShut {NoStop}%
\bibitem [{\citenamefont {Mukherjee}\ \emph {et~al.}(2007)\citenamefont
  {Mukherjee}, \citenamefont {Divakaran}, \citenamefont {Dutta},\ and\
  \citenamefont {Sen}}]{QKZteor-c}%
  \BibitemOpen
  \bibfield  {author} {\bibinfo {author} {\bibfnamefont {V.}~\bibnamefont
  {Mukherjee}}, \bibinfo {author} {\bibfnamefont {U.}~\bibnamefont
  {Divakaran}}, \bibinfo {author} {\bibfnamefont {A.}~\bibnamefont {Dutta}}, \
  and\ \bibinfo {author} {\bibfnamefont {D.}~\bibnamefont {Sen}},\ }\href
  {\doibase 10.1103/PhysRevB.76.174303} {\bibfield  {journal} {\bibinfo
  {journal} {Phys. Rev. B}\ }\textbf {\bibinfo {volume} {76}},\ \bibinfo
  {pages} {174303} (\bibinfo {year} {2007})}\BibitemShut {NoStop}%
\bibitem [{\citenamefont {Cucchietti}\ \emph {et~al.}(2007)\citenamefont
  {Cucchietti}, \citenamefont {Damski}, \citenamefont {Dziarmaga},\ and\
  \citenamefont {Zurek}}]{QKZteor-d}%
  \BibitemOpen
  \bibfield  {author} {\bibinfo {author} {\bibfnamefont {F.~M.}\ \bibnamefont
  {Cucchietti}}, \bibinfo {author} {\bibfnamefont {B.}~\bibnamefont {Damski}},
  \bibinfo {author} {\bibfnamefont {J.}~\bibnamefont {Dziarmaga}}, \ and\
  \bibinfo {author} {\bibfnamefont {W.~H.}\ \bibnamefont {Zurek}},\ }\href
  {\doibase 10.1103/PhysRevA.75.023603} {\bibfield  {journal} {\bibinfo
  {journal} {Phys. Rev. A}\ }\textbf {\bibinfo {volume} {75}},\ \bibinfo
  {pages} {023603} (\bibinfo {year} {2007})}\BibitemShut {NoStop}%
\bibitem [{\citenamefont {Cincio}\ \emph {et~al.}(2007)\citenamefont {Cincio},
  \citenamefont {Dziarmaga}, \citenamefont {Rams},\ and\ \citenamefont
  {Zurek}}]{QKZteor-e}%
  \BibitemOpen
  \bibfield  {author} {\bibinfo {author} {\bibfnamefont {L.}~\bibnamefont
  {Cincio}}, \bibinfo {author} {\bibfnamefont {J.}~\bibnamefont {Dziarmaga}},
  \bibinfo {author} {\bibfnamefont {M.~M.}\ \bibnamefont {Rams}}, \ and\
  \bibinfo {author} {\bibfnamefont {W.~H.}\ \bibnamefont {Zurek}},\ }\href
  {\doibase 10.1103/PhysRevA.75.052321} {\bibfield  {journal} {\bibinfo
  {journal} {Phys. Rev. A}\ }\textbf {\bibinfo {volume} {75}},\ \bibinfo
  {pages} {052321} (\bibinfo {year} {2007})}\BibitemShut {NoStop}%
\bibitem [{\citenamefont {Polkovnikov}\ and\ \citenamefont
  {Gritsev}(2008)}]{QKZteor-f}%
  \BibitemOpen
  \bibfield  {author} {\bibinfo {author} {\bibfnamefont {A.}~\bibnamefont
  {Polkovnikov}}\ and\ \bibinfo {author} {\bibfnamefont {V.}~\bibnamefont
  {Gritsev}},\ }\href {\doibase 10.1038/nphys963} {\bibfield  {journal}
  {\bibinfo  {journal} {Nat. Phys.}\ }\textbf {\bibinfo {volume} {4}},\
  \bibinfo {pages} {477} (\bibinfo {year} {2008})}\BibitemShut {NoStop}%
\bibitem [{\citenamefont {Sengupta}\ \emph {et~al.}(2008)\citenamefont
  {Sengupta}, \citenamefont {Sen},\ and\ \citenamefont {Mondal}}]{QKZteor-g}%
  \BibitemOpen
  \bibfield  {author} {\bibinfo {author} {\bibfnamefont {K.}~\bibnamefont
  {Sengupta}}, \bibinfo {author} {\bibfnamefont {D.}~\bibnamefont {Sen}}, \
  and\ \bibinfo {author} {\bibfnamefont {S.}~\bibnamefont {Mondal}},\ }\href
  {\doibase 10.1103/PhysRevLett.100.077204} {\bibfield  {journal} {\bibinfo
  {journal} {Phys. Rev. Lett.}\ }\textbf {\bibinfo {volume} {100}},\ \bibinfo
  {pages} {077204} (\bibinfo {year} {2008})}\BibitemShut {NoStop}%
\bibitem [{\citenamefont {Dziarmaga}\ \emph {et~al.}(2008)\citenamefont
  {Dziarmaga}, \citenamefont {Meisner},\ and\ \citenamefont
  {Zurek}}]{QKZteor-h}%
  \BibitemOpen
  \bibfield  {author} {\bibinfo {author} {\bibfnamefont {J.}~\bibnamefont
  {Dziarmaga}}, \bibinfo {author} {\bibfnamefont {J.}~\bibnamefont {Meisner}},
  \ and\ \bibinfo {author} {\bibfnamefont {W.~H.}\ \bibnamefont {Zurek}},\
  }\href {\doibase 10.1103/PhysRevLett.101.115701} {\bibfield  {journal}
  {\bibinfo  {journal} {Phys. Rev. Lett.}\ }\textbf {\bibinfo {volume} {101}},\
  \bibinfo {pages} {115701} (\bibinfo {year} {2008})}\BibitemShut {NoStop}%
\bibitem [{\citenamefont {De~Grandi}\ \emph {et~al.}(2010)\citenamefont
  {De~Grandi}, \citenamefont {Gritsev},\ and\ \citenamefont
  {Polkovnikov}}]{QKZteor-j}%
  \BibitemOpen
  \bibfield  {author} {\bibinfo {author} {\bibfnamefont {C.}~\bibnamefont
  {De~Grandi}}, \bibinfo {author} {\bibfnamefont {V.}~\bibnamefont {Gritsev}},
  \ and\ \bibinfo {author} {\bibfnamefont {A.}~\bibnamefont {Polkovnikov}},\
  }\href {\doibase 10.1103/PhysRevB.81.012303} {\bibfield  {journal} {\bibinfo
  {journal} {Phys. Rev. B}\ }\textbf {\bibinfo {volume} {81}},\ \bibinfo
  {pages} {012303} (\bibinfo {year} {2010})}\BibitemShut {NoStop}%
\bibitem [{\citenamefont {Pollmann}\ \emph {et~al.}(2010)\citenamefont
  {Pollmann}, \citenamefont {Mukerjee}, \citenamefont {Green},\ and\
  \citenamefont {Moore}}]{QKZteor-k}%
  \BibitemOpen
  \bibfield  {author} {\bibinfo {author} {\bibfnamefont {F.}~\bibnamefont
  {Pollmann}}, \bibinfo {author} {\bibfnamefont {S.}~\bibnamefont {Mukerjee}},
  \bibinfo {author} {\bibfnamefont {A.~G.}\ \bibnamefont {Green}}, \ and\
  \bibinfo {author} {\bibfnamefont {J.~E.}\ \bibnamefont {Moore}},\ }\href
  {\doibase 10.1103/PhysRevE.81.020101} {\bibfield  {journal} {\bibinfo
  {journal} {Phys. Rev. E}\ }\textbf {\bibinfo {volume} {81}},\ \bibinfo
  {pages} {020101} (\bibinfo {year} {2010})}\BibitemShut {NoStop}%
\bibitem [{\citenamefont {Damski}\ \emph {et~al.}(2011)\citenamefont {Damski},
  \citenamefont {Quan},\ and\ \citenamefont {Zurek}}]{QKZteor-l}%
  \BibitemOpen
  \bibfield  {author} {\bibinfo {author} {\bibfnamefont {B.}~\bibnamefont
  {Damski}}, \bibinfo {author} {\bibfnamefont {H.~T.}\ \bibnamefont {Quan}}, \
  and\ \bibinfo {author} {\bibfnamefont {W.~H.}\ \bibnamefont {Zurek}},\ }\href
  {\doibase 10.1103/PhysRevA.83.062104} {\bibfield  {journal} {\bibinfo
  {journal} {Phys. Rev. A}\ }\textbf {\bibinfo {volume} {83}},\ \bibinfo
  {pages} {062104} (\bibinfo {year} {2011})}\BibitemShut {NoStop}%
\bibitem [{\citenamefont {Zurek}(2013)}]{QKZteor-m}%
  \BibitemOpen
  \bibfield  {author} {\bibinfo {author} {\bibfnamefont {W.~H.}\ \bibnamefont
  {Zurek}},\ }\href {\doibase 10.1088/0953-8984/25/40/404209} {\bibfield
  {journal} {\bibinfo  {journal} {J. Phys. Condens. Matter}\ }\textbf {\bibinfo
  {volume} {25}},\ \bibinfo {pages} {404209} (\bibinfo {year}
  {2013})}\BibitemShut {NoStop}%
\bibitem [{\citenamefont {Sharma}\ \emph {et~al.}(2015)\citenamefont {Sharma},
  \citenamefont {Suzuki},\ and\ \citenamefont {Dutta}}]{QKZteor-n}%
  \BibitemOpen
  \bibfield  {author} {\bibinfo {author} {\bibfnamefont {S.}~\bibnamefont
  {Sharma}}, \bibinfo {author} {\bibfnamefont {S.}~\bibnamefont {Suzuki}}, \
  and\ \bibinfo {author} {\bibfnamefont {A.}~\bibnamefont {Dutta}},\ }\href
  {\doibase 10.1103/PhysRevB.92.104306} {\bibfield  {journal} {\bibinfo
  {journal} {Phys. Rev. B}\ }\textbf {\bibinfo {volume} {92}},\ \bibinfo
  {pages} {104306} (\bibinfo {year} {2015})}\BibitemShut {NoStop}%
\bibitem [{\citenamefont {Dutta}\ and\ \citenamefont
  {Dutta}(2017)}]{QKZteor-o}%
  \BibitemOpen
  \bibfield  {author} {\bibinfo {author} {\bibfnamefont {A.}~\bibnamefont
  {Dutta}}\ and\ \bibinfo {author} {\bibfnamefont {A.}~\bibnamefont {Dutta}},\
  }\href {\doibase 10.1103/PhysRevB.96.125113} {\bibfield  {journal} {\bibinfo
  {journal} {Phys. Rev. B}\ }\textbf {\bibinfo {volume} {96}},\ \bibinfo
  {pages} {125113} (\bibinfo {year} {2017})}\BibitemShut {NoStop}%
\bibitem [{\citenamefont {Sinha}\ \emph {et~al.}(2019)\citenamefont {Sinha},
  \citenamefont {Rams},\ and\ \citenamefont {Dziarmaga}}]{QKZteor-p}%
  \BibitemOpen
  \bibfield  {author} {\bibinfo {author} {\bibfnamefont {A.}~\bibnamefont
  {Sinha}}, \bibinfo {author} {\bibfnamefont {M.~M.}\ \bibnamefont {Rams}}, \
  and\ \bibinfo {author} {\bibfnamefont {J.}~\bibnamefont {Dziarmaga}},\ }\href
  {\doibase 10.1103/PhysRevB.99.094203} {\bibfield  {journal} {\bibinfo
  {journal} {Phys. Rev. B}\ }\textbf {\bibinfo {volume} {99}},\ \bibinfo
  {pages} {094203} (\bibinfo {year} {2019})}\BibitemShut {NoStop}%
\bibitem [{\citenamefont {Rams}\ \emph {et~al.}(2019)\citenamefont {Rams},
  \citenamefont {Dziarmaga},\ and\ \citenamefont {Zurek}}]{QKZteor-q}%
  \BibitemOpen
  \bibfield  {author} {\bibinfo {author} {\bibfnamefont {M.~M.}\ \bibnamefont
  {Rams}}, \bibinfo {author} {\bibfnamefont {J.}~\bibnamefont {Dziarmaga}}, \
  and\ \bibinfo {author} {\bibfnamefont {W.~H.}\ \bibnamefont {Zurek}},\ }\href
  {\doibase 10.1103/PhysRevLett.123.130603} {\bibfield  {journal} {\bibinfo
  {journal} {Phys. Rev. Lett.}\ }\textbf {\bibinfo {volume} {123}},\ \bibinfo
  {pages} {130603} (\bibinfo {year} {2019})}\BibitemShut {NoStop}%
\bibitem [{\citenamefont {Mathey}\ and\ \citenamefont
  {Diehl}(2020)}]{QKZteor-r}%
  \BibitemOpen
  \bibfield  {author} {\bibinfo {author} {\bibfnamefont {S.}~\bibnamefont
  {Mathey}}\ and\ \bibinfo {author} {\bibfnamefont {S.}~\bibnamefont {Diehl}},\
  }\href {\doibase 10.1103/physrevresearch.2.013150} {\bibfield  {journal}
  {\bibinfo  {journal} {Phys. Rev. Research}\ }\textbf {\bibinfo {volume}
  {2}},\ \bibinfo {pages} {013150} (\bibinfo {year} {2020})}\BibitemShut
  {NoStop}%
\bibitem [{\citenamefont {S.}\ and\ \citenamefont
  {Divakaran}(2019)}]{QKZteor-s}%
  \BibitemOpen
  \bibfield  {author} {\bibinfo {author} {\bibfnamefont {R.~B.}\ \bibnamefont
  {S.}}\ and\ \bibinfo {author} {\bibfnamefont {U.}~\bibnamefont {Divakaran}},\
  }\href@noop {} {\enquote {\bibinfo {title} {Adiabatic dynamics of
  quasiperiodic transverse ising model},}\ } (\bibinfo {year} {2019}),\ \Eprint
  {http://arxiv.org/abs/arXiv:1908.01959} {arXiv:1908.01959} \BibitemShut
  {NoStop}%
\bibitem [{\citenamefont {Sen}\ \emph {et~al.}(2008)\citenamefont {Sen},
  \citenamefont {Sengupta},\ and\ \citenamefont {Mondal}}]{QKZteor-t}%
  \BibitemOpen
  \bibfield  {author} {\bibinfo {author} {\bibfnamefont {D.}~\bibnamefont
  {Sen}}, \bibinfo {author} {\bibfnamefont {K.}~\bibnamefont {Sengupta}}, \
  and\ \bibinfo {author} {\bibfnamefont {S.}~\bibnamefont {Mondal}},\ }\href
  {\doibase 10.1103/PhysRevLett.101.016806} {\bibfield  {journal} {\bibinfo
  {journal} {Phys. Rev. Lett.}\ }\textbf {\bibinfo {volume} {101}},\ \bibinfo
  {pages} {016806} (\bibinfo {year} {2008})}\BibitemShut {NoStop}%
\bibitem [{\citenamefont {Shimizu}\ \emph
  {et~al.}(2018{\natexlab{a}})\citenamefont {Shimizu}, \citenamefont {Kuno},
  \citenamefont {Hirano},\ and\ \citenamefont {Ichinose}}]{Keita2018a}%
  \BibitemOpen
  \bibfield  {author} {\bibinfo {author} {\bibfnamefont {K.}~\bibnamefont
  {Shimizu}}, \bibinfo {author} {\bibfnamefont {Y.}~\bibnamefont {Kuno}},
  \bibinfo {author} {\bibfnamefont {T.}~\bibnamefont {Hirano}}, \ and\ \bibinfo
  {author} {\bibfnamefont {I.}~\bibnamefont {Ichinose}},\ }\href {\doibase
  10.1103/PhysRevA.97.033626} {\bibfield  {journal} {\bibinfo  {journal} {Phys.
  Rev. A}\ }\textbf {\bibinfo {volume} {97}},\ \bibinfo {pages} {033626}
  (\bibinfo {year} {2018}{\natexlab{a}})}\BibitemShut {NoStop}%
\bibitem [{\citenamefont {Shimizu}\ \emph
  {et~al.}(2018{\natexlab{b}})\citenamefont {Shimizu}, \citenamefont {Hirano},
  \citenamefont {Park}, \citenamefont {Kuno},\ and\ \citenamefont
  {Ichinose}}]{Keita2018b}%
  \BibitemOpen
  \bibfield  {author} {\bibinfo {author} {\bibfnamefont {K.}~\bibnamefont
  {Shimizu}}, \bibinfo {author} {\bibfnamefont {T.}~\bibnamefont {Hirano}},
  \bibinfo {author} {\bibfnamefont {J.}~\bibnamefont {Park}}, \bibinfo {author}
  {\bibfnamefont {Y.}~\bibnamefont {Kuno}}, \ and\ \bibinfo {author}
  {\bibfnamefont {I.}~\bibnamefont {Ichinose}},\ }\href {\doibase
  10.1103/PhysRevA.98.063603} {\bibfield  {journal} {\bibinfo  {journal} {Phys.
  Rev. A}\ }\textbf {\bibinfo {volume} {98}},\ \bibinfo {pages} {063603}
  (\bibinfo {year} {2018}{\natexlab{b}})}\BibitemShut {NoStop}%
\bibitem [{\citenamefont {Shimizu}\ \emph
  {et~al.}(2018{\natexlab{c}})\citenamefont {Shimizu}, \citenamefont {Hirano},
  \citenamefont {Park}, \citenamefont {Kuno},\ and\ \citenamefont
  {Ichinose}}]{Keita2018c}%
  \BibitemOpen
  \bibfield  {author} {\bibinfo {author} {\bibfnamefont {K.}~\bibnamefont
  {Shimizu}}, \bibinfo {author} {\bibfnamefont {T.}~\bibnamefont {Hirano}},
  \bibinfo {author} {\bibfnamefont {J.}~\bibnamefont {Park}}, \bibinfo {author}
  {\bibfnamefont {Y.}~\bibnamefont {Kuno}}, \ and\ \bibinfo {author}
  {\bibfnamefont {I.}~\bibnamefont {Ichinose}},\ }\href@noop {} {\bibfield
  {journal} {\bibinfo  {journal} {New Journal of Physics}\ }\textbf {\bibinfo
  {volume} {20}},\ \bibinfo {pages} {083006} (\bibinfo {year}
  {2018}{\natexlab{c}})}\BibitemShut {NoStop}%
\bibitem [{\citenamefont {Lamacraft}(2007)}]{Lamacraft2007}%
  \BibitemOpen
  \bibfield  {author} {\bibinfo {author} {\bibfnamefont {A.}~\bibnamefont
  {Lamacraft}},\ }\href {\doibase 10.1103/PhysRevLett.98.160404} {\bibfield
  {journal} {\bibinfo  {journal} {Phys. Rev. Lett.}\ }\textbf {\bibinfo
  {volume} {98}},\ \bibinfo {pages} {160404} (\bibinfo {year}
  {2007})}\BibitemShut {NoStop}%
\bibitem [{\citenamefont {Lee}(2009)}]{Lee2009}%
  \BibitemOpen
  \bibfield  {author} {\bibinfo {author} {\bibfnamefont {C.}~\bibnamefont
  {Lee}},\ }\href {\doibase 10.1103/PhysRevLett.102.070401} {\bibfield
  {journal} {\bibinfo  {journal} {Phys. Rev. Lett.}\ }\textbf {\bibinfo
  {volume} {102}},\ \bibinfo {pages} {070401} (\bibinfo {year}
  {2009})}\BibitemShut {NoStop}%
\bibitem [{\citenamefont {Sabbatini}\ \emph {et~al.}(2011)\citenamefont
  {Sabbatini}, \citenamefont {Zurek},\ and\ \citenamefont
  {Davis}}]{Sabbatini2011}%
  \BibitemOpen
  \bibfield  {author} {\bibinfo {author} {\bibfnamefont {J.}~\bibnamefont
  {Sabbatini}}, \bibinfo {author} {\bibfnamefont {W.~H.}\ \bibnamefont
  {Zurek}}, \ and\ \bibinfo {author} {\bibfnamefont {M.~J.}\ \bibnamefont
  {Davis}},\ }\href {\doibase 10.1103/PhysRevLett.107.230402} {\bibfield
  {journal} {\bibinfo  {journal} {Phys. Rev. Lett.}\ }\textbf {\bibinfo
  {volume} {107}},\ \bibinfo {pages} {230402} (\bibinfo {year}
  {2011})}\BibitemShut {NoStop}%
\bibitem [{\citenamefont {\ifmmode~\acute{S}\else \'{S}\fi{}wis\l{}ocki}\ \emph
  {et~al.}(2013)\citenamefont {\ifmmode~\acute{S}\else \'{S}\fi{}wis\l{}ocki},
  \citenamefont {Witkowska}, \citenamefont {Dziarmaga},\ and\ \citenamefont
  {Matuszewski}}]{Swistock2013}%
  \BibitemOpen
  \bibfield  {author} {\bibinfo {author} {\bibfnamefont {T.}~\bibnamefont
  {\ifmmode~\acute{S}\else \'{S}\fi{}wis\l{}ocki}}, \bibinfo {author}
  {\bibfnamefont {E.}~\bibnamefont {Witkowska}}, \bibinfo {author}
  {\bibfnamefont {J.}~\bibnamefont {Dziarmaga}}, \ and\ \bibinfo {author}
  {\bibfnamefont {M.}~\bibnamefont {Matuszewski}},\ }\href {\doibase
  10.1103/PhysRevLett.110.045303} {\bibfield  {journal} {\bibinfo  {journal}
  {Phys. Rev. Lett.}\ }\textbf {\bibinfo {volume} {110}},\ \bibinfo {pages}
  {045303} (\bibinfo {year} {2013})}\BibitemShut {NoStop}%
\bibitem [{\citenamefont {Saito}\ \emph {et~al.}(2013)\citenamefont {Saito},
  \citenamefont {Kawaguchi},\ and\ \citenamefont {Ueda}}]{Saito2013}%
  \BibitemOpen
  \bibfield  {author} {\bibinfo {author} {\bibfnamefont {H.}~\bibnamefont
  {Saito}}, \bibinfo {author} {\bibfnamefont {Y.}~\bibnamefont {Kawaguchi}}, \
  and\ \bibinfo {author} {\bibfnamefont {M.}~\bibnamefont {Ueda}},\ }\href@noop
  {} {\bibfield  {journal} {\bibinfo  {journal} {Journal of Physics: Condensed
  Matter}\ }\textbf {\bibinfo {volume} {25}},\ \bibinfo {pages} {404212}
  (\bibinfo {year} {2013})}\BibitemShut {NoStop}%
\bibitem [{\citenamefont {Witkowska}\ \emph {et~al.}(2013)\citenamefont
  {Witkowska}, \citenamefont {Dziarmaga}, \citenamefont
  {\ifmmode~\acute{S}\else \'{S}\fi{}wis\l{}ocki},\ and\ \citenamefont
  {Matuszewski}}]{Witkowska2013}%
  \BibitemOpen
  \bibfield  {author} {\bibinfo {author} {\bibfnamefont {E.}~\bibnamefont
  {Witkowska}}, \bibinfo {author} {\bibfnamefont {J.}~\bibnamefont
  {Dziarmaga}}, \bibinfo {author} {\bibfnamefont {T.}~\bibnamefont
  {\ifmmode~\acute{S}\else \'{S}\fi{}wis\l{}ocki}}, \ and\ \bibinfo {author}
  {\bibfnamefont {M.}~\bibnamefont {Matuszewski}},\ }\href {\doibase
  10.1103/PhysRevB.88.054508} {\bibfield  {journal} {\bibinfo  {journal} {Phys.
  Rev. B}\ }\textbf {\bibinfo {volume} {88}},\ \bibinfo {pages} {054508}
  (\bibinfo {year} {2013})}\BibitemShut {NoStop}%
\bibitem [{\citenamefont {Anquez}\ \emph {et~al.}(2016)\citenamefont {Anquez},
  \citenamefont {Robbins}, \citenamefont {Bharath}, \citenamefont
  {Boguslawski}, \citenamefont {Hoang},\ and\ \citenamefont
  {Chapman}}]{QKZexp-b}%
  \BibitemOpen
  \bibfield  {author} {\bibinfo {author} {\bibfnamefont {M.}~\bibnamefont
  {Anquez}}, \bibinfo {author} {\bibfnamefont {B.~A.}\ \bibnamefont {Robbins}},
  \bibinfo {author} {\bibfnamefont {H.~M.}\ \bibnamefont {Bharath}}, \bibinfo
  {author} {\bibfnamefont {M.}~\bibnamefont {Boguslawski}}, \bibinfo {author}
  {\bibfnamefont {T.~M.}\ \bibnamefont {Hoang}}, \ and\ \bibinfo {author}
  {\bibfnamefont {M.~S.}\ \bibnamefont {Chapman}},\ }\href {\doibase
  10.1103/PhysRevLett.116.155301} {\bibfield  {journal} {\bibinfo  {journal}
  {Phys. Rev. Lett.}\ }\textbf {\bibinfo {volume} {116}},\ \bibinfo {pages}
  {155301} (\bibinfo {year} {2016})}\BibitemShut {NoStop}%
\bibitem [{\citenamefont {Baumann}\ \emph {et~al.}(2011)\citenamefont
  {Baumann}, \citenamefont {Mottl}, \citenamefont {Brennecke},\ and\
  \citenamefont {Esslinger}}]{QKZexp-c}%
  \BibitemOpen
  \bibfield  {author} {\bibinfo {author} {\bibfnamefont {K.}~\bibnamefont
  {Baumann}}, \bibinfo {author} {\bibfnamefont {R.}~\bibnamefont {Mottl}},
  \bibinfo {author} {\bibfnamefont {F.}~\bibnamefont {Brennecke}}, \ and\
  \bibinfo {author} {\bibfnamefont {T.}~\bibnamefont {Esslinger}},\ }\href
  {\doibase 10.1103/PhysRevLett.107.140402} {\bibfield  {journal} {\bibinfo
  {journal} {Phys. Rev. Lett.}\ }\textbf {\bibinfo {volume} {107}},\ \bibinfo
  {pages} {140402} (\bibinfo {year} {2011})}\BibitemShut {NoStop}%
\bibitem [{\citenamefont {Clark}\ \emph {et~al.}(2016)\citenamefont {Clark},
  \citenamefont {Feng},\ and\ \citenamefont {Chin}}]{QKZexp-d}%
  \BibitemOpen
  \bibfield  {author} {\bibinfo {author} {\bibfnamefont {L.~W.}\ \bibnamefont
  {Clark}}, \bibinfo {author} {\bibfnamefont {L.}~\bibnamefont {Feng}}, \ and\
  \bibinfo {author} {\bibfnamefont {C.}~\bibnamefont {Chin}},\ }\href {\doibase
  10.1126/science.aaf9657} {\bibfield  {journal} {\bibinfo  {journal}
  {Science}\ }\textbf {\bibinfo {volume} {354}},\ \bibinfo {pages} {606}
  (\bibinfo {year} {2016})}\BibitemShut {NoStop}%
\bibitem [{\citenamefont {Chen}\ \emph {et~al.}(2011)\citenamefont {Chen},
  \citenamefont {White}, \citenamefont {Borries},\ and\ \citenamefont
  {DeMarco}}]{QKZexp-e}%
  \BibitemOpen
  \bibfield  {author} {\bibinfo {author} {\bibfnamefont {D.}~\bibnamefont
  {Chen}}, \bibinfo {author} {\bibfnamefont {M.}~\bibnamefont {White}},
  \bibinfo {author} {\bibfnamefont {C.}~\bibnamefont {Borries}}, \ and\
  \bibinfo {author} {\bibfnamefont {B.}~\bibnamefont {DeMarco}},\ }\href
  {\doibase 10.1103/PhysRevLett.106.235304} {\bibfield  {journal} {\bibinfo
  {journal} {Phys. Rev. Lett.}\ }\textbf {\bibinfo {volume} {106}},\ \bibinfo
  {pages} {235304} (\bibinfo {year} {2011})}\BibitemShut {NoStop}%
\bibitem [{\citenamefont {Braun}\ \emph {et~al.}(2015)\citenamefont {Braun},
  \citenamefont {Friesdorf}, \citenamefont {Hodgman}, \citenamefont
  {Schreiber}, \citenamefont {Ronzheimer}, \citenamefont {Riera}, \citenamefont
  {del Rey}, \citenamefont {Bloch}, \citenamefont {Eisert},\ and\ \citenamefont
  {Schneider}}]{QKZexp-f}%
  \BibitemOpen
  \bibfield  {author} {\bibinfo {author} {\bibfnamefont {S.}~\bibnamefont
  {Braun}}, \bibinfo {author} {\bibfnamefont {M.}~\bibnamefont {Friesdorf}},
  \bibinfo {author} {\bibfnamefont {S.~S.}\ \bibnamefont {Hodgman}}, \bibinfo
  {author} {\bibfnamefont {M.}~\bibnamefont {Schreiber}}, \bibinfo {author}
  {\bibfnamefont {J.~P.}\ \bibnamefont {Ronzheimer}}, \bibinfo {author}
  {\bibfnamefont {A.}~\bibnamefont {Riera}}, \bibinfo {author} {\bibfnamefont
  {M.}~\bibnamefont {del Rey}}, \bibinfo {author} {\bibfnamefont
  {I.}~\bibnamefont {Bloch}}, \bibinfo {author} {\bibfnamefont
  {J.}~\bibnamefont {Eisert}}, \ and\ \bibinfo {author} {\bibfnamefont
  {U.}~\bibnamefont {Schneider}},\ }\href {\doibase 10.1073/pnas.1408861112}
  {\bibfield  {journal} {\bibinfo  {journal} {Proc. Natl. Acad. Sci. U.S.A.}\
  }\textbf {\bibinfo {volume} {112}},\ \bibinfo {pages} {3641} (\bibinfo {year}
  {2015})}\BibitemShut {NoStop}%
\bibitem [{\citenamefont {Gardas}\ \emph {et~al.}(2018)\citenamefont {Gardas},
  \citenamefont {Dziarmaga}, \citenamefont {Zurek},\ and\ \citenamefont
  {Zwolak}}]{QKZexp-g}%
  \BibitemOpen
  \bibfield  {author} {\bibinfo {author} {\bibfnamefont {B.}~\bibnamefont
  {Gardas}}, \bibinfo {author} {\bibfnamefont {J.}~\bibnamefont {Dziarmaga}},
  \bibinfo {author} {\bibfnamefont {W.~H.}\ \bibnamefont {Zurek}}, \ and\
  \bibinfo {author} {\bibfnamefont {M.}~\bibnamefont {Zwolak}},\ }\href
  {\doibase 10.1038/s41598-018-22763-2} {\bibfield  {journal} {\bibinfo
  {journal} {Sci. Rep.}\ }\textbf {\bibinfo {volume} {8}},\ \bibinfo {pages}
  {4539} (\bibinfo {year} {2018})}\BibitemShut {NoStop}%
\bibitem [{\citenamefont {Meldgin}\ \emph {et~al.}(2016)\citenamefont
  {Meldgin}, \citenamefont {Ray}, \citenamefont {Russ}, \citenamefont {Chen},
  \citenamefont {Ceperley},\ and\ \citenamefont {DeMarco}}]{deMarco2}%
  \BibitemOpen
  \bibfield  {author} {\bibinfo {author} {\bibfnamefont {C.}~\bibnamefont
  {Meldgin}}, \bibinfo {author} {\bibfnamefont {U.}~\bibnamefont {Ray}},
  \bibinfo {author} {\bibfnamefont {P.}~\bibnamefont {Russ}}, \bibinfo {author}
  {\bibfnamefont {D.}~\bibnamefont {Chen}}, \bibinfo {author} {\bibfnamefont
  {D.~M.}\ \bibnamefont {Ceperley}}, \ and\ \bibinfo {author} {\bibfnamefont
  {B.}~\bibnamefont {DeMarco}},\ }\href {\doibase 10.1038/nphys3695} {\bibfield
   {journal} {\bibinfo  {journal} {Nat. Phys.}\ }\textbf {\bibinfo {volume}
  {12}},\ \bibinfo {pages} {646} (\bibinfo {year} {2016})}\BibitemShut
  {NoStop}%
\bibitem [{\citenamefont {Keesling}\ \emph {et~al.}(2019)\citenamefont
  {Keesling}, \citenamefont {Omran}, \citenamefont {Levine}, \citenamefont
  {Bernien}, \citenamefont {Pichler}, \citenamefont {Choi}, \citenamefont
  {Samajdar}, \citenamefont {Schwartz}, \citenamefont {Silvi}, \citenamefont
  {Sachdev}, \citenamefont {Zoller}, \citenamefont {Endres}, \citenamefont
  {Greiner}, \citenamefont {Vuletić},\ and\ \citenamefont {Lukin}}]{Lukin18}%
  \BibitemOpen
  \bibfield  {author} {\bibinfo {author} {\bibfnamefont {A.}~\bibnamefont
  {Keesling}}, \bibinfo {author} {\bibfnamefont {A.}~\bibnamefont {Omran}},
  \bibinfo {author} {\bibfnamefont {H.}~\bibnamefont {Levine}}, \bibinfo
  {author} {\bibfnamefont {H.}~\bibnamefont {Bernien}}, \bibinfo {author}
  {\bibfnamefont {H.}~\bibnamefont {Pichler}}, \bibinfo {author} {\bibfnamefont
  {S.}~\bibnamefont {Choi}}, \bibinfo {author} {\bibfnamefont {R.}~\bibnamefont
  {Samajdar}}, \bibinfo {author} {\bibfnamefont {S.}~\bibnamefont {Schwartz}},
  \bibinfo {author} {\bibfnamefont {P.}~\bibnamefont {Silvi}}, \bibinfo
  {author} {\bibfnamefont {S.}~\bibnamefont {Sachdev}}, \bibinfo {author}
  {\bibfnamefont {P.}~\bibnamefont {Zoller}}, \bibinfo {author} {\bibfnamefont
  {M.}~\bibnamefont {Endres}}, \bibinfo {author} {\bibfnamefont
  {M.}~\bibnamefont {Greiner}}, \bibinfo {author} {\bibfnamefont
  {V.}~\bibnamefont {Vuletić}}, \ and\ \bibinfo {author} {\bibfnamefont
  {M.~D.}\ \bibnamefont {Lukin}},\ }\href {\doibase 10.1038/s41586-019-1070-1}
  {\bibfield  {journal} {\bibinfo  {journal} {Nature}\ }\textbf {\bibinfo
  {volume} {568}},\ \bibinfo {pages} {207} (\bibinfo {year}
  {2019})}\BibitemShut {NoStop}%
\bibitem [{\citenamefont {Qiu}\ \emph {et~al.}(2020)\citenamefont {Qiu},
  \citenamefont {Liang}, \citenamefont {Yang}, \citenamefont {Yang},
  \citenamefont {Tian}, \citenamefont {Xu},\ and\ \citenamefont
  {Duan}}]{Qiu2020}%
  \BibitemOpen
  \bibfield  {author} {\bibinfo {author} {\bibfnamefont {L.-Y.}\ \bibnamefont
  {Qiu}}, \bibinfo {author} {\bibfnamefont {H.-Y.}\ \bibnamefont {Liang}},
  \bibinfo {author} {\bibfnamefont {Y.-B.}\ \bibnamefont {Yang}}, \bibinfo
  {author} {\bibfnamefont {H.-X.}\ \bibnamefont {Yang}}, \bibinfo {author}
  {\bibfnamefont {T.}~\bibnamefont {Tian}}, \bibinfo {author} {\bibfnamefont
  {Y.}~\bibnamefont {Xu}}, \ and\ \bibinfo {author} {\bibfnamefont {L.-M.}\
  \bibnamefont {Duan}},\ }\href@noop {} {\bibfield  {journal} {\bibinfo
  {journal} {Science Advances}\ }\textbf {\bibinfo {volume} {6}},\ \bibinfo
  {pages} {eaba7292} (\bibinfo {year} {2020})}\BibitemShut {NoStop}%
\bibitem [{\citenamefont {Kibble}\ and\ \citenamefont
  {Volovik}(1997)}]{inhomo_classical-a}%
  \BibitemOpen
  \bibfield  {author} {\bibinfo {author} {\bibfnamefont {T.~W.~B.}\
  \bibnamefont {Kibble}}\ and\ \bibinfo {author} {\bibfnamefont {G.~E.}\
  \bibnamefont {Volovik}},\ }\href {\doibase 10.1134/1.567332} {\bibfield
  {journal} {\bibinfo  {journal} {JEPT Lett.}\ }\textbf {\bibinfo {volume}
  {65}},\ \bibinfo {pages} {102} (\bibinfo {year} {1997})}\BibitemShut
  {NoStop}%
\bibitem [{\citenamefont {del Campo}\ \emph {et~al.}(2013)\citenamefont {del
  Campo}, \citenamefont {Kibble},\ and\ \citenamefont
  {Zurek}}]{inhomo_classical-f}%
  \BibitemOpen
  \bibfield  {author} {\bibinfo {author} {\bibfnamefont {A.}~\bibnamefont {del
  Campo}}, \bibinfo {author} {\bibfnamefont {T.~W.~B.}\ \bibnamefont {Kibble}},
  \ and\ \bibinfo {author} {\bibfnamefont {W.~H.}\ \bibnamefont {Zurek}},\
  }\href {\doibase 10.1088/0953-8984/25/40/404210} {\bibfield  {journal}
  {\bibinfo  {journal} {J. Phys. Condens. Matter}\ }\textbf {\bibinfo {volume}
  {25}},\ \bibinfo {pages} {404210} (\bibinfo {year} {2013})}\BibitemShut
  {NoStop}%
\bibitem [{\citenamefont {Sadhukhan}\ \emph {et~al.}(2019)\citenamefont
  {Sadhukhan}, \citenamefont {Sinha}, \citenamefont {A.~Francuz}, \citenamefont
  {Rams}, \citenamefont {Dziarmaga},\ and\ \citenamefont {Zurek}}]{sonic}%
  \BibitemOpen
  \bibfield  {author} {\bibinfo {author} {\bibfnamefont {D.}~\bibnamefont
  {Sadhukhan}}, \bibinfo {author} {\bibfnamefont {A.}~\bibnamefont {Sinha}},
  \bibinfo {author} {\bibfnamefont {J.~S.}\ \bibnamefont {A.~Francuz}},
  \bibinfo {author} {\bibfnamefont {M.~M.}\ \bibnamefont {Rams}}, \bibinfo
  {author} {\bibfnamefont {J.}~\bibnamefont {Dziarmaga}}, \ and\ \bibinfo
  {author} {\bibfnamefont {W.~H.}\ \bibnamefont {Zurek}},\ }\href@noop {}
  {\enquote {\bibinfo {title} {Sonic horizons and causality in the phase
  transition dynamics},}\ } (\bibinfo {year} {2019}),\ \Eprint
  {http://arxiv.org/abs/arXiv:1912.02815} {arXiv:1912.02815} \BibitemShut
  {NoStop}%
\bibitem [{\citenamefont {Kolodrubetz}\ \emph {et~al.}(2012)\citenamefont
  {Kolodrubetz}, \citenamefont {Clark},\ and\ \citenamefont
  {Huse}}]{KZscaling1}%
  \BibitemOpen
  \bibfield  {author} {\bibinfo {author} {\bibfnamefont {M.}~\bibnamefont
  {Kolodrubetz}}, \bibinfo {author} {\bibfnamefont {B.~K.}\ \bibnamefont
  {Clark}}, \ and\ \bibinfo {author} {\bibfnamefont {D.~A.}\ \bibnamefont
  {Huse}},\ }\href {\doibase 10.1103/PhysRevLett.109.015701} {\bibfield
  {journal} {\bibinfo  {journal} {Phys. Rev. Lett.}\ }\textbf {\bibinfo
  {volume} {109}},\ \bibinfo {pages} {015701} (\bibinfo {year}
  {2012})}\BibitemShut {NoStop}%
\bibitem [{\citenamefont {Chandran}\ \emph {et~al.}(2012)\citenamefont
  {Chandran}, \citenamefont {Erez}, \citenamefont {Gubser},\ and\ \citenamefont
  {Sondhi}}]{KZscaling2}%
  \BibitemOpen
  \bibfield  {author} {\bibinfo {author} {\bibfnamefont {A.}~\bibnamefont
  {Chandran}}, \bibinfo {author} {\bibfnamefont {A.}~\bibnamefont {Erez}},
  \bibinfo {author} {\bibfnamefont {S.~S.}\ \bibnamefont {Gubser}}, \ and\
  \bibinfo {author} {\bibfnamefont {S.~L.}\ \bibnamefont {Sondhi}},\ }\href
  {\doibase 10.1103/PhysRevB.86.064304} {\bibfield  {journal} {\bibinfo
  {journal} {Phys. Rev. B}\ }\textbf {\bibinfo {volume} {86}},\ \bibinfo
  {pages} {064304} (\bibinfo {year} {2012})}\BibitemShut {NoStop}%
\bibitem [{\citenamefont {Francuz}\ \emph {et~al.}(2016)\citenamefont
  {Francuz}, \citenamefont {Dziarmaga}, \citenamefont {Gardas},\ and\
  \citenamefont {Zurek}}]{Francuzetal}%
  \BibitemOpen
  \bibfield  {author} {\bibinfo {author} {\bibfnamefont {A.}~\bibnamefont
  {Francuz}}, \bibinfo {author} {\bibfnamefont {J.}~\bibnamefont {Dziarmaga}},
  \bibinfo {author} {\bibfnamefont {B.}~\bibnamefont {Gardas}}, \ and\ \bibinfo
  {author} {\bibfnamefont {W.~H.}\ \bibnamefont {Zurek}},\ }\href {\doibase
  10.1103/PhysRevB.93.075134} {\bibfield  {journal} {\bibinfo  {journal} {Phys.
  Rev. B}\ }\textbf {\bibinfo {volume} {93}},\ \bibinfo {pages} {075134}
  (\bibinfo {year} {2016})}\BibitemShut {NoStop}%
\bibitem [{\citenamefont {Bradley}\ \emph {et~al.}(2008)\citenamefont
  {Bradley}, \citenamefont {Gardiner},\ and\ \citenamefont
  {Davis}}]{Bradley2008}%
  \BibitemOpen
  \bibfield  {author} {\bibinfo {author} {\bibfnamefont {A.~S.}\ \bibnamefont
  {Bradley}}, \bibinfo {author} {\bibfnamefont {C.~W.}\ \bibnamefont
  {Gardiner}}, \ and\ \bibinfo {author} {\bibfnamefont {M.~J.}\ \bibnamefont
  {Davis}},\ }\href {\doibase 10.1103/PhysRevA.77.033616} {\bibfield  {journal}
  {\bibinfo  {journal} {Phys. Rev. A}\ }\textbf {\bibinfo {volume} {77}},\
  \bibinfo {pages} {033616} (\bibinfo {year} {2008})}\BibitemShut {NoStop}%
\bibitem [{\citenamefont {Blakie}\ \emph {et~al.}(2008)\citenamefont {Blakie},
  \citenamefont {Bradley}, \citenamefont {Davis}, \citenamefont {Ballagh},\
  and\ \citenamefont {Gardiner}}]{blakie2008dynamics}%
  \BibitemOpen
  \bibfield  {author} {\bibinfo {author} {\bibfnamefont {P.}~\bibnamefont
  {Blakie}}, \bibinfo {author} {\bibfnamefont {A.}~\bibnamefont {Bradley}},
  \bibinfo {author} {\bibfnamefont {M.}~\bibnamefont {Davis}}, \bibinfo
  {author} {\bibfnamefont {R.}~\bibnamefont {Ballagh}}, \ and\ \bibinfo
  {author} {\bibfnamefont {C.}~\bibnamefont {Gardiner}},\ }\href@noop {}
  {\bibfield  {journal} {\bibinfo  {journal} {Advances in Physics}\ }\textbf
  {\bibinfo {volume} {57}},\ \bibinfo {pages} {363} (\bibinfo {year}
  {2008})}\BibitemShut {NoStop}%
\bibitem [{\citenamefont {Proukakis}\ and\ \citenamefont
  {Jackson}(2008)}]{proukakis2008finite}%
  \BibitemOpen
  \bibfield  {author} {\bibinfo {author} {\bibfnamefont {N.~P.}\ \bibnamefont
  {Proukakis}}\ and\ \bibinfo {author} {\bibfnamefont {B.}~\bibnamefont
  {Jackson}},\ }\href@noop {} {\bibfield  {journal} {\bibinfo  {journal} {J.
  Phys. B: At. Mol. Opt. Phys}\ }\textbf {\bibinfo {volume} {41}},\ \bibinfo
  {pages} {203002} (\bibinfo {year} {2008})}\BibitemShut {NoStop}%
\bibitem [{\citenamefont {Rooney}\ \emph {et~al.}(2013)\citenamefont {Rooney},
  \citenamefont {Neely}, \citenamefont {Anderson},\ and\ \citenamefont
  {Bradley}}]{Rooney2013}%
  \BibitemOpen
  \bibfield  {author} {\bibinfo {author} {\bibfnamefont {S.~J.}\ \bibnamefont
  {Rooney}}, \bibinfo {author} {\bibfnamefont {T.~W.}\ \bibnamefont {Neely}},
  \bibinfo {author} {\bibfnamefont {B.~P.}\ \bibnamefont {Anderson}}, \ and\
  \bibinfo {author} {\bibfnamefont {A.~S.}\ \bibnamefont {Bradley}},\ }\href
  {\doibase 10.1103/PhysRevA.88.063620} {\bibfield  {journal} {\bibinfo
  {journal} {Phys. Rev. A}\ }\textbf {\bibinfo {volume} {88}},\ \bibinfo
  {pages} {063620} (\bibinfo {year} {2013})}\BibitemShut {NoStop}%
\bibitem [{\citenamefont {Proukakis}\ \emph {et~al.}(2013)\citenamefont
  {Proukakis}, \citenamefont {Gardiner}, \citenamefont {Davis},\ and\
  \citenamefont {Szymańska}}]{Proukakis13}%
  \BibitemOpen
  \bibinfo {editor} {\bibfnamefont {N.}~\bibnamefont {Proukakis}}, \bibinfo
  {editor} {\bibfnamefont {S.}~\bibnamefont {Gardiner}}, \bibinfo {editor}
  {\bibfnamefont {M.}~\bibnamefont {Davis}}, \ and\ \bibinfo {editor}
  {\bibfnamefont {M.}~\bibnamefont {Szymańska}},\ eds.,\ \href@noop {} {\emph
  {\bibinfo {title} {Quantum Gases: Finite Temperature and Non-Equilibrium
  Dynamics: 1 (Cold Atoms)}}}\ (\bibinfo  {publisher} {ICP},\ \bibinfo {year}
  {2013})\BibitemShut {NoStop}%
\bibitem [{\citenamefont {Berloff}\ \emph {et~al.}(2014)\citenamefont
  {Berloff}, \citenamefont {Brachet},\ and\ \citenamefont
  {Proukakis}}]{berloff_brachet_14}%
  \BibitemOpen
  \bibfield  {author} {\bibinfo {author} {\bibfnamefont {N.~G.}\ \bibnamefont
  {Berloff}}, \bibinfo {author} {\bibfnamefont {M.}~\bibnamefont {Brachet}}, \
  and\ \bibinfo {author} {\bibfnamefont {N.~P.}\ \bibnamefont {Proukakis}},\
  }\href {\doibase 10.1073/pnas.1312549111} {\bibfield  {journal} {\bibinfo
  {journal} {Proceedings of the National Academy of Sciences}\ }\textbf
  {\bibinfo {volume} {111}},\ \bibinfo {pages} {4675} (\bibinfo {year}
  {2014})}\BibitemShut {NoStop}%
\bibitem [{\citenamefont {Brewczyk}\ \emph {et~al.}(2007)\citenamefont
  {Brewczyk}, \citenamefont {Gajda},\ and\ \citenamefont
  {Rza{\.z}ewski}}]{Brewczyk2007}%
  \BibitemOpen
  \bibfield  {author} {\bibinfo {author} {\bibfnamefont {M.}~\bibnamefont
  {Brewczyk}}, \bibinfo {author} {\bibfnamefont {M.}~\bibnamefont {Gajda}}, \
  and\ \bibinfo {author} {\bibfnamefont {K.}~\bibnamefont {Rza{\.z}ewski}},\
  }\href@noop {} {\bibfield  {journal} {\bibinfo  {journal} {Journal of Physics
  B: Atomic, Molecular and Optical Physics}\ }\textbf {\bibinfo {volume}
  {40}},\ \bibinfo {pages} {R1} (\bibinfo {year} {2007})}\BibitemShut {NoStop}%
\bibitem [{\citenamefont {Stoof}\ and\ \citenamefont
  {Bijlsma}(2001)}]{Stoof01}%
  \BibitemOpen
  \bibfield  {author} {\bibinfo {author} {\bibfnamefont {H.~T.~C.}\
  \bibnamefont {Stoof}}\ and\ \bibinfo {author} {\bibfnamefont {M.~J.}\
  \bibnamefont {Bijlsma}},\ }\href {\doibase 10.1023/a:1017519118408}
  {\bibfield  {journal} {\bibinfo  {journal} {Journal of Low Temperature
  Physics}\ }\textbf {\bibinfo {volume} {124}},\ \bibinfo {pages} {431}
  (\bibinfo {year} {2001})}\BibitemShut {NoStop}%
\bibitem [{\citenamefont {Proukakis}(2003)}]{Proukakis03}%
  \BibitemOpen
  \bibfield  {author} {\bibinfo {author} {\bibfnamefont {N.}~\bibnamefont
  {Proukakis}},\ }\href@noop {} {\bibfield  {journal} {\bibinfo  {journal}
  {Las. Phys.}\ }\textbf {\bibinfo {volume} {13}},\ \bibinfo {pages} {527}
  (\bibinfo {year} {2003})}\BibitemShut {NoStop}%
\bibitem [{\citenamefont {Proukakis}\ \emph {et~al.}(2006)\citenamefont
  {Proukakis}, \citenamefont {Schmiedmayer},\ and\ \citenamefont
  {Stoof}}]{Proukakis_Schmiedmayer_2006}%
  \BibitemOpen
  \bibfield  {author} {\bibinfo {author} {\bibfnamefont {N.~P.}\ \bibnamefont
  {Proukakis}}, \bibinfo {author} {\bibfnamefont {J.}~\bibnamefont
  {Schmiedmayer}}, \ and\ \bibinfo {author} {\bibfnamefont {H.~T.~C.}\
  \bibnamefont {Stoof}},\ }\href {\doibase 10.1103/PhysRevA.73.053603}
  {\bibfield  {journal} {\bibinfo  {journal} {Phys. Rev. A}\ }\textbf {\bibinfo
  {volume} {73}},\ \bibinfo {pages} {053603} (\bibinfo {year}
  {2006})}\BibitemShut {NoStop}%
\bibitem [{\citenamefont {Cockburn}\ and\ \citenamefont
  {Proukakis}(2009)}]{Proukakis09}%
  \BibitemOpen
  \bibfield  {author} {\bibinfo {author} {\bibfnamefont {S.~P.}\ \bibnamefont
  {Cockburn}}\ and\ \bibinfo {author} {\bibfnamefont {N.~P.}\ \bibnamefont
  {Proukakis}},\ }\href {\doibase 10.1134/s1054660x09040057} {\bibfield
  {journal} {\bibinfo  {journal} {Laser Physics}\ }\textbf {\bibinfo {volume}
  {19}},\ \bibinfo {pages} {558} (\bibinfo {year} {2009})}\BibitemShut
  {NoStop}%
\bibitem [{\citenamefont {De}\ \emph {et~al.}(2014)\citenamefont {De},
  \citenamefont {Campbell}, \citenamefont {Price}, \citenamefont {Putra},
  \citenamefont {Anderson},\ and\ \citenamefont {Spielman}}]{De_2014}%
  \BibitemOpen
  \bibfield  {author} {\bibinfo {author} {\bibfnamefont {S.}~\bibnamefont
  {De}}, \bibinfo {author} {\bibfnamefont {D.~L.}\ \bibnamefont {Campbell}},
  \bibinfo {author} {\bibfnamefont {R.~M.}\ \bibnamefont {Price}}, \bibinfo
  {author} {\bibfnamefont {A.}~\bibnamefont {Putra}}, \bibinfo {author}
  {\bibfnamefont {B.~M.}\ \bibnamefont {Anderson}}, \ and\ \bibinfo {author}
  {\bibfnamefont {I.~B.}\ \bibnamefont {Spielman}},\ }\href {\doibase
  10.1103/PhysRevA.89.033631} {\bibfield  {journal} {\bibinfo  {journal} {Phys.
  Rev. A}\ }\textbf {\bibinfo {volume} {89}},\ \bibinfo {pages} {033631}
  (\bibinfo {year} {2014})}\BibitemShut {NoStop}%
\bibitem [{\citenamefont {Gallucci}\ and\ \citenamefont
  {Proukakis}(2016)}]{gallucci2016engineering}%
  \BibitemOpen
  \bibfield  {author} {\bibinfo {author} {\bibfnamefont {D.}~\bibnamefont
  {Gallucci}}\ and\ \bibinfo {author} {\bibfnamefont {N.}~\bibnamefont
  {Proukakis}},\ }\href@noop {} {\bibfield  {journal} {\bibinfo  {journal} {New
  J. Phys.}\ }\textbf {\bibinfo {volume} {18}},\ \bibinfo {pages} {025004}
  (\bibinfo {year} {2016})}\BibitemShut {NoStop}%
\bibitem [{\citenamefont {{Kobayashi, Michikazu}}\ and\ \citenamefont
  {{Cugliandolo, Leticia F.}}(2016)}]{Kobayashi_16a}%
  \BibitemOpen
  \bibfield  {author} {\bibinfo {author} {\bibnamefont {{Kobayashi,
  Michikazu}}}\ and\ \bibinfo {author} {\bibnamefont {{Cugliandolo, Leticia
  F.}}},\ }\href {\doibase 10.1209/0295-5075/115/20007} {\bibfield  {journal}
  {\bibinfo  {journal} {EPL}\ }\textbf {\bibinfo {volume} {115}},\ \bibinfo
  {pages} {20007} (\bibinfo {year} {2016})}\BibitemShut {NoStop}%
\bibitem [{\citenamefont {Kobayashi}\ and\ \citenamefont
  {Cugliandolo}(2016)}]{Kobayashi_16b}%
  \BibitemOpen
  \bibfield  {author} {\bibinfo {author} {\bibfnamefont {M.}~\bibnamefont
  {Kobayashi}}\ and\ \bibinfo {author} {\bibfnamefont {L.~F.}\ \bibnamefont
  {Cugliandolo}},\ }\href {\doibase 10.1103/PhysRevE.94.062146} {\bibfield
  {journal} {\bibinfo  {journal} {Phys. Rev. E}\ }\textbf {\bibinfo {volume}
  {94}},\ \bibinfo {pages} {062146} (\bibinfo {year} {2016})}\BibitemShut
  {NoStop}%
\bibitem [{\citenamefont {Eckel}\ \emph {et~al.}(2018)\citenamefont {Eckel},
  \citenamefont {Kumar}, \citenamefont {Jacobson}, \citenamefont {Spielman},\
  and\ \citenamefont {Campbell}}]{Eckel_2018}%
  \BibitemOpen
  \bibfield  {author} {\bibinfo {author} {\bibfnamefont {S.}~\bibnamefont
  {Eckel}}, \bibinfo {author} {\bibfnamefont {A.}~\bibnamefont {Kumar}},
  \bibinfo {author} {\bibfnamefont {T.}~\bibnamefont {Jacobson}}, \bibinfo
  {author} {\bibfnamefont {I.~B.}\ \bibnamefont {Spielman}}, \ and\ \bibinfo
  {author} {\bibfnamefont {G.~K.}\ \bibnamefont {Campbell}},\ }\href {\doibase
  10.1103/PhysRevX.8.021021} {\bibfield  {journal} {\bibinfo  {journal} {Phys.
  Rev. X}\ }\textbf {\bibinfo {volume} {8}},\ \bibinfo {pages} {021021}
  (\bibinfo {year} {2018})}\BibitemShut {NoStop}%
\bibitem [{\citenamefont {Comaron}\ \emph {et~al.}(2019)\citenamefont
  {Comaron}, \citenamefont {Larcher}, \citenamefont {Dalfovo},\ and\
  \citenamefont {Proukakis}}]{comaron2019quench}%
  \BibitemOpen
  \bibfield  {author} {\bibinfo {author} {\bibfnamefont {P.}~\bibnamefont
  {Comaron}}, \bibinfo {author} {\bibfnamefont {F.}~\bibnamefont {Larcher}},
  \bibinfo {author} {\bibfnamefont {F.}~\bibnamefont {Dalfovo}}, \ and\
  \bibinfo {author} {\bibfnamefont {N.~P.}\ \bibnamefont {Proukakis}},\ }\href
  {\doibase 10.1103/PhysRevA.100.033618} {\bibfield  {journal} {\bibinfo
  {journal} {Phys. Rev. A}\ }\textbf {\bibinfo {volume} {100}},\ \bibinfo
  {pages} {033618} (\bibinfo {year} {2019})}\BibitemShut {NoStop}%
\bibitem [{\citenamefont {Ota}\ \emph {et~al.}(2018)\citenamefont {Ota},
  \citenamefont {Larcher}, \citenamefont {Dalfovo}, \citenamefont {Pitaevskii},
  \citenamefont {Proukakis},\ and\ \citenamefont {Stringari}}]{Ota_2018}%
  \BibitemOpen
  \bibfield  {author} {\bibinfo {author} {\bibfnamefont {M.}~\bibnamefont
  {Ota}}, \bibinfo {author} {\bibfnamefont {F.}~\bibnamefont {Larcher}},
  \bibinfo {author} {\bibfnamefont {F.}~\bibnamefont {Dalfovo}}, \bibinfo
  {author} {\bibfnamefont {L.}~\bibnamefont {Pitaevskii}}, \bibinfo {author}
  {\bibfnamefont {N.~P.}\ \bibnamefont {Proukakis}}, \ and\ \bibinfo {author}
  {\bibfnamefont {S.}~\bibnamefont {Stringari}},\ }\href {\doibase
  10.1103/PhysRevLett.121.145302} {\bibfield  {journal} {\bibinfo  {journal}
  {Phys. Rev. Lett.}\ }\textbf {\bibinfo {volume} {121}},\ \bibinfo {pages}
  {145302} (\bibinfo {year} {2018})}\BibitemShut {NoStop}%
\bibitem [{\citenamefont {Comaron}\ \emph {et~al.}(2018)\citenamefont
  {Comaron}, \citenamefont {Dagvadorj}, \citenamefont {Zamora}, \citenamefont
  {Carusotto}, \citenamefont {Proukakis},\ and\ \citenamefont
  {Szyma\ifmmode~\acute{n}\else \'{n}\fi{}ska}}]{comaron_18}%
  \BibitemOpen
  \bibfield  {author} {\bibinfo {author} {\bibfnamefont {P.}~\bibnamefont
  {Comaron}}, \bibinfo {author} {\bibfnamefont {G.}~\bibnamefont {Dagvadorj}},
  \bibinfo {author} {\bibfnamefont {A.}~\bibnamefont {Zamora}}, \bibinfo
  {author} {\bibfnamefont {I.}~\bibnamefont {Carusotto}}, \bibinfo {author}
  {\bibfnamefont {N.~P.}\ \bibnamefont {Proukakis}}, \ and\ \bibinfo {author}
  {\bibfnamefont {M.~H.}\ \bibnamefont {Szyma\ifmmode~\acute{n}\else
  \'{n}\fi{}ska}},\ }\href {\doibase 10.1103/PhysRevLett.121.095302} {\bibfield
   {journal} {\bibinfo  {journal} {Phys. Rev. Lett.}\ }\textbf {\bibinfo
  {volume} {121}},\ \bibinfo {pages} {095302} (\bibinfo {year}
  {2018})}\BibitemShut {NoStop}%
\bibitem [{\citenamefont {Cockburn}\ and\ \citenamefont
  {Proukakis}(2012)}]{Cockburn_2012}%
  \BibitemOpen
  \bibfield  {author} {\bibinfo {author} {\bibfnamefont {S.~P.}\ \bibnamefont
  {Cockburn}}\ and\ \bibinfo {author} {\bibfnamefont {N.~P.}\ \bibnamefont
  {Proukakis}},\ }\href {\doibase 10.1103/PhysRevA.86.033610} {\bibfield
  {journal} {\bibinfo  {journal} {Phys. Rev. A}\ }\textbf {\bibinfo {volume}
  {86}},\ \bibinfo {pages} {033610} (\bibinfo {year} {2012})}\BibitemShut
  {NoStop}%
\bibitem [{Fer()}]{Ferrari_private}%
  \BibitemOpen
  \href@noop {} {}\bibinfo {note} {G. Ferrari and G. Lamporesi, private
  communication}\BibitemShut {NoStop}%
\bibitem [{\citenamefont {Rooney}\ \emph {et~al.}(2010)\citenamefont {Rooney},
  \citenamefont {Bradley},\ and\ \citenamefont {Blakie}}]{Rooney2010}%
  \BibitemOpen
  \bibfield  {author} {\bibinfo {author} {\bibfnamefont {S.~J.}\ \bibnamefont
  {Rooney}}, \bibinfo {author} {\bibfnamefont {A.~S.}\ \bibnamefont {Bradley}},
  \ and\ \bibinfo {author} {\bibfnamefont {P.~B.}\ \bibnamefont {Blakie}},\
  }\href {\doibase 10.1103/PhysRevA.81.023630} {\bibfield  {journal} {\bibinfo
  {journal} {Phys. Rev. A}\ }\textbf {\bibinfo {volume} {81}},\ \bibinfo
  {pages} {023630} (\bibinfo {year} {2010})}\BibitemShut {NoStop}%
\bibitem [{ncu()}]{ncut}%
  \BibitemOpen
  \href@noop {} {}\bibinfo {note} {Following Ref.~\cite{Rooney2010}, the cutoff
  occupation number in the context of the self-consistent HF estimation was
  taken as $n_{\rm cut}=2$.}\BibitemShut {Stop}%
\bibitem [{\citenamefont {Andersen}(2004)}]{Andersen2004}%
  \BibitemOpen
  \bibfield  {author} {\bibinfo {author} {\bibfnamefont {J.~O.}\ \bibnamefont
  {Andersen}},\ }\href {\doibase 10.1103/RevModPhys.76.599} {\bibfield
  {journal} {\bibinfo  {journal} {Rev. Mod. Phys.}\ }\textbf {\bibinfo {volume}
  {76}},\ \bibinfo {pages} {599} (\bibinfo {year} {2004})}\BibitemShut
  {NoStop}%
\bibitem [{\citenamefont {Giorgini}\ \emph {et~al.}(1996)\citenamefont
  {Giorgini}, \citenamefont {Pitaevskii},\ and\ \citenamefont
  {Stringari}}]{Giorgini1996}%
  \BibitemOpen
  \bibfield  {author} {\bibinfo {author} {\bibfnamefont {S.}~\bibnamefont
  {Giorgini}}, \bibinfo {author} {\bibfnamefont {L.~P.}\ \bibnamefont
  {Pitaevskii}}, \ and\ \bibinfo {author} {\bibfnamefont {S.}~\bibnamefont
  {Stringari}},\ }\href {\doibase 10.1103/PhysRevA.54.R4633} {\bibfield
  {journal} {\bibinfo  {journal} {Phys. Rev. A}\ }\textbf {\bibinfo {volume}
  {54}},\ \bibinfo {pages} {R4633} (\bibinfo {year} {1996})}\BibitemShut
  {NoStop}%
\bibitem [{\citenamefont {Bezett}\ and\ \citenamefont
  {Blakie}(2009)}]{Bezett2009}%
  \BibitemOpen
  \bibfield  {author} {\bibinfo {author} {\bibfnamefont {A.}~\bibnamefont
  {Bezett}}\ and\ \bibinfo {author} {\bibfnamefont {P.~B.}\ \bibnamefont
  {Blakie}},\ }\href {\doibase 10.1103/PhysRevA.79.033611} {\bibfield
  {journal} {\bibinfo  {journal} {Phys. Rev. A}\ }\textbf {\bibinfo {volume}
  {79}},\ \bibinfo {pages} {033611} (\bibinfo {year} {2009})}\BibitemShut
  {NoStop}%
\bibitem [{ave()}]{average_time}%
  \BibitemOpen
  \href@noop {} {}\bibinfo {note} {The averaging duration has been selected as
  $\approx0.72\hbar/\gamma\mu_f \approx 8$ms for the dynamical simulations,
  which is much longer than our numerical time discretization $dt$, and shorter
  than the timescale on which vortices and other excitations evolve, so that
  their dynamics can still be efficiently tracked~\cite{KZexp-w}. The longer
  averaging period of $100\hbar/\gamma\mu$ has been used for the equilibrium
  simulations.}\BibitemShut {Stop}%
\bibitem [{\citenamefont {Damle}\ \emph {et~al.}(1996)\citenamefont {Damle},
  \citenamefont {Senthil}, \citenamefont {Majumdar},\ and\ \citenamefont
  {Sachdev}}]{Damle1996}%
  \BibitemOpen
  \bibfield  {author} {\bibinfo {author} {\bibfnamefont {K.}~\bibnamefont
  {Damle}}, \bibinfo {author} {\bibfnamefont {T.}~\bibnamefont {Senthil}},
  \bibinfo {author} {\bibfnamefont {S.~N.}\ \bibnamefont {Majumdar}}, \ and\
  \bibinfo {author} {\bibfnamefont {S.}~\bibnamefont {Sachdev}},\ }\href
  {\doibase 10.1209/epl/i1996-00179-4} {\bibfield  {journal} {\bibinfo
  {journal} {Europhysics Letters ({EPL})}\ }\textbf {\bibinfo {volume} {36}},\
  \bibinfo {pages} {7} (\bibinfo {year} {1996})}\BibitemShut {NoStop}%
\bibitem [{\citenamefont {Dalfovo}\ \emph {et~al.}(1999)\citenamefont
  {Dalfovo}, \citenamefont {Giorgini}, \citenamefont {Pitaevskii},\ and\
  \citenamefont {Stringari}}]{Dalfovo1999}%
  \BibitemOpen
  \bibfield  {author} {\bibinfo {author} {\bibfnamefont {F.}~\bibnamefont
  {Dalfovo}}, \bibinfo {author} {\bibfnamefont {S.}~\bibnamefont {Giorgini}},
  \bibinfo {author} {\bibfnamefont {L.~P.}\ \bibnamefont {Pitaevskii}}, \ and\
  \bibinfo {author} {\bibfnamefont {S.}~\bibnamefont {Stringari}},\ }\href
  {\doibase 10.1103/RevModPhys.71.463} {\bibfield  {journal} {\bibinfo
  {journal} {Rev. Mod. Phys.}\ }\textbf {\bibinfo {volume} {71}},\ \bibinfo
  {pages} {463} (\bibinfo {year} {1999})}\BibitemShut {NoStop}%
\bibitem [{\citenamefont {Campostrini}\ \emph {et~al.}(2001)\citenamefont
  {Campostrini}, \citenamefont {Hasenbusch}, \citenamefont {Pelissetto},
  \citenamefont {Rossi},\ and\ \citenamefont {Vicari}}]{Campostrini2001}%
  \BibitemOpen
  \bibfield  {author} {\bibinfo {author} {\bibfnamefont {M.}~\bibnamefont
  {Campostrini}}, \bibinfo {author} {\bibfnamefont {M.}~\bibnamefont
  {Hasenbusch}}, \bibinfo {author} {\bibfnamefont {A.}~\bibnamefont
  {Pelissetto}}, \bibinfo {author} {\bibfnamefont {P.}~\bibnamefont {Rossi}}, \
  and\ \bibinfo {author} {\bibfnamefont {E.}~\bibnamefont {Vicari}},\ }\href
  {\doibase 10.1103/PhysRevB.63.214503} {\bibfield  {journal} {\bibinfo
  {journal} {Phys. Rev. B}\ }\textbf {\bibinfo {volume} {63}},\ \bibinfo
  {pages} {214503} (\bibinfo {year} {2001})}\BibitemShut {NoStop}%
\bibitem [{\citenamefont {Davis}\ and\ \citenamefont
  {Morgan}(2003)}]{Davis2003}%
  \BibitemOpen
  \bibfield  {author} {\bibinfo {author} {\bibfnamefont {M.~J.}\ \bibnamefont
  {Davis}}\ and\ \bibinfo {author} {\bibfnamefont {S.~A.}\ \bibnamefont
  {Morgan}},\ }\href {\doibase 10.1103/PhysRevA.68.053615} {\bibfield
  {journal} {\bibinfo  {journal} {Phys. Rev. A}\ }\textbf {\bibinfo {volume}
  {68}},\ \bibinfo {pages} {053615} (\bibinfo {year} {2003})}\BibitemShut
  {NoStop}%
\bibitem [{\citenamefont {Davis}\ and\ \citenamefont
  {Blakie}(2006)}]{Davis2006}%
  \BibitemOpen
  \bibfield  {author} {\bibinfo {author} {\bibfnamefont {M.~J.}\ \bibnamefont
  {Davis}}\ and\ \bibinfo {author} {\bibfnamefont {P.~B.}\ \bibnamefont
  {Blakie}},\ }\href {\doibase 10.1103/PhysRevLett.96.060404} {\bibfield
  {journal} {\bibinfo  {journal} {Phys. Rev. Lett.}\ }\textbf {\bibinfo
  {volume} {96}},\ \bibinfo {pages} {060404} (\bibinfo {year}
  {2006})}\BibitemShut {NoStop}%
\bibitem [{\citenamefont {Wright}\ \emph {et~al.}(2011)\citenamefont {Wright},
  \citenamefont {Proukakis},\ and\ \citenamefont
  {Davis}}]{wright_proukakis_11}%
  \BibitemOpen
  \bibfield  {author} {\bibinfo {author} {\bibfnamefont {T.~M.}\ \bibnamefont
  {Wright}}, \bibinfo {author} {\bibfnamefont {N.~P.}\ \bibnamefont
  {Proukakis}}, \ and\ \bibinfo {author} {\bibfnamefont {M.~J.}\ \bibnamefont
  {Davis}},\ }\href {\doibase 10.1103/PhysRevA.84.023608} {\bibfield  {journal}
  {\bibinfo  {journal} {Phys. Rev. A}\ }\textbf {\bibinfo {volume} {84}},\
  \bibinfo {pages} {023608} (\bibinfo {year} {2011})}\BibitemShut {NoStop}%
\bibitem [{\citenamefont {Bray}(2002)}]{POK}%
  \BibitemOpen
  \bibfield  {author} {\bibinfo {author} {\bibfnamefont {A.~J.}\ \bibnamefont
  {Bray}},\ }\href {\doibase 10.1080/00018730110117433} {\bibfield  {journal}
  {\bibinfo  {journal} {Adv. Phys.}\ }\textbf {\bibinfo {volume} {51}},\
  \bibinfo {pages} {481} (\bibinfo {year} {2002})}\BibitemShut {NoStop}%
\bibitem [{dat()}]{data_open}%
  \BibitemOpen
  \href {http://dx.doi.org/10.25405/data.ncl.12604721} {}\bibinfo {note} {I.-K.
  Liu, J. Dziarmaga, S.-C. Gou, F. Dalfovo,and N. P.Proukakis, Data supporting
  publication: Kibble-Zurek Dynamics in a Trapped Ultracold Bose Gas (2020),
  \href{http://dx.doi.org/10.25405/data.ncl.12604721}{
  {http://dx.doi.org/10.25405/data.ncl.12604721}}}\BibitemShut {NoStop}%
\bibitem [{\citenamefont {Larcher}(2018)}]{Fabrizio2018}%
  \BibitemOpen
  \bibfield  {author} {\bibinfo {author} {\bibfnamefont {F.}~\bibnamefont
  {Larcher}},\ }\href@noop {} { {\bibinfo {title} {Phd thesis,
  {U}niversity of {T}rento and {N}ewcastle {U}niversity}}\ } (\bibinfo {year}
  {2018})\BibitemShut {NoStop}%
\bibitem [{\citenamefont {Donner}\ \emph {et~al.}(2007)\citenamefont {Donner},
  \citenamefont {Ritter}, \citenamefont {Bourdel}, \citenamefont {Ottl},
  \citenamefont {Kohl},\ and\ \citenamefont {Esslinger}}]{Donner2007}%
  \BibitemOpen
  \bibfield  {author} {\bibinfo {author} {\bibfnamefont {T.}~\bibnamefont
  {Donner}}, \bibinfo {author} {\bibfnamefont {S.}~\bibnamefont {Ritter}},
  \bibinfo {author} {\bibfnamefont {T.}~\bibnamefont {Bourdel}}, \bibinfo
  {author} {\bibfnamefont {A.}~\bibnamefont {Ottl}}, \bibinfo {author}
  {\bibfnamefont {M.}~\bibnamefont {Kohl}}, \ and\ \bibinfo {author}
  {\bibfnamefont {T.}~\bibnamefont {Esslinger}},\ }\href {\doibase
  10.1126/science.1138807} {\bibfield  {journal} {\bibinfo  {journal}
  {Science}\ }\textbf {\bibinfo {volume} {315}},\ \bibinfo {pages} {1556}
  (\bibinfo {year} {2007})}\BibitemShut {NoStop}%
\end{thebibliography}
%


\end{document}